\documentclass[12pt,journal,draftcls,onecolumn]{IEEEtran}
\usepackage{amsmath,amssymb,mathrsfs}
\usepackage{epsfig,epsf,subfigure,graphicx,graphics}
\usepackage{url}

\newtheorem{lemma}{Lemma}[section]
\newtheorem{theorem}[lemma]{Theorem}

\newtheorem{corollary}[lemma]{Corollary}

\newtheorem{remark}[lemma]{Remark}
\newtheorem{claim}[lemma]{Claim}
\newtheorem{fact}[lemma]{Fact}

\newcommand{\SNR}{\mathsf{SNR}}
\newcommand{\INR}{\mathsf{INR}}
\newcommand{\C}{\mathsf{C}^\mathsf{B}}

\newcommand{\Var}{\mathrm{Var}}

\newcommand{\aaaa}{\mathrm{(a)}}
\newcommand{\bbbb}{\mathrm{(b)}}
\newcommand{\cccc}{\mathrm{(c)}}

\newcommand{\lp}{\left(}
\newcommand{\rp}{\right)}
\newcommand{\lb}{\left[}
\newcommand{\rb}{\right]}
\newcommand{\lbp}{\left\{}
\newcommand{\rbp}{\right\}}
\newcommand{\ul}{\underline}
\newcommand{\ol}{\overline}
\newcommand{\mcal}{\mathcal}

\newcommand{\what}{\widehat}
\newcommand{\wtild}{\widetilde}

\newcommand{\etal}{{\it et al.}}

\title{Interference Mitigation through Limited Transmitter Cooperation}
\author{
\authorblockN{I-Hsiang Wang and David N. C. Tse}\\
\authorblockA{Wireless Foundations\\
University of California at Berkeley,\\
Berkeley, California 94720, USA\\
\textsf{\{ihsiang, dtse\}@eecs.berkeley.edu}}
\thanks{This work was supported by National Science Foundation under grant \# CCF-0830796 and a gift from Qualcomm Corporate.}
}

\begin{document}
\maketitle
\begin{abstract}
Interference limits performance in wireless networks, and cooperation among receivers or transmitters can help mitigate interference by forming distributed MIMO systems. Earlier work \cite{WangTse_09} shows how limited receiver cooperation helps mitigate interference. The scenario with transmitter cooperation, however, is more difficult to tackle. In this paper we study the two-user Gaussian interference channel with conferencing transmitters to make progress towards this direction. We characterize the capacity region to within $6.5$ bits/s/Hz, regardless of channel parameters. Based on the constant-to-optimality result, we show that there is an interesting reciprocity between the scenario with conferencing transmitters and the scenario with conferencing receivers, and their capacity regions are within a constant gap to each other. Hence in the interference-limited regime, the behavior of the benefit brought by transmitter cooperation is the same as that by receiver cooperation.
\end{abstract}

\section{Introduction}\label{sec_Intro}
In modern wireless communication systems and wireless networks, interference has become the major factor that limits the performance. Interference arises whenever multiple transmitter-receiver pairs are present, and each receiver is only interested in retrieving information from its own transmitter. Due to the broadcast and superposition nature of wireless channels, one user's information-carrying signal causes interference to other users. \emph{Interference channel} is the simplest information theoretic model for studying this issue, where each transmitter (receiver) is assumed to be isolated from other transmitters (receivers). In various practical scenarios, however, they are not isolated, and \emph{cooperation} among transmitters or receivers can be induced. For example, in downlink cellular systems, base stations are connected via infrastructure backhaul networks. 

In our previous work \cite{WangTse_09}, we have studied the two-user Gaussian interference channel with conferencing receivers to understand how limited receiver cooperation helps mitigate interference. We propose good coding strategies, prove tight outer bounds, and characterize the capacity region to within $2$ bits/s/Hz. Based upon the constant-gap-to-optimality result, we identify two regions regarding the gain from receiver cooperation: linear and saturation regions. In the linear region, receiver cooperation is \emph{efficient}, in the sense that the growth of user data rate is roughly linear with respect to the capacity of receiver-cooperative links. The gain in this region is the \emph{degrees-of-freedom} gain that distributed MIMO systems provide. In the saturation region, receiver cooperation is \emph{inefficient} in the sense that the growth of user data rate becomes saturated as one increases the rate in receiver-cooperative links. The gain is the \emph{power} gain of at most a constant number of bits, independent of the channel strength. Furthermore, until saturation the degree-of-freedom gain is either \emph{one cooperation bit buys one more bit} or \emph{two cooperation bits buy one more bit}. 

In this paper, we study its reciprocal problem, Gaussian interference channel with conferencing transmitters, to investigate how limited transmitter cooperation helps mitigate interference. A natural cooperative strategy between transmitter is that, prior to each block of transmission, two transmitters hold a conference to tell each other part of their messages. Hence the messages are classified into two kinds: (1) \emph{cooperative} messages, which are those known to both transmitters due to the conference, and (2) \emph{noncooperative} ones, which are those unknown to the other transmitter since the cooperative link capacities are finite. On the other hand, messages can also be classified based on their target receivers: (1) \emph{common} messages, which are those aimed at both receivers, and (2) \emph{private} ones, which are those aimed at its own receiver. Hence in total there are four kinds of messages for each user, and seven codes for the whole system\footnote{There is only one cooperative common code carrying both cooperative common messages.}. Now the question is, how do we encode these messages?

Generally speaking, Gaussian interference channel with transmitter cooperation is more difficult to tackle than Gaussian interference channel with receiver cooperation. Take the following extreme case. When transmitters can cooperate in an unlimited fashion, the scenario reduces to MIMO Gaussian broadcast channel. When receivers can cooperate in an unlimited fashion, the scenario reduces to MIMO Gaussian multiple access channel. The capacity region of the latter is fully characterized in the 70's \cite{Ahlswede_71} \cite{Liao_72}, while that of the former has not been solved until recently \cite{WeingartenSteinberg_06}. This is due to difficulties both in achievability and outer bounds.

Similar phenomenon arises between Gaussian interference channels with conferencing transmitters and Gaussian interference channels with conferencing receivers. Compared with the scenario with conferencing receivers \cite{WangTse_09} where each user just has two kinds of messages (common and private), in the scenario with conferencing transmitters not only does the message structure in the strategy become more complicated due to the collaboration among transmitters, but it is also more difficult to prove the outer bounds since the transmitters are potentially correlated. In order to overcome the difficulties, we first study an auxiliary problem in the \emph{linear deterministic} setting \cite{AvestimehrDiggavi_07} \cite{AvestimehrDiggavi_09}. We first characterize the capacity region of the linear deterministic interference channel with conferencing transmitters, and then make use of the intuition there to design good coding strategies and prove outer bounds in the Gaussian scenario. Eventually the proposed strategy in the Gaussian setting is a simple superposition of a pair of \emph{noncooperative} common and private codewords and a pair of \emph{cooperative} common and private codewords. For the noncooperative part, Han-Kobayashi scheme \cite{HanKobayashi_81} is employed, and the common-private split is such that the private interference is at or below the noise level at the unintended receiver \cite{EtkinTse_07}. For the cooperative part, we use a simple linear beamforming strategy for encoding the private messages, superimposed upon the common codewords. By choosing the power split and beamforming vectors cleverly, the strategy achieves the capacity region universally to within $6.5$ bits, regardless of channel parameters. The $6.5$-bit gap is the worst-case gap which can be loose in some regimes, and it is vanishingly small at high $\SNR$ when compared to the capacity. 

With the constant-gap-to-optimality result, we observe an interesting uplink-downlink reciprocity between the scenario with conferencing receivers and the scenario with conferencing transmitters: for the original and reciprocal channels, the capacity regions are within a constant gap to each other. Hence the fundamental gain from transmitter cooperation at high $\SNR$ is the same as that from receiver cooperation \cite{WangTse_09}.

\subsection*{Related Works}
Conferencing among transmitters is first studied by Willems \cite{Willems_83} in the context of \emph{multiple access channels}, where the capacity region is characterized. The capacity of Gaussian MAC with conferencing transmitters, however, has not been characterized explicitly in a computable form until recently by Bross \etal \cite{BrossLapidoth_08}, where the authors show that the optimization on auxiliary random variables can be reduced to finding optimal Gaussian input distribution. On the other hand, the extension to compound MAC has been done by Mari\'{c} \etal \cite{MaricYates_07}.

Works on Gaussian \emph{interference channel} with transmitter cooperation can be roughly divided into two categories. One set of works investigate cooperation in interference channels with a set-up where the cooperative links share the same band as the links in the interference channel. H$\o$st-Madsen \cite{Host-Madsen_06} proposes cooperative strategies based on decode-forward, compress-forward, and dirty paper coding, and derives the achievable rates. The recent work by Prabhakaran \etal \cite{PrabhakaranViswanathSRC_09} characterizes the sum capacity of Gaussian interference channels with reciprocal in-band transmitter cooperation to within a constant gap. The other set of works focus on conferencing transmitters, that is, cooperative links are orthogonal to each other as well as the links in the interference channel. Some works are dedicated to achievable rates. Cao \etal \cite{CaoChen_07} derive an achievable rate region based on superposition coding and dirty paper coding. Some works consider special cases of the channel. One such special case attracting particularly broad interest is the \emph{cognitive interference channel}, where one of the transmitters (the cognitive user) is assumed to have full knowledge about the other's transmission (the primary user). It is equivalent to the case where transmitter cooperation is unidirectional and unlimited. As for cognitive interference channel, Mari\'{c} \etal \cite{MaricYates_07} characterize the capacity region in strong interference regime. Wu \etal \cite{WuVishwanath_07} and Jovi\v{c}i\'{c} \etal \cite{JovicicViswanath_09} independently characterize the capacity region when the interference at the primary receiver is weak. Very recently, Rini \etal \cite{RiniTuninetti_10} characterize the capacity region to within $1.87$ bits universally, regardless of channel parameters. On the other hand, works on the case with limited cooperative capacities are not rich in the literature. Bagheri \etal \cite{BagheriMotahari_09} investigate symmetric Gaussian interference channel with unidirectional limited transmitter cooperation, and characterize the sum capacity to within two bits. 

Our main contribution in this paper is characterizing the capacity \emph{region} of Gaussian interference channel with conferencing transmitters to within a constant gap for \emph{arbitrary} channel strength and cooperative link capacities. The rest of the paper is organized as follows. After we formulate the problem in Section \ref{sec_Formulation}, we investigate the auxiliary linear deterministic channel in Section \ref{sec_LDC}. Then we carry the intuitions and techniques to solve the original problem in Section \ref{sec_GIC} and characterize the capacity region to within a constant gap. In Section \ref{sec_Reciprocity} we discuss the interesting uplink-downlink reciprocity.

\section{Problem Formulation}\label{sec_Formulation}
\subsection{Channel Model}
\begin{figure}[htbp]
{\center
\includegraphics[width=4in]{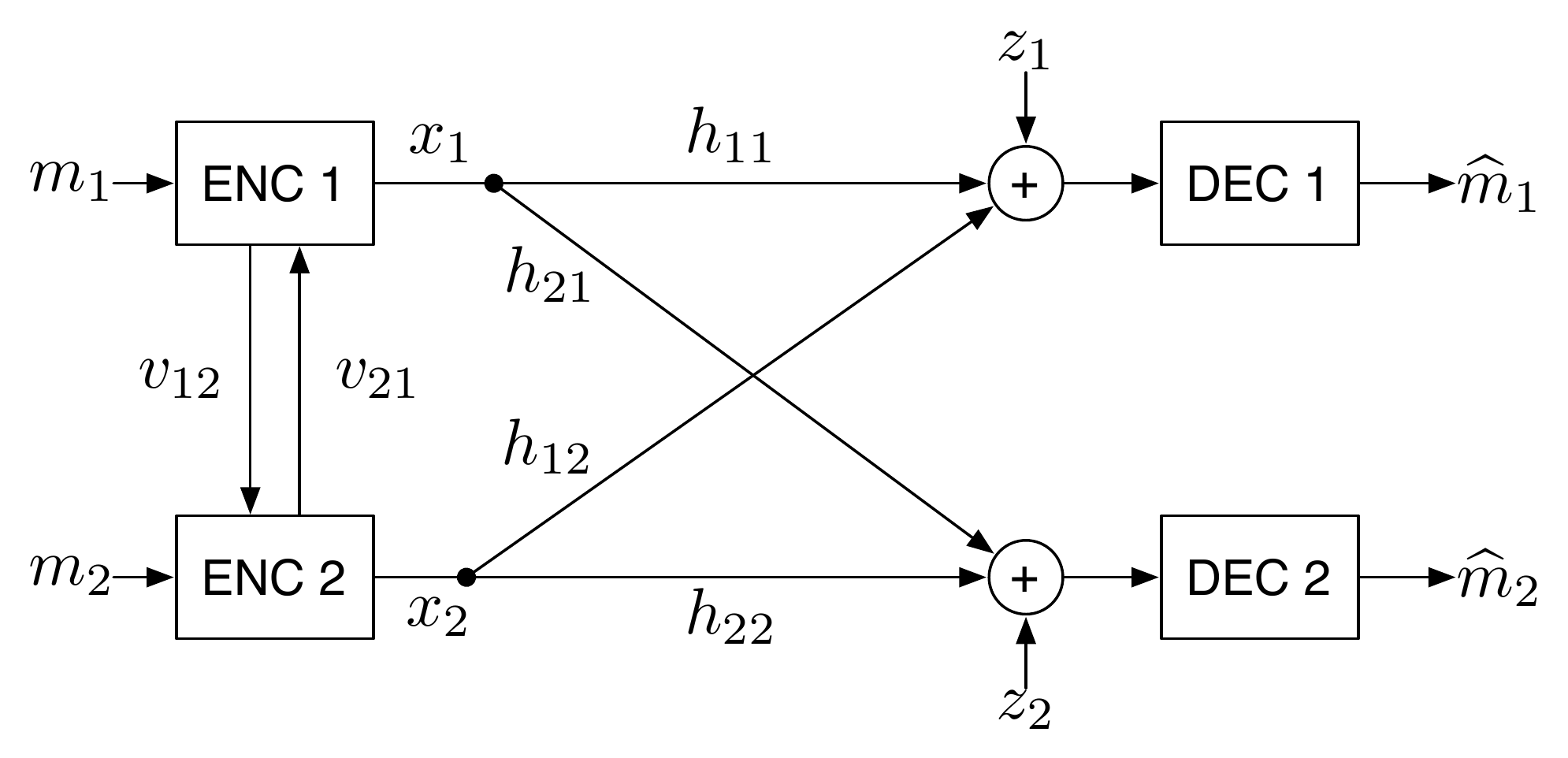}
\caption{Channel Model}
\label{fig_ChModel}
}
\end{figure}

The Gaussian interference channel with conferencing transmitters is depicted in Fig. \ref{fig_ChModel}.

The links among transmitters and receivers are modeled as the \emph{normalized} Gaussian interference channel:
\begin{align}
y_1 = h_{11}x_1 + h_{12}x_2 + z_1,\ 
y_2 = h_{21}x_1 + h_{22}x_2 + z_2,
\end{align}
where the additive noise processes $\{z_i[k]\}$, ($i=1,2$), are independent $\mathcal{CN}(0,1)$, i.i.d. over time. In this paper, we use $[.]$ to denote time indices. Transmitter $i$ intends to convey message $m_i$ to receiver $i$ by encoding it into a block codeword $\{x_i[k]\}_{k=1}^N$, with transmit power constraints
\begin{align} 
\frac{1}{N}\sum_{k=1}^N\big\lvert x_i[k] \big\lvert^2 \le 1,\ i=1,2,
\end{align}
for arbitrary block length $N$. Note that outcome of the encoder depends on both messages. Messages $m_1,m_2$ are independent. Define channel parameters
\begin{align}
\SNR_i := |h_{ii}|^2,\ \INR_i := |h_{ij}|^2,\ i,j=1,2,\ i\ne j.
\end{align}

The cooperative links between transmitters are noiseless with finite capacity $\C_{ij}$ from transmitter $i$ to $j$. Encoding must satisfy causality constraints: for any time index $k=1,2,\ldots, N$, $v_{ij}[k]$ is only a function of $\{m_i, v_{ji}[1], \ldots, v_{ji}[k-1]\}$.

\subsection{Notations}
We summarize below the notations used in the rest of this paper.
\begin{itemize}
\item
For a real number $a$, $(a)^+ := \max(a,0)$ denotes its positive part.
\item
For a real number $a$, $\lfloor a \rfloor$ denotes the closest integer that is not greater than $a$.
\item
For sets $A,B \subseteq \mathbb{R}^k$ in $k$-dimensional space, $A \oplus B := \{a+b: a\in A, b\in B\}$ denotes the direct sum of $A$ and $B$. 
\item
With a little abuse of notations, for $x,y\in\mathbb{F}_q$, $x\oplus y$ denotes the modulo-$q$ sum of $x$ and $y$.
\item
Unless specified, all the logarithms $\log(.)$ is of base 2.
\end{itemize}

\section{Linear Deterministic Interference Channel with Conferencing Transmitters}\label{sec_LDC}
As discussed in Section \ref{sec_Intro}, we shall first study an auxiliary problem, \emph{linear deterministic} interference channel with conferencing transmitters, to overcome the complications both in achievability and outer bounds. 

\begin{figure}[htbp]
{\center
\subfigure[Channel Model]{\includegraphics[width=2.5in]{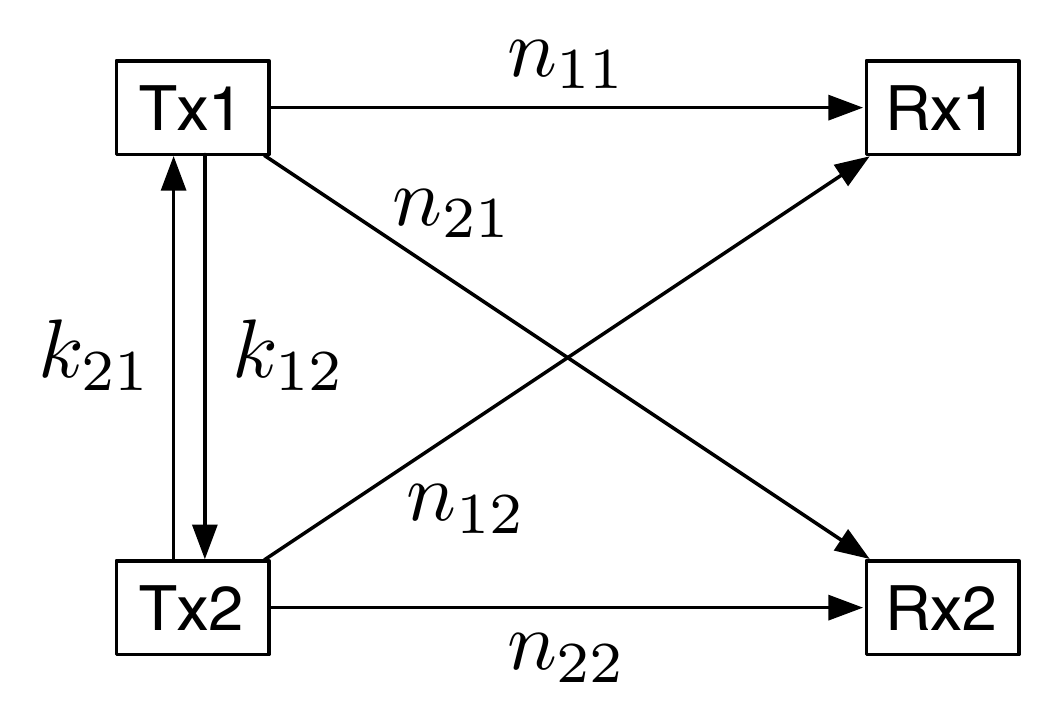}}
\subfigure[Example Channel]{\includegraphics[width=2.5in]{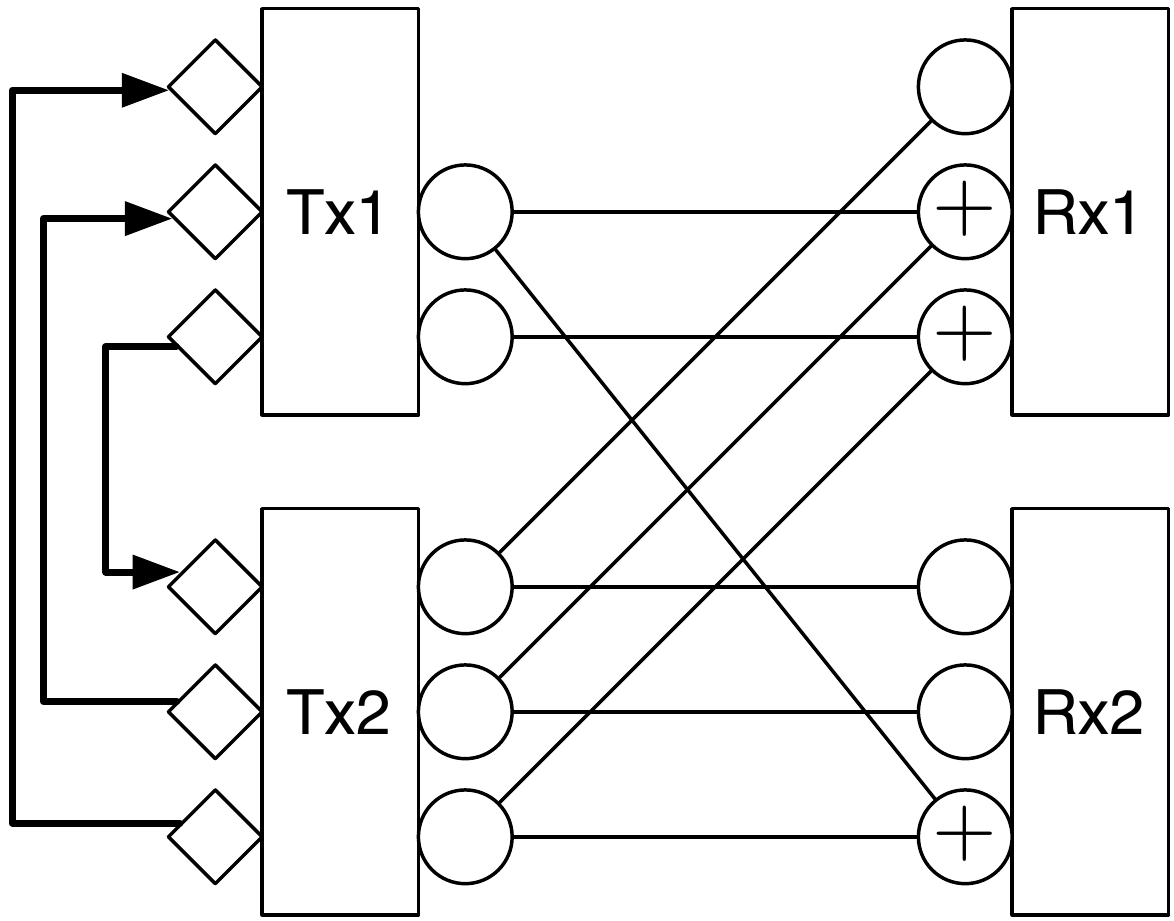}}
\caption{Linear Deterministic Interference Channel with Conferencing Transmitters.}
\label{fig_LDCModel}
}
\end{figure}

The corresponding linear deterministic channel (LDC) is parametrized by nonnegative integers $n_{11}$, $n_{21}$, $n_{22}$, $n_{12}$, $k_{12}$, and $k_{21}$, where 
\begin{align}
n_{ij} := \lp \lfloor \log |h_{ij}|^2 \rfloor \rp^+,\ i,j\in\{1,2\}
\end{align}
correspond to the channel gains in logarithmic-two scale, and 
\begin{align}
k_{12} := \lfloor \C_{12} \rfloor,\ k_{21} := \lfloor \C_{21} \rfloor
\end{align}
correspond to the cooperative link capacities. An illustration is depicted in Fig. \ref{fig_LDCModel}(a) along with an example in Fig. \ref{fig_LDCModel}(b). Each circle or diamond represents a bit. The bit emitting from a single circle at transmitters will broadcast noiselessly through the edges to the circles at receivers. Multiple incoming bits at a circle are summed up using modulo-two addition and produce a single received bit. The diamonds represent the bits exchanged between transmitters. In Fig. \ref{fig_LDCModel}(b), Tx1 can send one bit to Tx2, and Tx2 can send two bits to Tx1. For more details about this model, we point the readers to reference \cite{AvestimehrDiggavi_07} \cite{AvestimehrDiggavi_09} \cite{BreslerTse_08}.

The following theorem characterizes the capacity region of this channel.

\begin{theorem}\label{thm_LDCRegion}
Nonnegative $\lp R_1,R_2\rp$ is achievable if and only if it satisfies the following:
\begin{align}
R_1 &\le \min\Big\{ \max\lp n_{11}, n_{12}\rp,\ n_{11}+k_{12}\Big\}\\
R_2 &\le \min\Big\{ \max\lp n_{22}, n_{21}\rp,\ n_{22}+k_{21}\Big\}\\
R_1+R_2 &\le \lp n_{11}-n_{21}\rp^+ + \max\lp n_{22}, n_{21}\rp + k_{12} \label{eq_SumBound1} \\
R_1+R_2 &\le \lp n_{22}-n_{12}\rp^+ + \max\lp n_{11}, n_{12}\rp + k_{21} \label{eq_SumBound2} \\
R_1+R_2 &\le \max\lbp n_{12}, \lp n_{11}-n_{21}\rp^+ \rbp + \max\lbp n_{21}, \lp n_{22}-n_{12}\rp^+ \rbp + k_{12} + k_{21} \label{eq_SumBound3} \\
R_1+R_2 &\le \lbp\begin{array}{ll}
\max\lbp n_{11}+n_{22}, n_{12}+n_{21}\rbp, &\textrm{if }n_{11}+n_{22} \ne n_{12}+n_{21}\\
\max\lbp n_{11},n_{12},n_{21},n_{22}\rbp, &\textrm{if }n_{11}+n_{22} = n_{12}+n_{21}
\end{array}\right. \label{eq_SumBound4} \\
2R_1+R_2 &\le \max\lp n_{11}, n_{12}\rp + \max\lbp n_{21}, \lp n_{22}-n_{12}\rp^+ \rbp + \lp n_{11}-n_{21}\rp^+ + k_{12} + k_{21} \label{eq_SlopeTwoBound1}\\
R_1+2R_2 &\le \max\lp n_{22}, n_{21}\rp + \max\lbp n_{12}, \lp n_{11}-n_{21}\rp^+ \rbp + \lp n_{22}-n_{12}\rp^+ + k_{21} + k_{12}\\
2R_1+R_2 &\le n_{21} + \max\lbp n_{11}+\lp n_{22}-n_{21}\rp^+, n_{12}\rbp + \lp n_{11}-n_{21}\rp^+ + k_{12} \label{eq_SlopeTwoBound2}\\
R_1+2R_2 &\le n_{12} + \max\lbp n_{22}+\lp n_{11}-n_{12}\rp^+, n_{21}\rbp + \lp n_{22}-n_{12}\rp^+ + k_{21}
\end{align} 
\end{theorem}

\subsection{Motivating Examples}
Before going into technical details of proving the achievability and outer bounds, we first give an example to motivate the scheme as well as the outer bounds. The first example channel is depicted in Fig. \ref{fig_LDCModel}(b), where $n_{11}=2, n_{12}=3, n_{21}=1, n_{22}=3, k_{12}=1, k_{21}=2$. 

\begin{figure}[htbp]
{\center
\subfigure[Achieving $R_1=2, R_2=3$]{\includegraphics[width=3in]{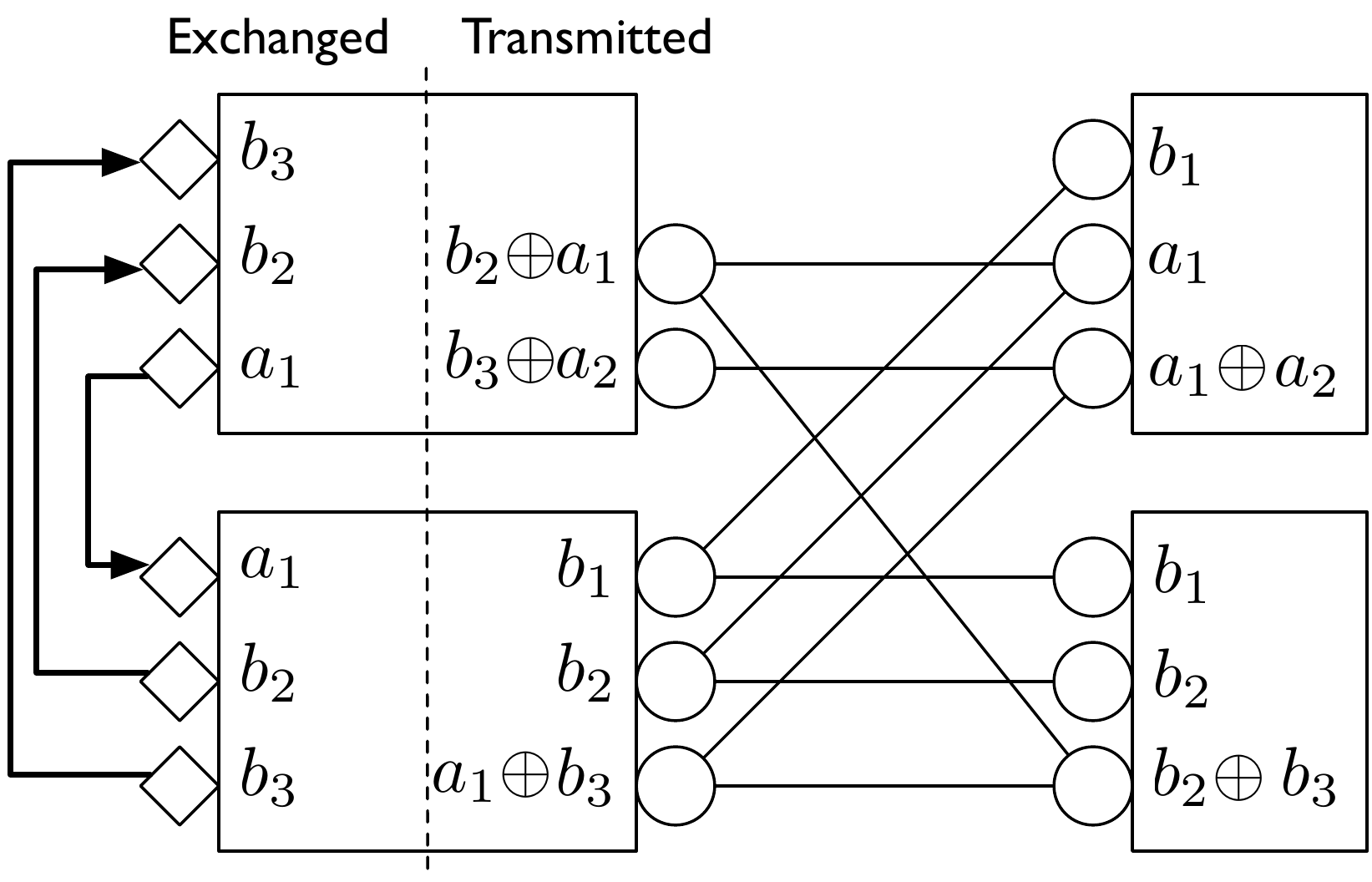}}
\subfigure[Achieving $R_1=3, R_2=1$]{\includegraphics[width=3in]{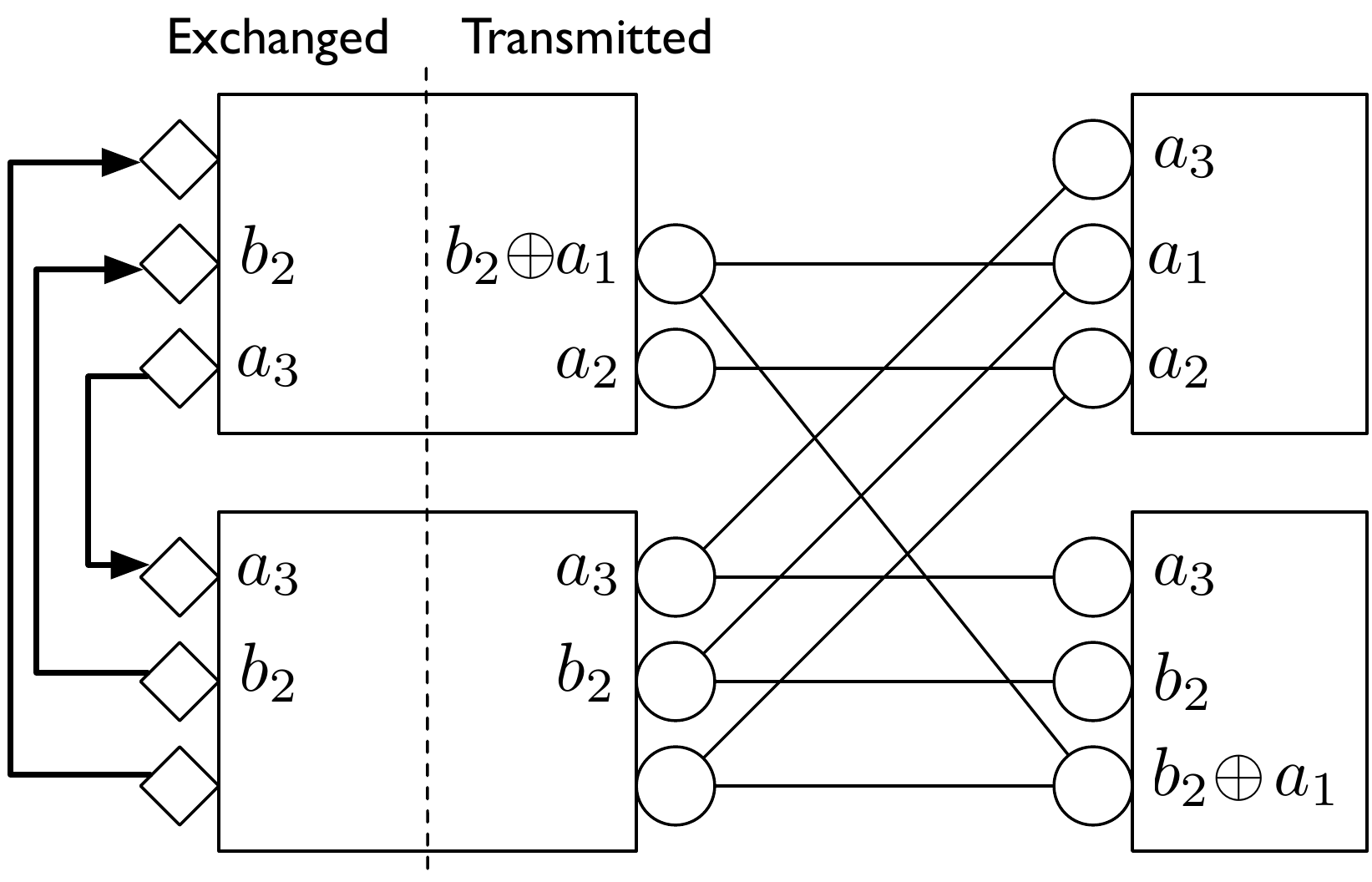}}
\caption{Coding Strategies for Example Channel in Fig. \ref{fig_LDCModel}}
\label{fig_LDC_Code}
}
\end{figure}

\subsubsection{Achievability}
First consider its achievability. To achieve the rate point $\lp R_1, R_2\rp = (2,3)$, one simple strategy is depicted in Fig. \ref{fig_LDC_Code}(a). In this coding scheme, we identify the message structure in Table \ref{table_Message1}.
\begin{table}[htdp]
\caption{Message Structure in Fig. \ref{fig_LDC_Code}(a)}
\begin{center}
\begin{tabular}{|c|c|c|c|}
\hline
Cooperative common & Cooperative private & Noncooperative common & Noncooperative private\\
\hline 
None & $a_1$ and $\lp b_2, b_3\rp$ & $b_1$ & $a_2$\\
\hline
\end{tabular}
\label{table_Message1}
\end{center}
\end{table}
Note that transmitter 2 sends $b_2$ and $b_3$ to transmitter 1 so that it can carry out proper precoding to null out interference $b_2$ and $b_3$ at receiver 1. Similarly transmitter 1 sends $a_1$ to transmitter 2 so that it can null out interference $a_1$ at receiver 2.

On the other hand, to achieve the rate point $\lp R_1, R_2\rp=(3,1)$, one simple strategy is depicted in Fig. \ref{fig_LDC_Code}(b). In this coding scheme, we identify the message structure in Table \ref{table_Message2}.
\begin{table}[htdp]
\caption{Message Structure in Fig. \ref{fig_LDC_Code}(b)}
\begin{center}
\begin{tabular}{|c|c|c|c|}
\hline
Cooperative common & Cooperative private & Noncooperative common & Noncooperative private\\
\hline 
$a_3$ & $b_2$ & $a_1$ & $a_2$\\
\hline
\end{tabular}
\label{table_Message2}
\end{center}
\end{table}
Note that to support a third bit $a_3$ for user 1, it has to occupy the topmost circle level at transmitter 2 and both receivers, since the direct link from transmitter 1 to receiver 1 has only two levels. 
Hence, receiver 2 inevitably will decode bit $a_3$, which is then classified as cooperative common. From this example we see that \emph{cooperative common} messages are needed, and it should occupy the levels that appear at both receivers cleanly. For the cooperative private parts, the example suggests that one should design precoding cleverly such that interference is nulled out at the unintended receiver. Based upon these intuitions, we propose an explicit scheme in Section \ref{subsec_Scheme}.

In the above example, we can see that the use of the cooperative links is two-fold: (1) null out interference, as in Fig. \ref{fig_LDC_Code}(a), or (2) relay additional bits, as the link from transmitter 1 to 2 in Fig. \ref{fig_LDC_Code}(b). This observation is also useful in motivating outer bounds.

\subsubsection{Fundamental Tradeoff on $2R_1+R_2$ and $R_1+2R_2$}
For outer bounds, the main difference from the interference channel \emph{without} cooperation \cite{El-GamalCosta_82} \cite{BreslerTse_08} is that there are two different types of bounds on $2R_1+R_2$ (and $R_1+2R_2$ correspondingly). Below we demonstrate the two different types of fundamental tradeoff on $2R_1+R_2$ through two examples.

\begin{figure}[htbp]
{\center
\subfigure[Achieving $R_1=4, R_2=2$]{\includegraphics[width=3in]{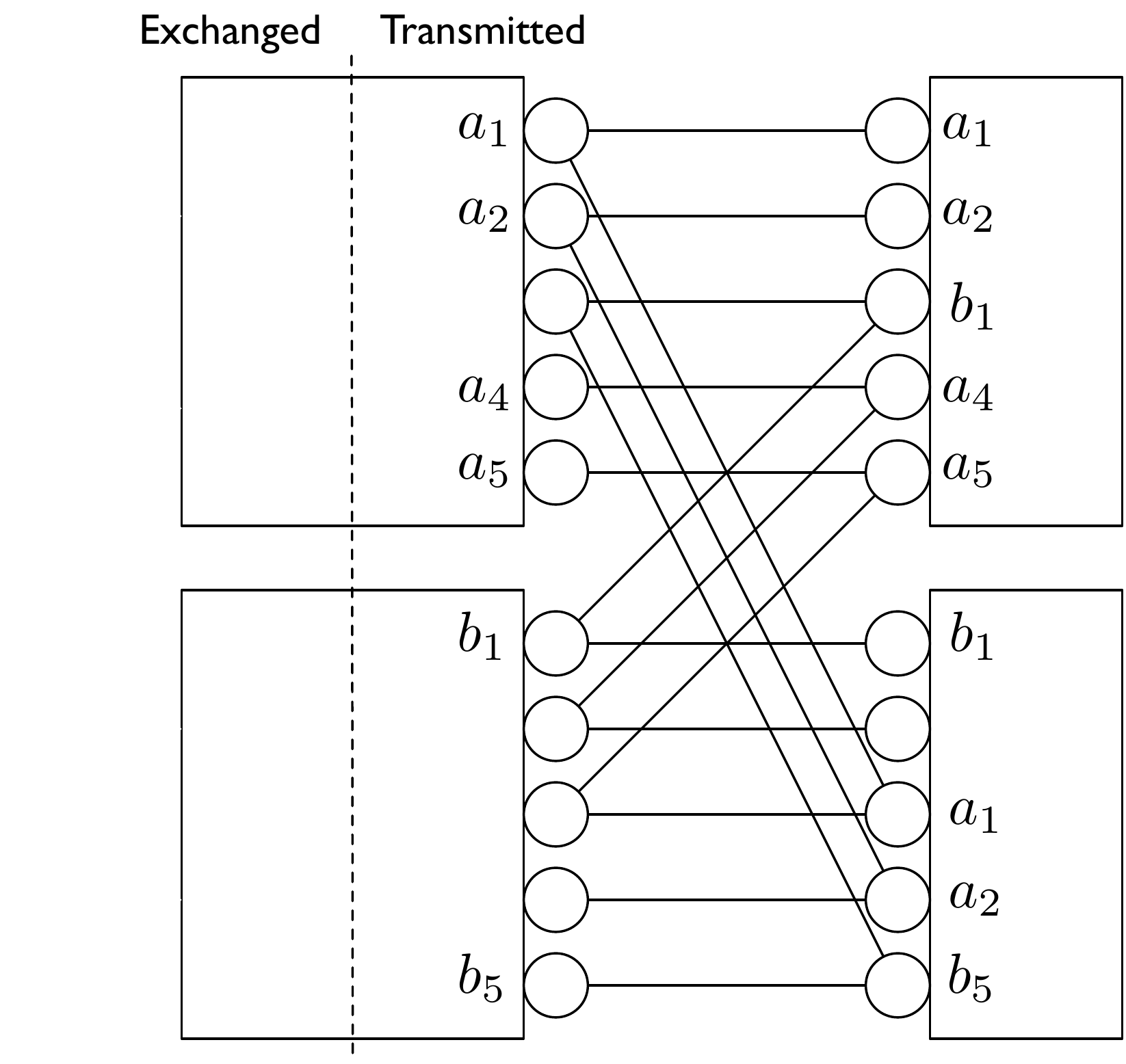}}
\subfigure[Achieving $R_1=5, R_2=0$]{\includegraphics[width=3in]{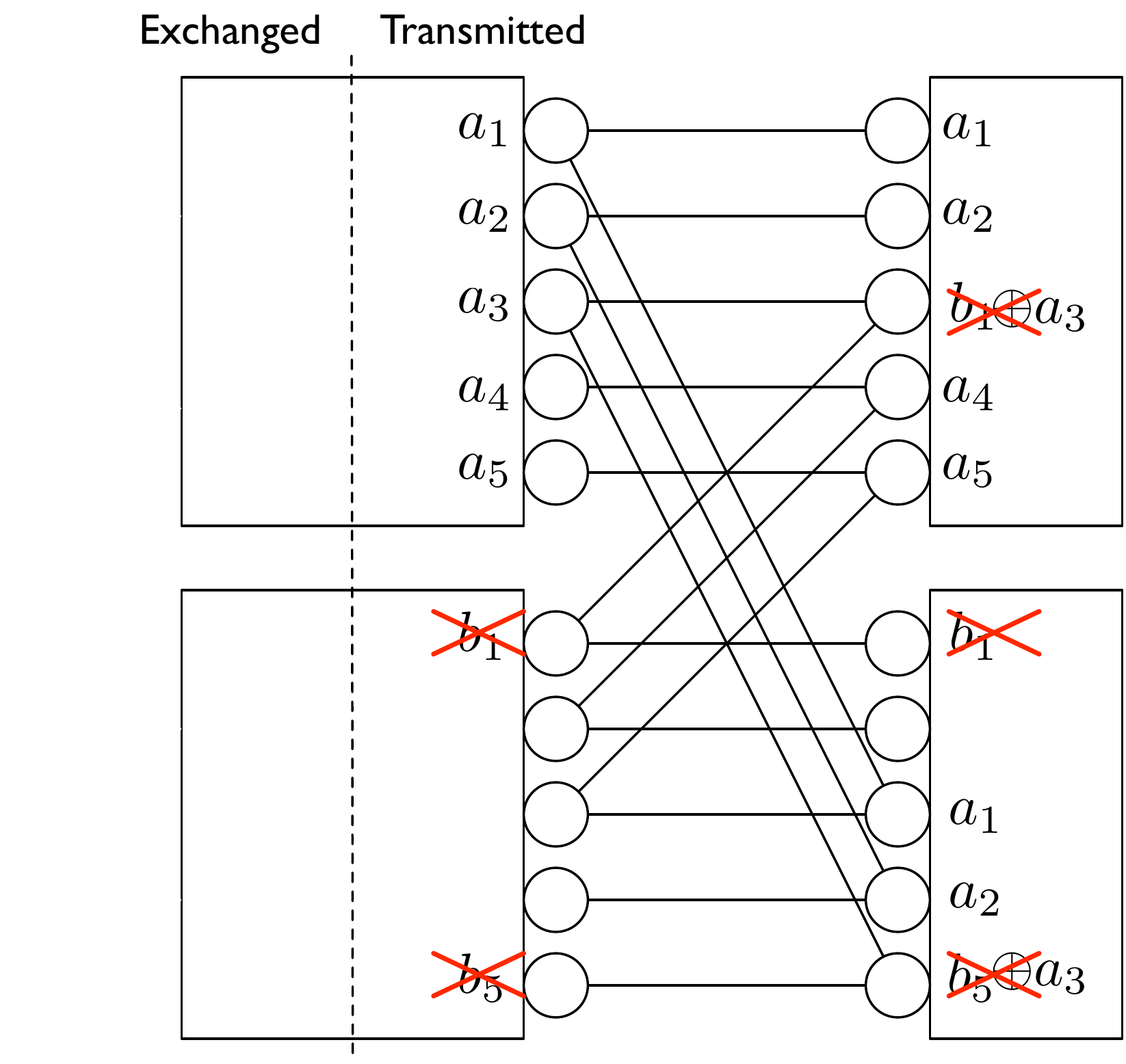}}
\caption{Example Channel without Transmitter Coopertaion: Tradoff from $(R_1,R_2) = (4,2)$ to $(5,0)$}
\label{fig_LDC_Tradeoff}
}
\end{figure}

\begin{figure}[htbp]
{\center
\subfigure[Achieving $R_1=4, R_2=4$]{\includegraphics[width=3in]{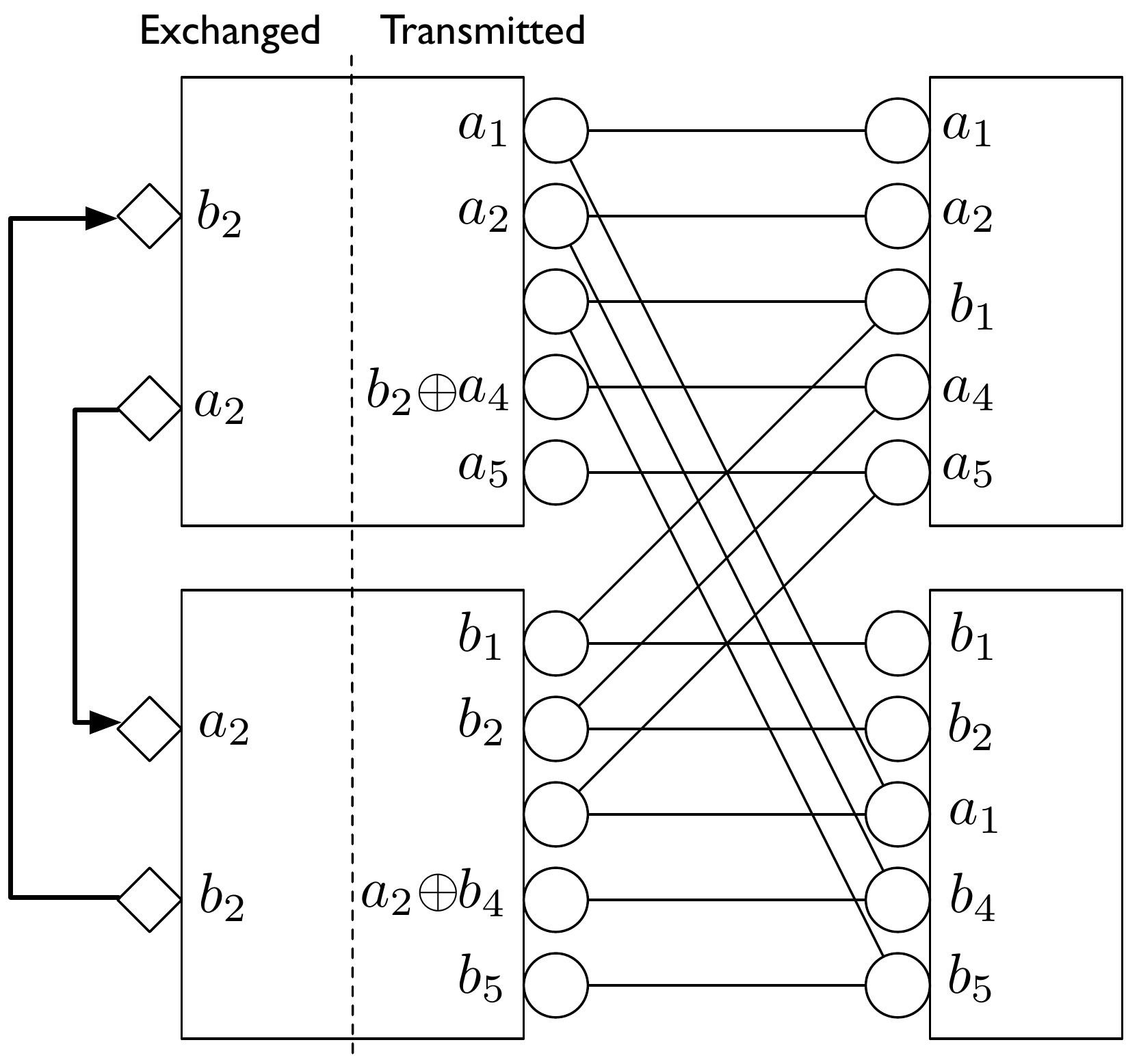}}
\subfigure[Achieving $R_1=5, R_2=2$]{\includegraphics[width=3in]{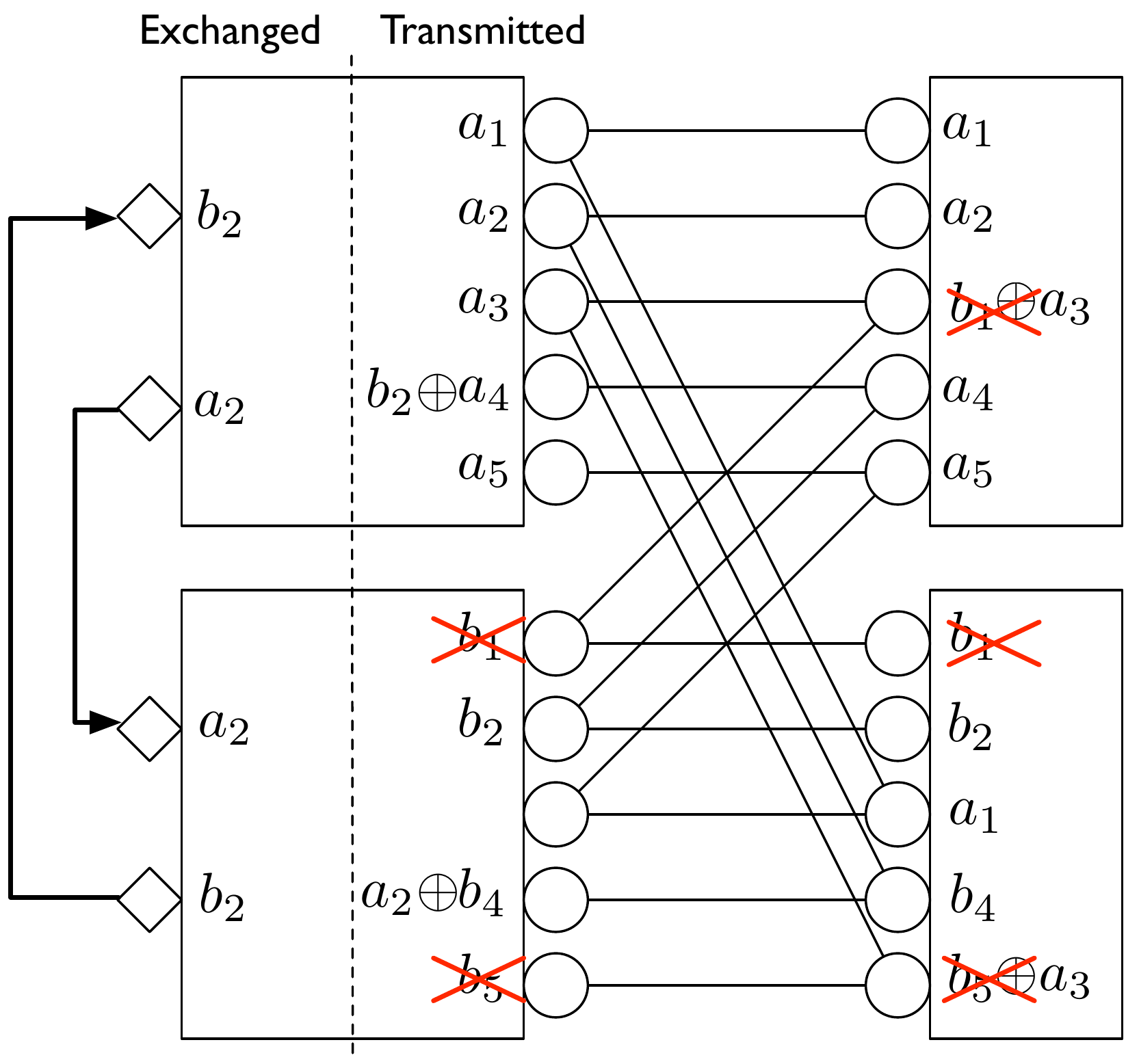}}
\caption{Example Channel with Transmitter Coopertaion: Tradoff from $(R_1,R_2) = (4,4)$ to $(5,2)$}
\label{fig_LDC_Tradeoff2}
}
\end{figure}

The first type of tradeoff does not involve the information that flows in the cooperative links. Consider the example channel with $n_{11}=n_{22} = 5$, $n_{12}=n_{21}=3$, $k_{12}=k_{21}=1$. We first consider the case \emph{without} cooperation. Two corner points of the capacity region is $(R_1, R_2) = (4,2)$ and $(5,0)$, and the optimal strategies are depicted in Fig. \ref{fig_LDC_Tradeoff}(a) and (b) respectively. To enhance user 1's rate from $4$ to $5$ bits, the bit $a_3$ has to be turned on and causes collisions at the third level at receiver 1 and the fifth level at receiver 2. Transmitter 2 then has to turn off bit $b_1$ to avoid destroying bit $a_3$, and $b_5$ cannot be decoded since it is corrupted by $a_3$. Now consider the case \emph{with} cooperation. The two corner points of the capacity region is $(R_1, R_2) = (4,4)$ and $(5,2)$, and the optimal strategies are depicted in Fig. \ref{fig_LDC_Tradeoff2}(a) and (b) respectively. Note that to enhance user 1's rate from $4$ to $5$ bits, again the bit $a_3$ has to be turned on and again causes collisions at the same places as in the case without cooperation. Note that the bits exchanged in the cooperative links remains the same, and hence the information that flows in the cooperative links is not involved in this tradeoff. Furthermore, the tradeoff is qualitatively the same as that in the case \emph{without} cooperation. Later we will see that this type of outer bound on $2R_1+R_2$ can be generalized from the $2R_1+R_2$ bound in deterministic interference channel \emph{without} cooperation \cite{El-GamalCosta_82}, and the proof technique is quite similar.

The second type of tradeoff is a new phenomenon in interference channel with cooperation, and it involves the information that flows in the cooperative links. Consider the example channel in Fig. \ref{fig_LDCModel}(b). The two rate points $\lp R_1,R_2\rp = (2,3)$ and $\lp R_1,R_2\rp = (3,1)$ are on the boundary of the capacity region. To enhance user 1's rate from $2$ to $3$ bits, since the number of levels from transmitter 1 to receiver 1 is only $2$, the third bit $a_3$ has to be relayed from transmitter 2 to receiver 1. Hence, the topmost level at transmitter 2 has to be occupied by information exclusively for user 1, that is, $a_3$, and at receiver 2 the topmost level is no longer available for user 2. On the other hand, since the cooperative link from transmitter 1 to transmitter 2 is now occupied by $a_3$, the opportunity of nulling out the interference at the third level at receiver 2 is eliminated. As a consequence, the only available level for user 2 at receiver 2 is the second level, and user 2 has to back off its rate from $3$ to $1$. Note that the key difference from the first type of tradeoff is that, at the rate point $\lp R_1,R_2\rp = (2,3)$ the cooperative link from transmitter 1 to 2 is used for \emph{nulling out interference}, while at $\lp R_1,R_2\rp = (3,1)$ it is used for \emph{relaying additional bits}. Hence, the information that flows in the cooperative links \emph{is} involved in this tradeoff, and the tradeoff is qualitatively different from that in the case \emph{without} cooperation. As we will show later, to prove this type of outer bound on $2R_1+R_2$, we need to develop a new technique for giving side information to the receivers.

\subsection{Outer Bounds}
To prove the converse part of Theorem \ref{thm_LDCRegion}, instead of giving full details of the proof\footnote{We will provide full details when we deal with the Gaussian problem.}, here we describe the techniques used in the proof. These techniques will be reused for proving outer bounds in the Gaussian problem.

{\flushleft 1) Bounds on $R_1$ and $R_2$:} These bounds are straightforward cut-set bounds.

{\flushleft 2) Bounds on $R_1+R_2$:} Bound \eqref{eq_SumBound4} is a standard cut-set bound. 

Bound \eqref{eq_SumBound1} is obtained by providing side information $\lp m_2, v_{12}^N\rp$ to receiver 1 so that receiver 1 is not interfered by transmitter 2 at all. This leads to the part $\lp n_{11}-n_{21}\rp^+ + \max\lp n_{22}, n_{21}\rp$, which is identical to the Z-channel bound in interference channel without cooperation. Giving the side information enhances the sum rate by at most $H\lp v_{12}^N|m_2\rp \le Nk_{12}$ bits. Similar arguments works for bound \eqref{eq_SumBound2}. 

Bound \eqref{eq_SumBound3} is obtained by providing side information $\lp v_{12}^N, v_{21}^N, s_1^N\rp$ and $\lp v_{12}^N, v_{21}^N, s_2^N\rp$ to receiver 1 and 2 respectively, where $s_1^N$ denotes the interference caused by transmitter 1 at receiver 2 (and vice versa for $s_2^N$). Giving side information $\lp v_{12}^N, v_{21}^N\rp$ to both receivers enhances the sum rate by at most $H\lp v_{12}^N, v_{21}^N\rp \le N(k_{12}+k_{21})$ bits. Finally, we are able to prove the bound by making use of the Markov relations observed first in \cite{Willems_83} for MAC with conferencing transmitters: $m_i - \lp v_{12}^N,v_{21}^N\rp - x_j^N$, for $(i,j)=(1,2)$ or $(2,1)$.


{\flushleft 3) Bounds on $2R_1+R_2$ and $R_1+2R_2$}: By symmetry we shall focus on the bounds on $2R_1+R_2$.
 
For linear deterministic interference channel \emph{without} cooperation, the outer bound on $2R_1+R_2$ is proved by first creating a copy of receiver 1 and then giving proper side information to these three receivers \cite{El-GamalCosta_82}. The side information structure is the following: give side information $\lp x_2^N, s_1^N\rp$ to one of the two receiver 1's and side information $s_2^N$ to receiver 2.

As discussed in the previous section, there are two types of tradeoff on $2R_1+R_2$. They correspond to bound \eqref{eq_SlopeTwoBound1} and bound \eqref{eq_SlopeTwoBound2} respectively. Bound \eqref{eq_SlopeTwoBound1} can be obtained via a similar technique as that in \cite{El-GamalCosta_82}. The side information structure is the following: give side information $\lp m_2, s_1^N, v_{12}^N, v_{21}^N\rp$ to one of the two receiver 1's, and $\lp s_1^N, v_{12}^N, v_{21}^N\rp$ to receiver 2. The role of $\lp m_2, v_{12}^N\rp$ is the same as $x_2^N$, and the additional side information $\lp v_{12}^N, v_{21}^N\rp$ is to make the transmitters conditionally independent. We then make use of the above Markov property to complete the proof.

\begin{figure}[htbp]
{\center
\includegraphics[width=3in]{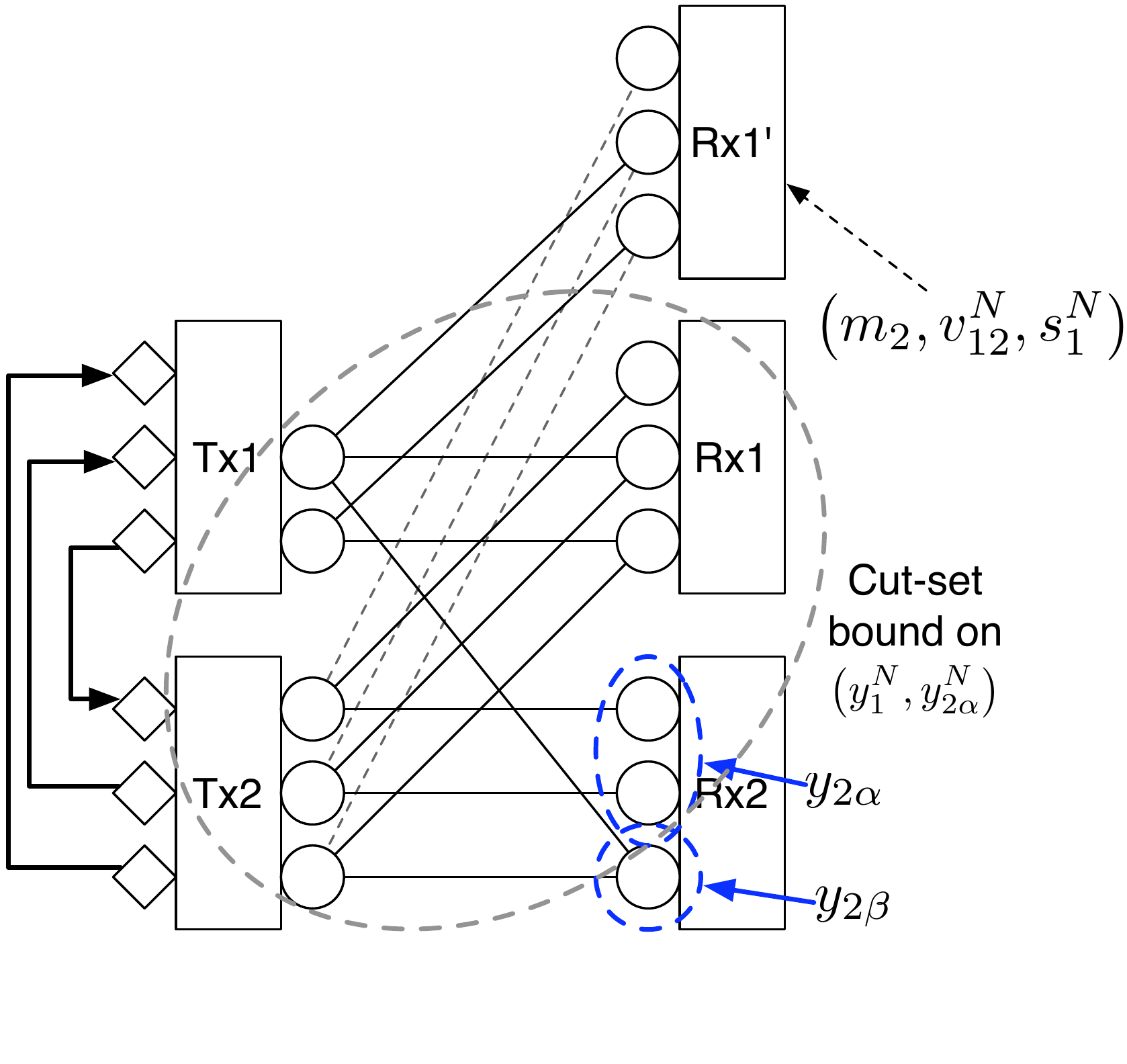}
\caption{Side Information Structure for Outer Bound \eqref{eq_SlopeTwoBound2}}
\label{fig_LDC_Converse}
}
\end{figure}

Bound \eqref{eq_SlopeTwoBound2}, which corresponds to the second type of tradeoff discussed earlier, is obtained by splitting receiver 2's signal into two parts: $y_2^N = \lp y_{2\alpha}^N,y_{2\beta}^N\rp$, where $y^N_{2\alpha}$ is the part of transmitter 2's signal that is not corrupted by $s_1^N$, the interference from transmitter 1. Then we apply a cut-set bound argument on one of the two receiver 1's and $y^N_{2\alpha}$, and provide side information $\lp m_2, v_{12}^N, s_1^N\rp$ to the other receiver 1. Fig. \ref{fig_LDC_Converse} provides an illustration. Since this kind of side information structure has not been reported in literature, we detail the proof below:

\begin{proof}
If $(R_1,R_2)$ is achievable, by Fano's inequality,
\begin{align}
&N\lp 2R_1+R_2 - \epsilon_N\rp \le 2I\lp m_1; y_1^N\rp +  I\lp m_2; y_2^N\rp\\
&\le I\lp m_1; y_1^N, s_1^N | m_2, v_{12}^N\rp + I\lp m_1; v_{12}^N|m_2\rp + I\lp m_1; y_1^N\rp + I\lp m_2; y_{2\alpha}^N, y_{2\beta}^N\rp\\
&= H\lp x_1^N, s_1^N| m_2, v_{12}^N\rp + I\lp m_1; y_1^N\rp + I\lp m_2; y_{2\alpha}^N\rp + I\lp m_2; y_{2\beta}^N| y_{2\alpha}^N\rp\\&\quad + I\lp m_1; v_{12}^N|m_2\rp\\
&\overset{\aaaa}{\le} H\lp x_1^N, s_1^N| m_2, v_{12}^N\rp + I\lp m_1,m_2; y_1^N,y_{2\alpha}^N\rp + I\lp m_2,v_{12}^N; y_{2\beta}^N| y_{2\alpha}^N\rp\\&\quad + H\lp v_{12}^N |m_2\rp\\
&\overset{\bbbb}{\le} H\lp x_1^N, s_1^N| m_2, v_{12}^N\rp + H\lp y_1^N,y_{2\alpha}^N\rp\\&\quad + H\lp y_{2\beta}^N \rp - H\lp y_{2\beta}^N |y_{2\alpha}^N, m_2, v_{12}^N \rp + H\lp v_{12}^N\rp\\
&\overset{\cccc}{=} H\lp x_1^N, s_1^N| m_2, v_{12}^N\rp + H\lp y_1^N,y_{2\alpha}^N\rp\\&\quad + H\lp y_{2\beta}^N \rp - H\lp s_1^N | m_2, v_{12}^N \rp + H\lp v_{12}^N\rp\\
&= H\lp x_1^N| s_1^N, m_2, v_{12}^N\rp + H\lp y_1^N,y_{2\alpha}^N\rp + H\lp y_{2\beta}^N \rp + H\lp v_{12}^N\rp\\
&\le N\Big\{ \lp n_{11}-n_{21}\rp^+ + \max\lbp n_{11}+\lp n_{22}-n_{21}\rp^+ , n_{12}\rbp + n_{21} + k_{12}\Big\},
\end{align}
where $\epsilon_N\rightarrow 0$ as $N\rightarrow \infty$. (a) is due to a simple fact that $I\lp m_1; y_1^N\rp + I\lp m_2; y_{2\alpha}^N\rp \le I\lp m_1,m_2; y_1^N,y_{2\alpha}^N\rp$. (b) is due to that conditioning reduces entropy. (c) holds since $y_{2\alpha}^N$ is a function of $\lp m_2,v_{12}^N\rp$.
\end{proof}

Let us revisit the example in Fig. \ref{fig_LDCModel}(b) and demonstrate that bound \eqref{eq_SlopeTwoBound2} is active. Plugging the channel parameters into Theorem \ref{thm_LDCRegion}, we see that \emph{without} bound \eqref{eq_SlopeTwoBound2}, the region is 
\begin{align}
R_1 \le 3,\ R_2\le 3,\ R_1+R_2 \le 5,
\end{align}
and the rate point $\lp 3,1\rp$ is not on its boundary. In this example, $y_{2\alpha}$ spans the topmost two levels at receiver 2. Hence, $H\lp y_1^N, y_{2\alpha}^N\rp \le 4N$, and $2R_1+R_2 \le 1+4+1+1=7$ which is active in the capacity region:
\begin{align}
R_1 \le 3,\ R_2\le 3,\ R_1+R_2 \le 5,\ 2R_1+R_2 \le 7.
\end{align}

\subsection{Achievability via Linear Reciprocity}\label{subsec_LinearScheme}
Unlike the linear deterministic interference channel with conferencing receivers, it is not straightforward to directly show that linear strategies achieves the capacity in the case with conferencing transmitters. We can overcome this by using \emph{linear reciprocity} of linear deterministic networks \cite{RajaPrabhakaran_09} and prove the achievability part of Theorem \ref{thm_LDCRegion}. We sketch the idea of the proof as follows.


First it is not hard to show that linear strategies are optimal for the reciprocal channel, that is, linear deterministic interference channel with conferencing receivers. In such linear strategies, each user modulates its information bits (message) onto the transmit signal vector via a linear transformation. Each receiver, serving as a relay, linearly transforms its received signal and sends it to the other receiver through the finite-capacity link. Since the channel is linear and deterministic, the exchanged signals between receivers are again linear transformations of the transmit information bits. Finally, each receiver solves all its received linear equations of the transmit information bits (one set from the other receiver and the other from the transmitters) and recovers its desired message. Note that the decoding process is again a linear transformation. By choosing these linear transformations (encoding, relaying, and decoding) properly, the scheme achieves the capacity.

Next by linear reciprocity, we immediately shows that the capacity region of the reciprocal channel (linear deterministic interference channel with conferencing receivers) is an achievable region of the original channel. The strategy is again linear. Each transmitter sends a linear transformation of its information bits to the other transmitter through the finite-capacity link. Then it sends out a linear transformation of the received bits from the other transmitter and its own information bits to the receivers. Finally, each receiver solves the linear equations it receives to recover its desired message. 
It remains to show that this region coincides with that given in Theorem \ref{thm_LDCRegion}, which is a straightforward calculation. 

Note that in such linear strategies, there is no need to split the messages at the transmitters, and the decoding process at the receivers can be viewed as \emph{treating interference as noise}. This is first observed in Lecture Notes 6 in \cite{El-GamalKim_10} for linear deterministic interference channels without cooperation. This implies that the complicated message structure described in Section \ref{sec_Intro} is not necessary for linear deterministic interference channel with conferencing receivers or transmitters.

To this end, there are two paths towards constructing good coding strategies in the Gaussian scenario. The first approach is deriving structured lattice strategies based on the capacity-achieving linear strategies of the corresponding linear deterministic channels. This approach, however, requires an explicit description of linear transformations in the capacity-achieving linear strategies for the LDC. The second approach is deriving Gaussian random coding strategies, which is the conventional approach for additive white Gaussian noise networks. In this paper, we will take the second approach. For this purpose, however, the proof of achievability via linear reciprocity does not give much insight. Below we give an alternative proof of achievability, which provides guidelines for designing good Gaussian random coding strategies in the Gaussian interference channel with conferencing transmitters.

\subsection{Alternative Proof of Achievability}\label{subsec_Scheme}
To get a better handle to deal with the the design of good Gaussian random coding schemes, we propose a general coding strategy that applies both to linear deterministic channel (LDC) and Gaussian channel. The strategy is based on Marton's coding scheme for general broadcast channels \cite{Marton_79} and superposition coding. It is described as follows: (Notations: subscript $o$ stands for \emph{cooperative common}, subscript $h$ stands for \emph{cooperative private}, subscript $c$ stands for \emph{noncooperative common}, and subscript $p$ stands for \emph{noncooperative private}.)

\begin{itemize}
\item [1)] First, generate the cooperative common vector codeword $\ul{x}_o^N\lp m_{1o}, m_{2o}\rp$ according to $p\lp \ul{x}_o^N\rp = \prod_{k=1}^Np\lp \ul{x}_o[k]\rp$. Denote $m_o := \lp m_{1o}, m_{2o}\rp$.
\item [2)] Second, for each $m_o$, generate the cooperative vector codeword $\ul{x}_{oh}^N\lp m_{1h}, m_{2h}, m_o\rp$ based on Marton's coding scheme according to conditional distribution $p\lp \ul{x}_{oh}^N, u_1^N,u_2^N|\ul{x}_o^N(m_o)\rp = \prod_{k=1}^Np\lp \ul{x}_{oh}[k], u_1[k],u_2[k]|\ul{x}_o(m_o)[k]\rp$, where the auxiliary codewords are $u_1^N\lp \wtild{m}_{1h}, m_o\rp$ and $u_2^N\lp \wtild{m}_{2h}, m_o\rp$.
\item [3)] Third, at transmitter $i$, $i=1,2$, generate the noncooperative common codeword $x_{ic}^N\lp m_{ic}\rp$ according to distribution $p\lp x_{ic}^N\rp = \prod_{k=1}^Np\lp x_{ic}[k]\rp$.
\item [4)] Fourth, at transmitter $i$, for each $m_{ic}$ generate the noncooperative codeword $x_{icp}^N\lp m_{ip}, m_{ic}\rp$ according to $p\lp x_{icp}^N| x_{ic}^N\lp m_{ic}\rp \rp = \prod_{k=1}^Np\lp x_{icp}[k]| x_{ic}\lp m_{ic}\rp[k] \rp$.
\item [5)] Finally, superimpose these two codewords to form the transmit codewords:
\begin{align}
x^N_1\lp m_{o}, m_{1h},m_{2h}, m_{1c}, m_{1p}\rp &= \ul{x}^N_{oh}[1] + x^N_{1cp}\\
x^N_2\lp m_{o}, m_{1h},m_{2h}, m_{2c}, m_{2p}\rp &= \ul{x}^N_{oh}[2] + x^N_{2cp}.
\end{align}
\end{itemize}

\begin{remark}
Note that in Step 4), say at transmitter 1, we can use Gelfand-Pinsker coding (dirty paper coding) to generate noncooperative private codeword so that it can be protected against known interference at transmitter 1, which is caused by the cooperative private auxiliary codeword of the other user, that is, $u_2$. Throughout the paper, however, we will choose $\lp u_1, u_2\rp$ cleverly such that the effect of $u_2$ is zero-forced \emph{exactly} in LDC and \emph{approximately} in the Gaussian setting, and hence Gelfand-Pinsker coding does not provide significant improvement.
\end{remark}

For decoding, receiver 1 looks for a unique $\lp m_o, \wtild{m}_{1h}, m_{1c}, m_{1p}\rp$ such that 
\begin{align}
\lp y_1^N, \ul{x}^N_o\lp m_o\rp, u_1^N\lp \wtild{m}_{1h}, m_o\rp, x_{1c}^N\lp m_{1c}\rp, x_{1cp}^N\lp m_{1p}, m_{1c}\rp, x_{2c}^N\lp \what{m}_{2c}\rp\rp
\end{align}
is jointly typical, for some $\what{m}_{2c}$. Receiver 2 uses the same decoding rule with index 1 and 2 exchanged.

Based on the above strategy, we have the following coding theorem:
\begin{theorem}\label{thm_GeneralAcieve}
Nonnegative rate tuple $\lp R_{1o},R_{1h},R_{1c},R_{1p}, R_{2o},R_{2h},R_{2c},R_{2p},\rp$ is achievable if it satisfies the following for some nonnegative $(\wtild{R}_{1h}, \wtild{R}_{2h})$: ( denote $R_o:=R_{1o}+ R_{2o}$ )

{\flushleft \ul{Constraints at Receiver 1}}:
\begin{align}
R_{1p} &\le I\lp x_{1cp}; y_1| x_{1c},x_{2c}, u_1, \ul{x}_o\rp\\
\wtild{R}_{1h} &\le I\lp u_1; y_1| x_{1cp}, x_{1c}, x_{2c}, \ul{x}_o\rp\\
\wtild{R}_{1h} + R_{1p} &\le I\lp u_1, x_{1cp};y_1| x_{1c}, x_{2c}, \ul{x}_o \rp\\
R_{2c} + R_{1p} &\le I\lp x_{2c}, x_{1cp};y_1| x_{1c}, u_1, \ul{x}_o \rp\\
R_{1c} + R_{1p} &\le I\lp x_{1c}, x_{1cp};y_1| x_{2c}, u_1, \ul{x}_o \rp\\
R_{2c} + \wtild{R}_{1h} &\le I\lp x_{2c}, u_1;y_1| x_{1cp}, x_{1c}, \ul{x}_o \rp\\
R_{2c} + \wtild{R}_{1h} + R_{1p} &\le I\lp x_{2c}, u_1, x_{1cp};y_1| x_{1c}, \ul{x}_o \rp\\
R_{1c} + \wtild{R}_{1h} + R_{1p} &\le I\lp x_{1c}, x_{1cp}, u_1;y_1| x_{2c}, \ul{x}_o \rp\\
R_{1c} + R_{2c} + R_{1p} &\le I\lp x_{1c}, x_{1cp}, x_{2c};y_1| u_1, \ul{x}_o \rp\\
R_{1c} + R_{2c} + \wtild{R}_{1h} + R_{1p} &\le I\lp x_{1c}, x_{1cp}, x_{2c}, u_1;y_1| \ul{x}_o \rp
\end{align}
\begin{align}
R_o + \wtild{R}_{1h} &\le I\lp \ul{x}_o, u_1; y_1| x_{1cp}, x_{1c}, x_{2c}\rp\\
R_o + \wtild{R}_{1h} + R_{1p} &\le I\lp \ul{x}_o, u_1, x_{1cp};y_1| x_{1c}, x_{2c}\rp\\
R_o + R_{2c} + \wtild{R}_{1h} &\le I\lp \ul{x}_o, x_{2c}, u_1;y_1| x_{1cp}, x_{1c} \rp\\
R_o + R_{2c} + \wtild{R}_{1h} + R_{1p} &\le I\lp \ul{x}_o, x_{2c}, u_1, x_{1cp};y_1| x_{1c} \rp\\
R_o + R_{1c} + \wtild{R}_{1h} + R_{1p} &\le I\lp \ul{x}_o, x_{1c}, x_{1cp}, u_1;y_1| x_{2c} \rp\\
R_o + R_{1c} + R_{2c} + \wtild{R}_{1h} + R_{1p} &\le I\lp \ul{x}_o, x_{1c}, x_{1cp}, x_{2c}, u_1;y_1 \rp
\end{align}

{\flushleft \ul{Constraints at Receiver 2}}: Above with index 1 and 2 exchanged.

{\flushleft \ul{Constraints at Transmitters}}:
\begin{align}
R_{1h} &\le \wtild{R}_{1h}\\ R_{2h} &\le \wtild{R}_{2h}\\
R_{1o} + R_{1h} &\le \C_{12}\\ R_{2o} + R_{2h} &\le \C_{21}\\
\wtild{R}_{1h} + \wtild{R}_{2h} - R_{1h} -R_{2h} &\ge I\lp u_1; u_2| \ul{x}_o\rp,
\end{align}
for some $(u_1,u_2) - \ul{x}_{oh} - (y_1,y_2)$, $x_1=x_{oh}[1]+x_{1cp}$, $x_2=x_{oh}[2]+x_{2cp}$, and
\begin{align}
&p\lp \ul{x}_{oh}, \ul{x}_o, u_1, u_2, x_{1c}, x_{1cp}, x_{2c}, x_{2cp}\rp\\
&= p\lp \ul{x}_o\rp p\lp \ul{x}_{oh}, u_1,u_2|\ul{x}_o\rp p\lp x_{1c}\rp p\lp x_{1cp}|x_{1c}\rp p\lp x_{2c}\rp p\lp x_{2cp}|x_{2c}\rp.
\end{align}
\end{theorem}
\begin{proof}
The proof is quite straightforward. It involves standard error probability analysis of superposition coding and Marton's coding scheme, and hence is omitted here. Note we have in total 5 independent messages to be decoded at each receiver, and hence in general there should be $2^5-1=31$ inequalities. However, say at receiver 1, decoding $m_{2c}$ incorrectly is not accounted as an error. Furthermore due to the superposition coding of $\wtild{m}_{1h}$ upon $\ul{x}^N_{o}$ and the superposition coding of $m_{1p}$ upon $x^N_{1c}$, we remove the inequality on $R_{2c}$ and the $2^3+2^3-2=14$ inequalities involving $R_{o}$ but not $\wtild{R}_{1h}$ or involving $R_{1c}$ but not $R_{1p}$. Hence in total we have $31-1-14=16$ inequalities at each receiver.
\end{proof}

Below we show that with proper choices of $p\lp \ul{x}_{oh}, \ul{x}_o, u_1, u_2, x_{1c}, x_{1cp}, x_{2c}, x_{2cp}\rp$, the above coding strategy can achieve the capacity region of LDC. We shall distinguish into two cases: (1) system transfer matrix is full-rank, and (2) system transfer matrix is not full-rank. 

\subsubsection{System matrix is full-rank: $n_{11}+n_{22} \ne n_{12}+n_{21}$}
In this case, for the cooperative part, we shall set $\ul{x}_o$ to be running over all transmit levels, and choose $\lp u_1, u_2\rp | \ul{x}_o \overset{\mathrm{d}}{=} \lp y_{1h}, y_{2h}\rp$ occupying the following numbers of least significant bits (LSB) at receiver 1 and 2 respectively:
\begin{align}
g_1 &= \max\lbp n_{11} - \lp n_{21}-n_{22}\rp^+, n_{12} - \lp n_{22}-n_{21}\rp^+\rbp, \label{eq_g1} \\
g_2 &= \max\lbp n_{22} - \lp n_{12}-n_{11}\rp^+, n_{21} - \lp n_{11}-n_{12}\rp^+\rbp. \label{eq_g2}
\end{align}
Then we choose $\ul{x}_h$ occupying the levels at transmitters so that it results in $\lp y_{1h}, y_{2h}\rp$ at receivers. The cooperative codeword is generated according to the distribution of $\ul{x}_{oh} \overset{\rm{d}}{=} \ul{x}_o+\ul{x}_h$. The addition here is bit-wise modulo-two. We observe the following:

\begin{claim}\label{claim_LDC1}
Under i.i.d. Bernoulli half inputs, $I\lp u_1; u_2 | \ul{x}_o\rp = 0$ with the above choice if $n_{11}+n_{22} \ne n_{12}+n_{21}$, and hence $u_1$ and $u_2$ are independent conditioned on $\ul{x}_o$.
\end{claim}
\begin{proof}
For the case $\lbp n_{11}\ge n_{12}, n_{22}\ge n_{21}\rbp$, $(g_1,g_2)$ becomes
\begin{align}
g_1 &= \max\lbp n_{11} - \lp n_{21}-n_{22}\rp^+, n_{12} - \lp n_{22}-n_{21}\rp^+\rbp = n_{11}\\
g_2 &= \max\lbp n_{22} - \lp n_{12}-n_{11}\rp^+, n_{21} - \lp n_{11}-n_{12}\rp^+\rbp = n_{22},
\end{align}
and the rank of the full system transfer matrix is $n_{11}+n_{22} = g_1+g_2$. Hence, under i.i.d. Bernoulli half inputs, 
\begin{align}
I\lp u_1;u_2| \ul{x}_o\rp &= H\lp u_1| \ul{x}_o\rp + H\lp u_2| \ul{x}_o\rp - H\lp u_1,u_2| \ul{x}_o\rp\\
&= n_{11} + n_{22} - \lp n_{11}+n_{22}\rp = 0.
\end{align}

Similar argument works for the case $\lbp n_{11}\le n_{12}, n_{22}\le n_{21}\rbp$. 

For the case $\lbp n_{11}\le n_{12}, n_{22}\ge n_{21}\rbp$, $(g_1,g_2)$ becomes
\begin{align}
g_1 = \max\lp n_{11}, n_{12}+n_{21}-n_{22}\rp,\ g_2 = \max\lp n_{11}+n_{22}-n_{12}, n_{21}\rp,
\end{align}
and the rank of the transfer matrix is $g_1+g_2$ again, since the subsystem (lies in the original system with $\lp y_{1h}, y_{2h}\rp$ as output and the corresponding levels at transmitters as input) has channel parameters
\begin{align}
n'_{11} = n_{11}, n'_{21}=n_{21}, n'_{12}= g_1, n'_{22} = g_2.
\end{align}
If $n_{11}+n_{22} > n_{12}+n_{21}$, then $g_1 = n_{11}$ and $g_2 = n_{11}+n_{22}-n_{12}$, and hence
\begin{align}
n'_{11}=n'_{12}, n'_{22} > n'_{21}.
\end{align}
Similarly, if $n_{11}+n_{22} < n_{12}+n_{21}$, then $g_1 = n_{12}+n_{21}-n_{22}$ and $g_2 = n_{21}$, and hence
\begin{align}
n'_{11} < n'_{12}, n'_{22} = n'_{21}.
\end{align}

Similar argument works for the case $\lbp n_{11}\ge n_{12}, n_{22}\le n_{21}\rbp$.

Therefore, under i.i.d. Bernoulli half inputs, $u_1$ and $u_2$ are independent conditioned on $\ul{x}_o$ from the analysis of the previous cases.
\end{proof}

For the noncooperative part, we set $x_{1cp} \overset{\rm{d}}{=} x_{1c} + x_{1p}$ such that $x_{1c}$ and $x_{1p}$ are independent. $x_{1c}$ is allowed to occupy all levels at transmitter 1, while $x_{1p}$ is allowed to occupy only the $\lp n_{11} - n_{21}\rp^+$ LSB's. The addition here is bit-wise modulo-two. The same design is applied to user 2. Note that with the above choice of $y_{2h}$ in \eqref{eq_g2}, there are no levels of $y_{2h}$ showing up at receiver 1 and hence no interference from $u_2$. Similar situation happens at receiver 2. Now take all inputs to be i.i.d. Bernoulli half across levels, we obtain a set of achievable rates from Theorem \ref{thm_GeneralAcieve}. After Fourier-Motzkin elimination, we show that the achievable region coincides with the region given in Theorem \ref{thm_LDCRegion}. 

\begin{lemma}\label{lem_LDC_FR}
The above strategy achieves the region given in Theorem \ref{thm_LDCRegion} when $n_{11}+n_{22}\ne n_{12}+n_{21}$.
\end{lemma}
\begin{proof}
The details are left in Appendix \ref{app_pf_LDCachieve}.
\end{proof}

\subsubsection{System matrix is not full-rank: $n_{11}+n_{22} = n_{12}+n_{21}$}
In this case, for the cooperative part, we shall again set $\ul{x}_o$ to be running over all transmit levels. The difference lies in the \emph{cooperative private} part. Here we also choose $\lp u_1, u_2\rp| \ul{x}_o \overset{\mathrm{d}}{=} \lp y_{1h}, y_{2h}\rp$, but occupying the following numbers of LSB's at receiver 1 and 2 respectively:
\begin{align}
g_1&=\lp n_{11} - n_{21}\rp^+,\
g_2=\lp n_{22} - n_{12}\rp^+.
\end{align}

\begin{claim}
Under i.i.d. Bernoulli half inputs, $I\lp u_1; u_2 | \ul{x}_o\rp = 0$ with the above choice if $n_{11}+n_{22} = n_{12}+n_{21}$.
\end{claim}
\begin{proof}
Since $y_{ih}$ only occupies levels that appear at receiver $i$ but do not appear at the other receiver, for $i=1,2$, hence they are conditionally independent given $\ul{x}_o$ under Bernoulli half i.i.d. inputs.
\end{proof}

For the noncooperative part, we use the same scheme as the previous case. Now take all inputs to be i.i.d. Bernoulli half across levels, we obtain achievable rates from Theorem \ref{thm_GeneralAcieve}. After Fourier-Motzkin elimination, we have the following lemma:

\begin{lemma}\label{lem_LDC_NFR}
The above strategy achieves the region given in Theorem \ref{thm_LDCRegion} when $n_{11}+n_{22} = n_{12}+n_{21}$.
\end{lemma}
\begin{proof}
The details are left in Appendix \ref{app_pf_LDCachieve}.
\end{proof}


We conclude this section with the following remark.
\begin{remark}[Implications about Gaussian Problem]\label{remark_Implications}
The two numbers $g_1$ and $g_2$ provide clues in determining the power allocated to the cooperative private codewords and the design of beamforming vectors in the Gaussian scenario. Take user 1 as an example. When $n_{11}+n_{22}\ne n_{12}+n_{21}$,
\begin{align}
g_1 &= \max\lbp n_{11} - \lp n_{21}-n_{22}\rp^+, n_{12} - \lp n_{22}-n_{21}\rp^+\rbp\\
&= \max\lbp n_{11}+n_{22} , n_{12}+n_{21}\rbp - \max\lbp n_{22}, n_{21}\rbp,
\end{align}
which corresponds to $\frac{|h_{11}h_{22}-h_{12}h_{21}|^2}{1+\SNR_2+\INR_2}$. On the other hand, when $n_{11}+n_{22} = n_{12}+n_{21}$, 
\begin{align}
g_1 = \lp n_{11}-n_{21}\rp^+ = \lp n_{12}-n_{22}\rp^+,
\end{align} 
which corresponds to $\frac{\SNR_1}{\INR_2} = \frac{\INR_1}{\SNR_2} = \frac{\SNR_1+\INR_1}{\SNR_2+\INR_2}$. This implies that the power of $u_1$ conditioned on $\ul{x}_o$ should be proportional to 
\begin{align}
\frac{|h_{11}h_{22}-h_{12}h_{21}|^2+\SNR_1+\INR_1}{1+\SNR_2+\INR_2} = \frac{|h_{11}h_{22}-h_{12}h_{21}|^2+|h_{11}|^2+|h_{22}|^2}{1+\SNR_2+\INR_2},
\end{align}
and that the beamforming vector should be a combination of zero-forcing and matched-filter vectors.
\end{remark}

\section{Gaussian Interference Channel with Conferencing Transmitters}\label{sec_GIC}

With the full understanding in linear deterministic interference channel with conferencing transmitters, now we have enough clues to crack the original Gaussian problem. As for the outer bounds, we shall mimic the genie-aided techniques and the structure of side informations in the proofs to develop outer bounds for the Gaussian interference channel with conferencing transmitters. As for the achievability, we shall mimic the choice of auxiliary random variables and level allocation to construct good schemes in the Gaussian scenario. Moreover, the achievable rate regions obtained prior to Fourier-Motzkin elimination can be made equivalent symbolically, and hence the proof of achieving approximate capacity in the Gaussian channel follows closely to the proof of achieving exact capacity in the linear deterministic channel. Although the Gaussian interference channel with conferencing transmitters and its corresponding linear deterministic channel are strongly related in coding strategies, proof of achievability, and outer bounds, unlike the two-user Gaussian interference channel \cite{BreslerTse_08}, their capacity regions are \emph{not} within a constant gap. Similar situations happen in MIMO channel, Gaussian relay networks \cite{AvestimehrDiggavi_09}, and Gaussian interference channel with conferencing receivers \cite{WangTse_09}, where explicit/implicit MIMO structures lie in the channel model.

Our main result is summarized in the following lemma and theorem.
\begin{lemma}[Outer Bounds]\label{OutBd_GIC}
If $\lp R_1, R_2\rp$ is achievable, it satisfies the following:
\begin{align}
R_1 &\le \min\lbp \log\lp 1+\SNR_1\rp +\C_{12}, \log\lp 1+\SNR_1+\INR_1+2\sqrt{\SNR_1\INR_1}\rp\rbp \label{bd_R1}\\
R_2 &\le \min\lbp \log\lp 1+\SNR_2\rp +\C_{21}, \log\lp 1+\SNR_2+\INR_2+2\sqrt{\SNR_2\INR_2}\rp\rbp \label{bd_R2}\\
R_1+R_2 &\le \log\lp 1+\frac{\SNR_1}{1+\INR_2}\rp + \log\lp 1+\SNR_2+\INR_2+2\sqrt{\SNR_2\INR_2}\rp + \C_{12} \label{bd_Sum1}\\
R_1+R_2 &\le \log\lp 1+\frac{\SNR_2}{1+\INR_1}\rp + \log\lp 1+\SNR_1+\INR_1+2\sqrt{\SNR_1\INR_1}\rp + \C_{21}\label{bd_Sum2}\\
R_1+R_2 &\le \lbp\begin{array}{l}
\log\lp 1+\frac{\SNR_1+2\sqrt{\SNR_1\INR_1}}{1+\INR_2} + \INR_1\rp\\
+\log\lp 1+\frac{\SNR_2+2\sqrt{\SNR_2\INR_2}}{1+\INR_1} + \INR_2\rp
\end{array}\rbp +\C_{12}+\C_{21}\label{bd_Sum3}\\
R_1+R_2 &\le \log\lp \begin{array}{l}1+\SNR_1+\INR_1+\SNR_2+\INR_2+2\sqrt{\SNR_1\INR_1}\\+2\sqrt{\SNR_2\INR_2}+|h_{11}h_{22}-h_{12}h_{21}|^2\end{array}\rp\label{bd_Sum4}\\
2R_1+R_2 &\le \lbp\begin{array}{l}\log\lp 1+\SNR_1+\INR_1+2\sqrt{\SNR_1\INR_1}\rp + \log\lp1+\frac{\SNR_1}{1+\INR_2}\rp\\ + \log\lp 1+\frac{\SNR_2+2\sqrt{\SNR_2\INR_2}}{1+\INR_1}+\INR_2\rp + \C_{12}+\C_{21}\end{array}\rbp\label{bd_SlopeTwo1}\\
R_1+2R_2 &\le \lbp\begin{array}{l}\log\lp 1+\SNR_2+\INR_2+2\sqrt{\SNR_2\INR_2}\rp + \log\lp1+\frac{\SNR_2}{1+\INR_1}\rp\\ + \log\lp 1+\frac{\SNR_1+2\sqrt{\SNR_1\INR_1}}{1+\INR_2}+\INR_1\rp + \C_{21}+\C_{12}\end{array}\rbp\label{bd_SlopeHalf1}\\
2R_1+R_2 &\le \lbp\begin{array}{l} \log\lp \begin{array}{l}1+\SNR_1+\INR_1+\SNR_2+\INR_2+\SNR_1\SNR_2\\+\INR_1\INR_2+\SNR_1\INR_2 + 2\lp1+\INR_2\rp\sqrt{\SNR_1\INR_1}\end{array}\rp\\
+\log\lp 1+\frac{\SNR_1}{1+\INR_2}\rp + 1 + \C_{12}\end{array}\rbp\label{bd_SlopeTwo2}\\
R_1+2R_2 &\le \lbp\begin{array}{l} \log\lp \begin{array}{l}1+\SNR_+\INR_1+\SNR_2+\INR_2+\SNR_1\SNR_2\\+\INR_1\INR_2+\SNR_2\INR_1 + 2\lp1+\INR_1\rp\sqrt{\SNR_2\INR_2}\end{array}\rp\\
+\log\lp 1+\frac{\SNR_2}{1+\INR_1}\rp + 1 + \C_{21}\end{array}\rbp\label{bd_SlopeHalf2}
\end{align}
\end{lemma}

\begin{theorem}[Constant Gap to Capacity]\label{thm_Gap}
The outer bounds in Lemma \ref{OutBd_GIC} is within $\log90 \approx 6.5$ bits per user to the capacity region.
\end{theorem}

\subsection{Outer Bounds}
Details of the proof of Lemma \ref{OutBd_GIC} are left in Appendix \ref{app_PfOutBd}. It follows closely to the techniques we develop in the proofs of LDC outer bounds. The only twist is how to mimic the proof of bound \eqref{eq_SlopeTwoBound2}, which is a new type of outer bound that does not appear in the case without cooperation. It corresponds to bound \eqref{bd_SlopeTwo2} here. Recall that in the proof there, we split receiver 2's signal into two parts: $y_2^N = \lp y_{2\alpha}^N, y_{2\beta}^N\rp$, where $y^N_{2\alpha}$ is the part of transmitter 2's signal that is not corrupted by $s_1^N$, the interference from transmitter 1. Such split is not possible in the Gaussian channel due to additive noise and carry-over in real addition. As shown in Appendix \ref{app_PfOutBd}, we will overcome this by providing the following side information to receiver 2:
\begin{align}
\wtild{y}_2^N := h_{22}x_2^N + \wtild{z}_2^N, 
\end{align}
where $\wtild{z}_2 \sim \mcal{CN}\lp 0, 1+\INR_2\rp$, i.i.d. over time and is independent of everything else. This mimics the signal $y_{2\alpha}^N$ in LDC, and helps us prove the $2R_1+R_2$ outer bound.

\subsection{Coding Strategy and Achievable Rates}
We shall employ the coding strategy proposed in Section \ref{subsec_Scheme}. The analysis in the linear deterministic setting suggests that, for the cooperative private messages, in the Gaussian setting one may choose its bearing auxiliary random variables $u_1$ and $u_2$ to be conditionally independent given $\ul{x}_o$. This implies that a simple linear beamforming strategy is sufficient. On the other hand, the interference should be zero-forced approximately. Based on this observation, we implement the following strategy.

For the cooperative common signal, recall that in the LDC we allow $\ul{x}_o$ to run over all transmit levels. To mimic it, in the Gaussian setting we choose $\ul{x}_o$ to be Gaussian with zero mean and a covariance matrix which has diagonal entries (values of transmit power) that are comparable with the total transmit power. For simplicity, we choose the covariance matrix to be diagonal:
\begin{align}
K_{\ul{x}_o} = \mathrm{diag}\lp Q_{1o}, Q_{2o}\rp,\ Q_{io} = 1/4,\ i=1,2.
\end{align}
Here the value $1/4$ is just a heuristic choice such that the transmit power constraints will be satisfied.

For the cooperative private signal, from the discussion in Remark \ref{remark_Implications}, we shall make it a superposition of \emph{zero-forcing} vectors 
\begin{align}
\ul{v}_{1z} &= \lb\begin{array}{c} h_{22}\\ -h_{21}\end{array}\rb,\ \ul{v}_{2z} = \lb\begin{array}{c} -h_{12}\\ h_{11}\end{array}\rb 
\end{align}
and \emph{matched-filter} vectors
\begin{align}
\ul{v}_{1m} &= \lb\begin{array}{c} h_{11}^*\\ h_{12}^*\end{array}\rb,\ \ul{v}_{2m} = \lb\begin{array}{c} h_{21}^*\\ h_{22}^*\end{array}\rb.
\end{align}
For the auxiliary random variables $u_1$ and $u_2$, we make them distributed as identical copies of user 1 and user 2's desired cooperative signals received at receiver 1 and 2 respectively. For example, $u_1$ would be the sum of the transmit cooperative common signal and user 1's cooperative private signal projected onto the channel vector $[h_{11}\ h_{12}]$.

Hence we choose $\lp \ul{x}_{oh}, \ul{x}_o, u_1, u_2\rp$ be jointly Gaussian such that 
\begin{align}
\ul{x}_{oh} &\overset{\rm{d}}{=} \ul{x}_o + \underset{\ul{x}_h}{\underbrace{w_{1z}\ul{v}_{1z} + w_{2z}\ul{v}_{2z} + w_{1m}\ul{v}_{1m} + w_{2m}\ul{v}_{2m}}}\\
u_1 &\overset{\rm{d}}{=} [h_{11}\ h_{12}]\lp \ul{x}_o + \ul{v}_{1z}w_{1z} + \ul{v}_{1m}w_{1m}\rp\\ 
u_2 &\overset{\rm{d}}{=} [h_{21}\ h_{22}]\lp \ul{x}_o + \ul{v}_{2z}w_{2z} + \ul{v}_{2m}w_{2m}\rp,
\end{align}
where 
$w_{1z}$, $w_{2z}$, $w_{1m}$, and $w_{2m}$ are independent Gaussians and independent of everything else, with variances $\theta_{1z}$, $\theta_{2z}$, $\theta_{1m}$, and $\theta_{2m}$ respectively. Their values are chosen such that  the total transmit power constraint will be met and the conditional variances of $u_1$ and $u_2$ conditioned on $\ul{x}_o$ behave as we predicted in Remark \ref{remark_Implications}. With this guideline, we choose 
\begin{align}
\theta_{1z} &= \frac{1/4}{\lp1+\SNR_2+\INR_2\rp} & \theta_{1m} &= \frac{1/4}{\lp\SNR_1+\INR_1\rp\lp1+\SNR_2+\INR_2\rp}\\
\theta_{2z} &= \frac{1/4}{\lp1+\SNR_1+\INR_1\rp} & \theta_{2m} &= \frac{1/4}{\lp\SNR_2+\INR_2\rp\lp1+\SNR_1+\INR_1\rp}.
\end{align}
Again, the factor $1/4$ is just a heuristic choice such that the transmit power constraints will be satisfied.

For the noncooperative part, we set $x_{ic} \sim \mcal{CN}\lp 0, Q_{ic}\rp$, where $Q_{ic} = 1/4 - Q_{ip}$, for $i=1,2$. $x_{icp} \overset{\rm{d}}{=} x_{ic}+x_{ip}$, where $x_{ip} \sim \mcal{CN}\lp 0,Q_{ip}\rp$ is independent of $x_{ic}$ and $Q_{ip} = \min\lp 1/4, 1/\INR_j\rp$, for $\lp i,j\rp = (1,2)$ or $(2,1)$. The choice of $Q_{ip}$ is such that the interference caused by the other user's \emph{noncooperative} private signal is at or below the noise level at the receiver.

At this stage, we shall check that the total transmit power constraint is met with the above heuristic choices of factors. We only need to show that the power for $\ul{x}_{h}$ at each transmitter is at most $1/2$, which is pretty straightforward\footnote{We have $Q_{1h} = \frac{\SNR_2 + \frac{\SNR_1}{\SNR_1+\INR_1}}{4\lp 1+\SNR_2+\INR_2\rp} +  \frac{\INR_1 + \frac{\INR_2}{\SNR_2+\INR_2}}{4\lp 1+\SNR_1+\INR_1\rp} \le \frac{1}{2}$, and vice versa for $Q_{2h}$.}.

Note that the variances of $u_1$ and $u_2$ conditioned on $\ul{x}_o$ are
\begin{align}
K_{u_1|\ul{x}_o} &= \frac{|h_{11}h_{22}-h_{12}h_{21}|^2+\SNR_1+\INR_1}{4\lp1+\SNR_2+\INR_2\rp},\\
K_{u_2|\ul{x}_o} &= \frac{|h_{11}h_{22}-h_{12}h_{21}|^2+\SNR_2+\INR_2}{4\lp1+\SNR_1+\INR_1\rp},
\end{align}
matching our prediction in Remark \ref{remark_Implications}.

With this encoding, the interference caused by the other user's \emph{cooperative} private signal should be nulled out approximately, that is, its variance is at or below the noise level. To see this, the received signals are jointly distributed with $\lp \ul{x}_o, u_1, u_2, x_{1c}, x_{1cp}, x_{2c}, x_{2cp}\rp$ such that
\begin{align}
y_1 &\overset{\rm{d}}{=} u_1 + \what{z}_1 + h_{11}x_{1cp} + h_{12}x_{2cp} + z_1\\
y_1 &\overset{\rm{d}}{=} u_2 + \what{z}_2 + h_{21}x_{1cp} + h_{22}x_{2cp} + z_2,
\end{align}
where the interferences caused by undesired cooperative private signals are
\begin{align}
\what{z}_1 &= \lp h_{11}h_{21}^* + h_{12}h_{22}^*\rp w_{2m},\ 
\what{z}_2 = \lp h_{21}h_{11}^* + h_{22}h_{12}^*\rp w_{1m},
\end{align}
at receiver 1 and 2 respectively. Note that the variance of these terms are upper bounded by a constant, since
\begin{align}
&|h_{11}h_{21}^* + h_{12}h_{22}^*|^2 = |h_{21}h_{11}^* + h_{22}h_{12}^*|^2\\
&= \lp\SNR_1+\INR_1\rp\lp\SNR_2+\INR_2\rp - |h_{11}h_{22} - h_{12}h_{21}|^2\\
&\le \lp\SNR_1+\INR_1\rp\lp\SNR_2+\INR_2\rp.
\end{align}
Hence, 
\begin{align}
\sigma_1^2 &:= \Var\lp \what{z}_1\rp = \frac{|h_{11}h_{21}^* + h_{12}h_{22}^*|^2}{4\lp\SNR_1+\INR_1\rp\lp1+\SNR_2+\INR_2\rp} \le \frac{1}{4}\\
\sigma_2^2 &:= \Var\lp \what{z}_2\rp = \frac{|h_{21}h_{11}^* + h_{22}h_{12}^*|^2}{4\lp1+\SNR_1+\INR_1\rp\lp\SNR_2+\INR_2\rp} \le \frac{1}{4},
\end{align}
and in effect the interference is nulled out approximately.

\begin{remark}
When the cooperative link capacities are sufficiently large and the channel becomes a two-user Gaussian MIMO broadcast channel with two transmit antennas and single receive antenna at each receiver, the proposed scheme in Theorem \ref{thm_GeneralAcieve} is capacity-achieving. Dirty paper coding among cooperative private messages is needed to achieve the capacity of Gaussian MIMO broadcast channel \emph{exactly} \cite{WeingartenSteinberg_06}, that is, $u_1$ and $u_2$ is not independent conditioned on $\ul{x}_o$ and $\ul{x}_o$ is made zero. As shown in Appendix \ref{app_GICAchieve} and \ref{app_PfThmGap}, however, linear beamforming strategies along with superposition coding suffice to achieve the capacity \emph{approximately}. We conjecture that dirty paper coding among cooperative private messages will lead to a better rate region and smaller gap to the outer bounds, while the procedure of computing the achievable region becomes complicated.
\end{remark}

We have designed a coding strategy and its configuration which met the observation and intuition from the analysis of LDC, and it turns out that it achieves the capacity to within a constant gap. This completes the proof of Theorem \ref{thm_Gap}. The proof is broken into two parts: (1) the computation of the achievable rate region, and (2) the evaluation of the gap among inner and outer bounds. Details are left in Appendix \ref{app_GICAchieve} and Appendix \ref{app_PfThmGap} respectively.


\section{Uplink-Downlink Reciprocity}\label{sec_Reciprocity}
Recall that in Section \ref{subsec_LinearScheme} we have demonstrated the reciprocity between linear deterministic interference channel with conferencing receivers and linear deterministic interference channel with conferencing transmitters. In this section, we show that a similar reciprocity holds in the Gaussian case.

\begin{figure}[htbp]
{\center
\subfigure[Original Gaussian Downlink Scenario]{\includegraphics[width=3in]{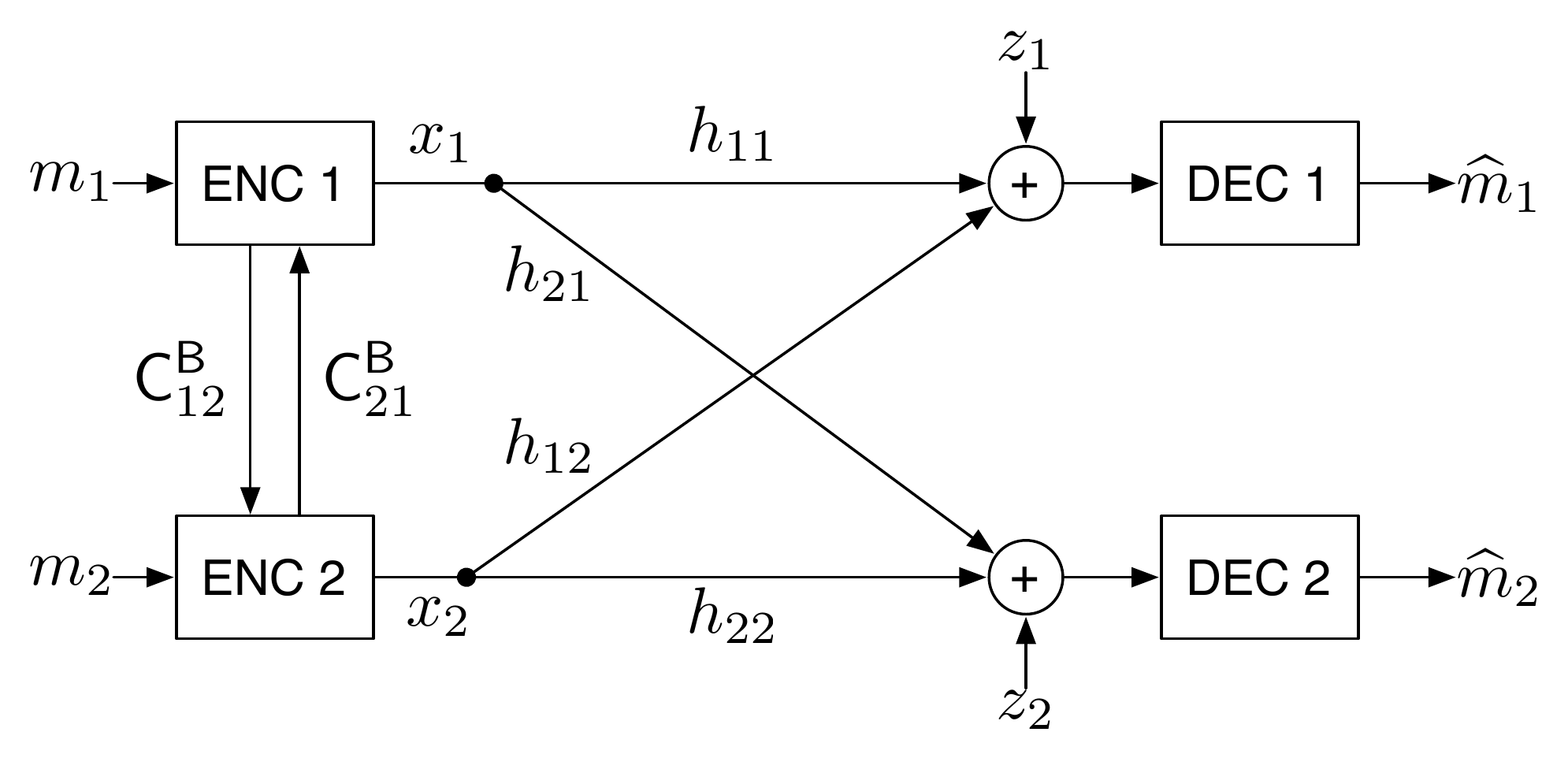}}
\subfigure[Original LDC Downlink Scenario]{\includegraphics[width=2in]{LDC_dl.pdf}}
\subfigure[Reciprocal Gaussian Uplink Scenario]{\includegraphics[width=3in]{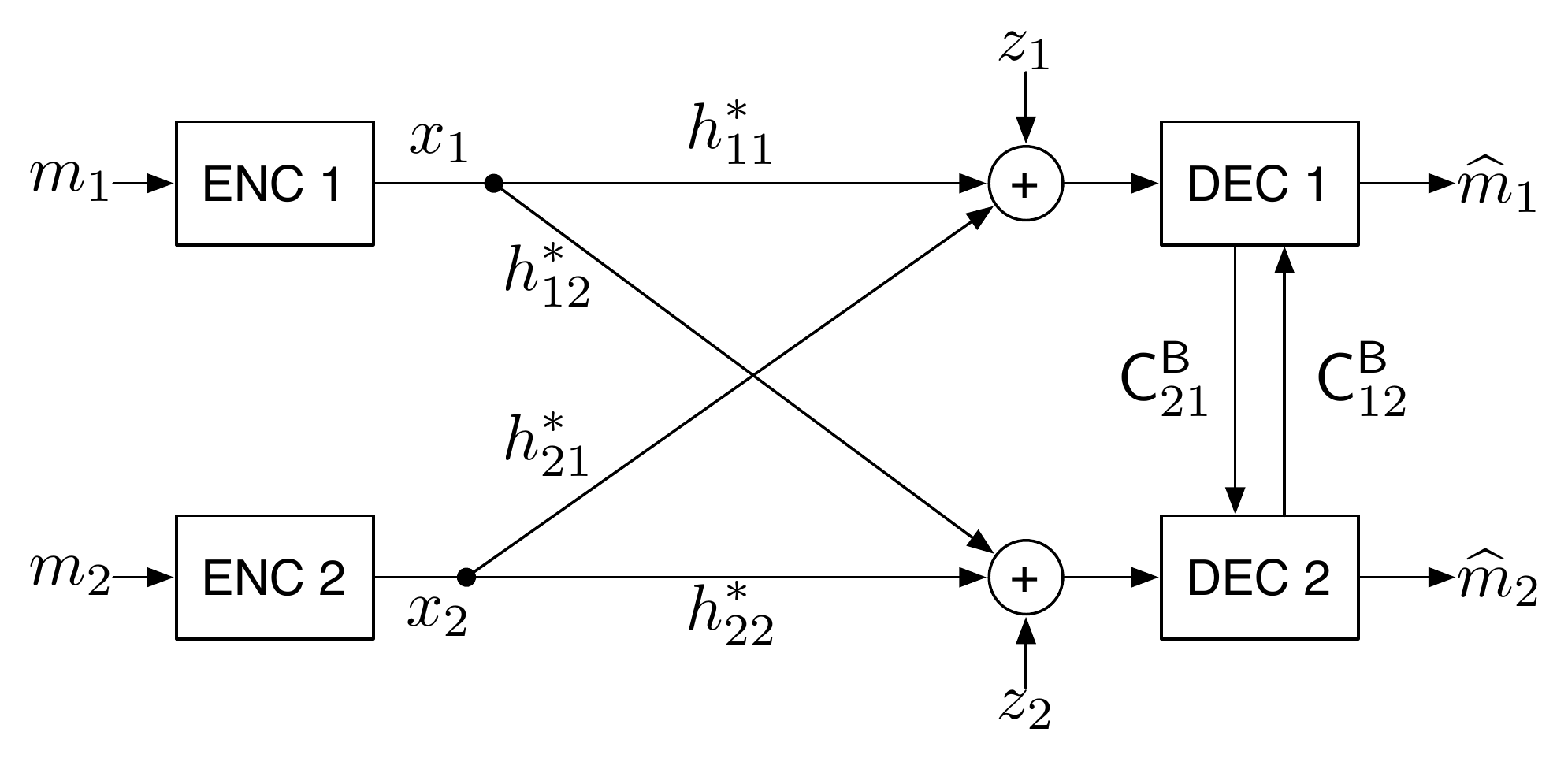}}
\subfigure[Reciprocal LDC Uplink Scenario]{\includegraphics[width=2in]{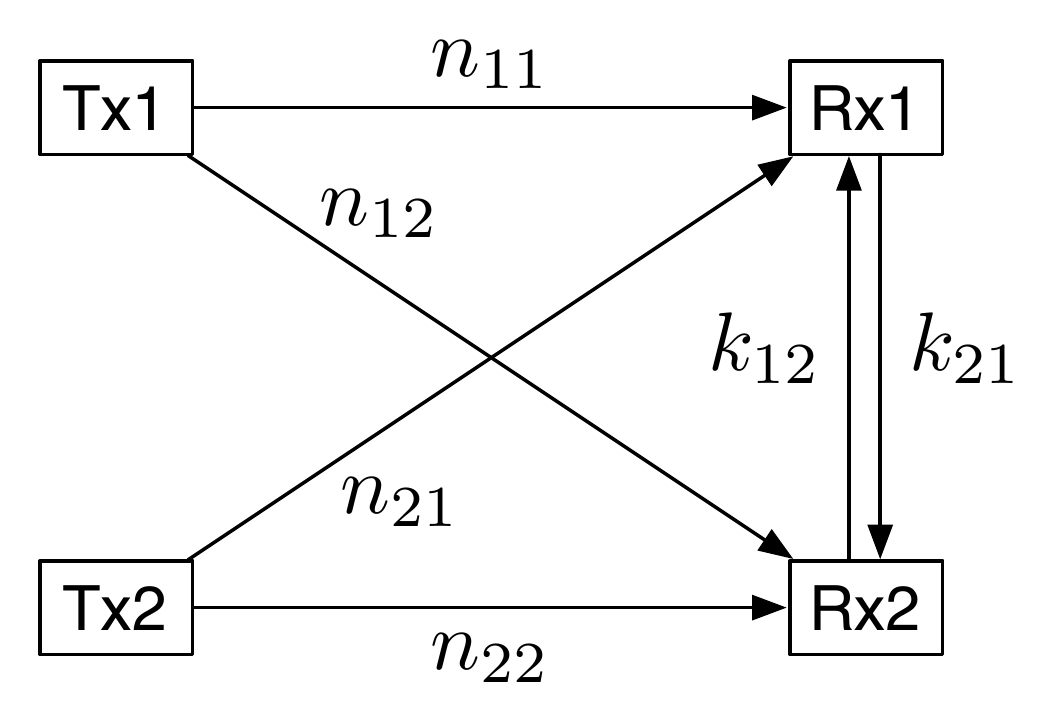}}
\caption{Uplink-Downlink Reciprocity}
\label{fig_Duality}
}
\end{figure}

For the channel described in Section \ref{sec_Formulation}, we define its \emph{reciprocal} channel as the Gaussian interference channel with conferencing receivers \cite{WangTse_09} with the 2-by-2 channel matrix
\begin{align}
\lb\begin{array}{cc}
h_{11}& h_{12}\\
h_{21}& h_{22}
\end{array}\rb ^H = 
\lb\begin{array}{cc}
h^*_{11}& h^*_{21}\\
h^*_{12}& h^*_{22}
\end{array}\rb
\end{align}
and cooperative link capacities $\C_{21}$ \emph{from receiver 1 to 2} and $\C_{12}$ \emph{from receiver 2 to 1}. Note that for the reciprocal channel, the channel matrix is the Hermitian of the original one and the cooperative link capacities are swapped. Motivated by backhaul cooperation in cellular networks where cooperation is among base stations, we term the interference channel with conferencing receivers the \emph{uplink} scenario, and the interference channel with conferencing transmitters the \emph{downlink} scenario. The original downlink and the reciprocal uplink scenarios are depicted in Fig. \ref{fig_Duality}.

\begin{theorem}\label{thm_Reciprocity}
The capacity regions of the original and the reciprocal channels are within a constant number of bits, regardless of channel parameters.
\end{theorem}
\begin{proof}
Details are left in Appendix \ref{app_PfReciprocity}.
\end{proof}

 

The reciprocity implies immediately that the gain from transmitter cooperation shares the same characteristics as that from receiver cooperation, that is, the degree-of-freedom gain is either one bit or half a bit per cooperation bit until saturation, and the power gain is at most a constant no matter how large the cooperative link capacities are after saturation.

\begin{remark}
As mentioned in Section \ref{subsec_LinearScheme}, there is an \emph{exact} reciprocity between the linear deterministic downlink scenario and the uplink scenario.
Not only are the capacity regions of the original and the reciprocal channel the same, but the capacity-achieving linear schemes are also reciprocal. 
On the other hand, for the Gaussian downlink scenario and the uplink scenario, combining the results in this paper and \cite{WangTse_09}, it seems such reciprocity in the proposed strategies does not exist, since the message structures are different. Although the strategies proposed in this paper and \cite{WangTse_09} are not reciprocal, we conjecture that such reciprocity may be obtained via structured lattice strategies derived from capacity-achieving linear schemes of the corresponding linear deterministic channels. Such conversion has been applied successfully in \cite{BreslerParekh_08} to construct lattice coding strategies for many-to-one and one-to-many Gaussian interference channels.
\end{remark}



\section*{Acknowledgement}
The authors would like to thank Prof. Emre Telatar for sharing his Fourier-Motzkin elimination program.

\bibliographystyle{ieeetr}

\newpage

\appendices

\section{Proof of Achievability in Theorem \ref{thm_LDCRegion}}\label{app_pf_LDCachieve}
\subsection{Proof of Lemma \ref{lem_LDC_FR}}

Plugging in the configuration, we have the following achievable rates from Theorem \ref{thm_GeneralAcieve}: for some nonnegative $(\wtild{R}_{1h}, \wtild{R}_{2h})$, (notations are listed in Table \ref{table_Notations})

{\flushleft \ul{Constraints at Receiver 1}}:
\begin{align}
R_{1p} &\le p_1\\ 
\wtild{R}_{1h} &\le g_1 & R_o + \wtild{R}_{1h} &\le m_1\\
\wtild{R}_{1h} + R_{1p} &\le g_1 & R_o + \wtild{R}_{1h} + R_{1p} &\le m_1\\
R_{2c} + R_{1p} &\le t_1\\ 
R_{1c} + R_{1p} &\le n_{11}\\ 
R_{2c} + \wtild{R}_{1h} &\le s_1 & R_o + R_{2c} + \wtild{R}_{1h} &\le m_1\\
R_{2c} + \wtild{R}_{1h} + R_{1p} &\le s_1 & R_o + R_{2c} + \wtild{R}_{1h} + R_{1p} &\le m_1\\
R_{1c} + \wtild{R}_{1h} + R_{1p} &\le l_1 & R_o + R_{1c} + \wtild{R}_{1h} + R_{1p} &\le m_1\\
R_{1c} + R_{2c} + R_{1p} &\le m_1\\ 
R_{1c} + R_{2c} + \wtild{R}_{1h} + R_{1p} &\le m_1 & R_o + R_{1c} + R_{2c} + \wtild{R}_{1h} + R_{1p} &\le m_1
\end{align}

{\flushleft \ul{Constraints at Receiver 2}}: Above with index 1 and 2 exchanged.

\begin{table}[htdp]
\caption{Notations}
\begin{center}
\begin{tabular}{|c|c|c|c|c|}
\hline
$p_1$& $t_1$& $m_1$& $l_1$& $s_1$\\
\hline
$\lp n_{11}-n_{21}\rp^+$ & $\max\lp n_{12}, p_1\rp$ & $\max\lp n_{11}, n_{12}\rp$ & $\max\lp n_{11}, g_1\rp$ & $\max\lp n_{12}, g_1\rp$\\
\hline \hline
$p_2$& $t_2$& $m_2$& $l_2$& $s_2$\\
\hline
$\lp n_{22}-n_{12}\rp^+$ & $\max\lp n_{21}, p_2\rp$ & $\max\lp n_{22}, n_{21}\rp$ & $\max\lp n_{22}, g_2\rp$ & $\max\lp n_{21}, g_2\rp$\\
\hline
\end{tabular}
\end{center}
\label{table_Notations}
\end{table}

{\flushleft \ul{Constraints at Transmitters}}:
\begin{align}
R_{1h} &\le \wtild{R}_{1h}\\
R_{2h} &\le \wtild{R}_{2h}\\
R_{1o}+ R_{2o} &= R_o\\
R_{1o} + R_{1h} &\le k_{12}\\
R_{2o} + R_{2h} &\le k_{21}\\
\wtild{R}_{1h} + \wtild{R}_{2h} - R_{1h} -R_{2h} &\ge 0
\end{align}

After Fourier-Motzkin elimination, we have the following achievable rates and identify all redundant terms. The claims used below to show the redundancy are proved in the end of this section.

{\flushleft (1) $R_1$ and $R_2$}:
\begin{align}
R_1 &\le n_{11} + k_{12} & & & R_2 &\le n_{22} + k_{21}\\
R_1 &\le m_1 & & & R_2 &\le m_2\\
R_1 &\le p_1+t_2 + k_{12} & & \textrm{redundant} & R_2 &\le p_2+t_1 + k_{21} & & \textrm{redundant}\\
R_1 &\le p_1+s_2 + k_{12} & & \textrm{redundant} & R_2 &\le p_2+s_1 + k_{21} & & \textrm{redundant}\\
R_1 &\le g_1+t_2 + k_{12} & & \textrm{redundant} & R_2 &\le g_2+t_1 + k_{21} & & \textrm{redundant}\\
R_1 &\le g_1+s_2 + k_{12} & & \textrm{redundant} & R_2 &\le g_2+s_1 + k_{21} & & \textrm{redundant}.
\end{align}

To show the redundancy, we need to prove the following claim
\begin{claim}\label{claim_LDC2}
$ $
\begin{itemize}
\item $p_1+t_2 \ge n_{11}$, $p_2+t_1 \ge n_{22}$
\item $g_1\ge p_1$, $g_2\ge p_2$
\item $s_1 \ge t_1$, $s_2 \ge t_2$
\end{itemize}
\end{claim}

{\flushleft (2) $R_1+R_2$}:
\begin{align}
R_1+R_2 &\le t_1+t_2 + k_{12}+k_{21}\\
R_1+R_2 &\le p_1+m_2 + k_{12}\\
R_1+R_2 &\le p_2+m_1 + k_{21}\\
R_1+R_2 &\le g_1+m_2\\
R_1+R_2 &\le g_2+m_1\\
R_1+R_2 &\le s_1+t_2 + k_{12}+k_{21} & & \textrm{redundant}\\
R_1+R_2 &\le s_2+t_1 + k_{12}+k_{21} & & \textrm{redundant}\\
R_1+R_2 &\le s_1+s_2 + k_{12}+k_{21} & & \textrm{redundant}
\end{align}

{\flushleft (3) $2R_1+R_2$ and $R_1+2R_2$}:
\begin{align}
2R_1+R_2 &\le p_1+m_1+t_2 + k_{12}+k_{21}\\
2R_1+R_2 &\le g_1+m_1+t_2 + k_{12}+k_{21}& & \textrm{redundant}\\
2R_1+R_2 &\le p_1+m_1+s_2 + k_{12}\\
2R_1+R_2 &\le g_1+m_1+s_2 + k_{12}& & \textrm{redundant}\\
2R_1+R_2 &\le p_1+l_1+m_2 + k_{12}\\
R_1+2R_2 &\le p_2+m_2+t_1 + k_{12}+k_{21}\\
R_1+2R_2 &\le g_2+m_2+t_1 + k_{12}+k_{21}& & \textrm{redundant}\\
R_1+2R_2 &\le p_2+m_2+s_1 + k_{21}\\
R_1+2R_2 &\le g_2+m_2+s_1 + k_{21}& & \textrm{redundant}\\
R_1+2R_2 &\le p_2+l_2+m_1 + k_{21}.
\end{align}

{\flushleft (4) $2R_1+2R_2$}:
\begin{align}
2R_1+2R_2 &\le p_1+s_1+t_2+m_2 + k_{12}+k_{21}& & \textrm{redundant}\\
2R_1+2R_2 &\le l_1+t_1+p_2+m_2 + k_{12}+k_{21}& & \textrm{redundant}\\
2R_1+2R_2 &\le p_1+s_1+s_2+m_2 + k_{12}+k_{21}& & \textrm{redundant}\\
2R_1+2R_2 &\le l_1+t_1+g_2+m_2 + k_{12}+k_{21}& & \textrm{redundant}\\
2R_1+2R_2 &\le t_1+m_1+p_2+s_2 + k_{12}+k_{21}& & \textrm{redundant}\\
2R_1+2R_2 &\le s_1+m_1+p_2+s_2 + k_{12}+k_{21}& & \textrm{redundant}\\
2R_1+2R_2 &\le p_1+m_1+l_2+t_2 + k_{12}+k_{21}& & \textrm{redundant}\\
2R_1+2R_2 &\le g_1+m_1+l_2+t_2 + k_{12}+k_{21}& & \textrm{redundant}
\end{align}

To show the redundancy, we need to prove the following claim:
\begin{claim}\label{claim_LDC3}
$ $
\begin{itemize}
\item $s_1+t_2 \ge p_2 + m_1$; $s_2+t_1 \ge p_1 + m_2$
\item $l_1+t_1 \ge p_1+m_1$; $l_2+t_2 \ge p_2+m_2$
\end{itemize}
\end{claim}

After removing the redundant terms, we have the following achievable region for LDC when $n_{11}+n_{22} \ne n_{12}+n_{21}$:
\begin{align}
R_1 &\le \min \lbp n_{11} + k_{12}, m_1\rbp\\
R_2 &\le \min\lbp n_{22} + k_{21}, m_2\rbp\\
R_1+R_2 &\le \min\lbp g_1+m_2, g_2+m_1\rbp\\
R_1+R_2 &\le t_1+t_2+k_{12}+k_{21}\\
R_1+R_2 &\le \min\lbp p_1 + m_2 + k_{12}, p_2 + m_1 + k_{21}\rbp\\
2R_1+R_2 &\le \min\lbp p_1 + l_1 + m_2 + k_{12}, p_1 + s_2 + m_1 + k_{12}\rbp\\
2R_1+R_2 &\le p_1 + m_1 + t_2 + k_{12}+k_{21}\\
R_1+2R_2 &\le \min\lbp p_2 + l_2 + m_1 + k_{21}, p_2 + s_1 + m_2 + k_{21}\rbp\\
R_1+2R_2 &\le p_2 + m_2 + t_1 + k_{21}+k_{12}.
\end{align}

To show that the above achievable region coincides with the rate region given in Theorem \ref{thm_LDCRegion}, the following facts are crucial:
\begin{claim}\label{claim_LDC4}
$ $
\begin{itemize}
\item $g_1+m_2 = g_2+m_1 = \max\lp n_{11}+n_{22}, n_{12}+n_{21}\rp$.
\item $s_2 + m_1 =  l_1 + m_2$; $s_1 + m_2 =  l_2 + m_1$
\end{itemize}
\end{claim}

With these facts, referring to Table \ref{table_Notations}, and checking with the outer bounds, we complete the proof.

\subsection{Proof of Lemma \ref{lem_LDC_NFR}}

We have the following achievable rates: for some nonnegative $(\wtild{R}_{1h}, \wtild{R}_{2h})$,

{\flushleft \ul{Constraints at Transmitters}}:
\begin{align}
R_{1h} &\le \wtild{R}_{1h}\\
R_{2h} &\le \wtild{R}_{2h}\\
R_{1o}+ R_{2o} &= R_o\\
R_{1o} + R_{1h} &\le k_{12}\\
R_{2o} + R_{2h} &\le k_{21}\\
\wtild{R}_{1h} + \wtild{R}_{2h} - R_{1h} -R_{2h} &\ge 0
\end{align}

{\flushleft \ul{Constraints at Receiver 1}}:
\begin{align}
R_{1p} &\le p_1\\ 
\wtild{R}_{1h} &\le p_1 & R_o + \wtild{R}_{1h} &\le m_1\\
\wtild{R}_{1h} + R_{1p} &\le p_1 & R_o + \wtild{R}_{1h} + R_{1p} &\le m_1\\
R_{2c} + R_{1p} &\le t_1\\ 
R_{1c} + R_{1p} &\le n_{11}\\ 
R_{2c} + \wtild{R}_{1h} &\le t_1 & R_o + R_{2c} + \wtild{R}_{1h} &\le m_1\\
R_{2c} + \wtild{R}_{1h} + R_{1p} &\le t_1 & R_o + R_{2c} + \wtild{R}_{1h} + R_{1p} &\le m_1\\
R_{1c} + \wtild{R}_{1h} + R_{1p} &\le n_{11} & R_o + R_{1c} + \wtild{R}_{1h} + R_{1p} &\le m_1\\
R_{1c} + R_{2c} + R_{1p} &\le m_1\\ 
R_{1c} + R_{2c} + \wtild{R}_{1h} + R_{1p} &\le m_1 & R_o + R_{1c} + R_{2c} + \wtild{R}_{1h} + R_{1p} &\le m_1
\end{align}

{\flushleft \ul{Constraints at Receiver 2}}: Above with index 1 and 2 exchanged.

After Fourier-Motzkin elimination and removing redundant terms based on facts derived in the previous analysis, we have the following achievable region:
\begin{align}
R_1 &\le \min\lbp m_1, n_{11} + k_{12}\rbp\\
R_2 &\le \min\lbp m_2, n_{22} + k_{21}\rbp\\
R_1+R_2 &\le \min\lbp p_1+m_2, p_2+m_1, t_1+t_2+k_{12}+k_{21}\rbp\\
2R_1+R_2 &\le p_1 + t_2 + m_1 + k_{12} & &\textrm{redundant}\\
R_1+2R_2 &\le p_2 + t_1 + m_2 + k_{21} & &\textrm{redundant},
\end{align}
which coincides with the outer bounds. To prove this, we need the following facts:
\begin{claim}\label{claim_LDC5}
$ $
\begin{itemize}
\item $p_1+m_2 = p_2+m_1 = \max\lp n_{11}, n_{22}, n_{12}, n_{21}\rp$
\item $p_1+t_2 = p_2+n_{11}$; $p_2+t_1 = p_1+n_{22}$
\end{itemize}
\end{claim}

With the first fact $p_1+m_2 = p_2+m_1 = \max\lp n_{11}, n_{22}, n_{12}, n_{21}\rp$, we show that the sum rate inner bound coincides the outer bound. With the second fact $p_1+t_2 = p_2+n_{11}$, we show that the $2R_1+R_2$ inner bound is redundant. Similarly the $R_1+2R_2$ inner bound is also redundant. This completes the proof.

\subsection{Proof of the Claims}
\subsection*{Proof of Claim \ref{claim_LDC2}}
\begin{itemize}
\item $p_1+t_2 \ge n_{11}$, $p_2+t_1 \ge n_{22}$
\begin{proof}
\begin{align}
p_1+t_2 &\ge \lp n_{11}-n_{21}\rp^+ + n_{21} \ge n_{11}\\
p_2+t_1 &\ge \lp n_{22}-n_{12}\rp^+ + n_{12} \ge n_{22}.
\end{align}
\end{proof}
\item $g_1\ge p_1$, $g_2\ge p_2$
\begin{proof}
\begin{align}
g_1 &= \max\lbp n_{11} - \lp n_{21}-n_{22}\rp^+, n_{12} - \lp n_{22}-n_{21}\rp^+\rbp\\
&\ge n_{11} - \lp n_{21}-n_{22}\rp^+ \ge n_{11} - n_{21}.
\end{align}
On the other hand, $g_1 \ge 0$. Hence, $g_1 \ge \lp n_{11}-n_{21}\rp^+ = p_1$. Similarly $g_2 \ge p_2$.
\end{proof}
\item $s_1 \ge t_1$, $s_2 \ge t_2$
\begin{proof}
\begin{align}
s_1 = \max\lp n_{12}, g_1\rp \ge \max\lp n_{12}, p_1\rp = t_1,
\end{align}
since $g_1 \ge p_1$. Similarly $s_2 \ge t_2$.
\end{proof}
\end{itemize}

\subsection*{Proof of Claim \ref{claim_LDC3}}
\begin{itemize}
\item $s_1+t_2 \ge p_2 + m_1$; $s_2+t_1 \ge p_1 + m_2$
\begin{proof}
If $n_{21}\le n_{22}$,
\begin{align}
s_1+t_2 = \max\lbp n_{12}, n_{11} - \lp n_{21}-n_{22}\rp^+\rbp + t_2 = m_1 + t_2 \ge m_1+p_2.
\end{align}

If $n_{21} > n_{22}$ and $n_{22}\le n_{12}$,
\begin{align}
s_1+t_2 &= \max\lbp n_{12}, n_{11}+n_{22} - n_{21}\rbp + n_{21}\\
&= \max\lbp n_{12}+n_{21}, n_{11}+n_{22} \rbp \\
&\ge 0 + \max\lp n_{11}, n_{12}\rp = p_2 + m_1. 
\end{align}

If $n_{21} > n_{22}$ and $n_{22}> n_{12}$,
\begin{align}
s_1+t_2 &= \max\lbp n_{12}, n_{11}+n_{22} - n_{21}\rbp + n_{21}\\
&= \max\lbp n_{12}+n_{21}, n_{11}+n_{22} \rbp \\
&\ge \max\lbp n_{11}+n_{22} - n_{12}, n_{21}\rbp \ge \max\lbp n_{11}+n_{22} - n_{12}, n_{22}\rbp\\
&\ge n_{22} - n_{12} + \max\lp n_{11}, n_{12}\rp = p_2 + m_1. 
\end{align}

Hence, $s_1+t_2 \ge p_2 + m_1$. Similarly, $s_2+t_1 \ge p_1 + m_2$.
\end{proof}

\item $l_1+t_1 \ge p_1+m_1$; $l_2+t_2 \ge p_2+m_2$
\begin{proof}
If $n_{21} \ge n_{22}$,
\begin{align}
l_1 + t_1 = \max\lbp n_{11}, n_{12}-\lp n_{22}-n_{21}\rp^+\rbp + t_1 = m_1 + t_1 \ge m_1 + p_1.
\end{align}

If $n_{21} < n_{22}$ and $n_{11} \ge n_{12}$,
\begin{align}
p_1 + m_1 = p_1 + n_{11} \le t_1 + l_1.
\end{align}

If $n_{21} < n_{22}$ and $n_{11} < n_{12}$, then $m_1 = t_1 = n_{12}$, hence
\begin{align}
p_1 + m_1 = p_1 + t_1 \le n_{11} + t_1 \le  l_1+ t_1.
\end{align}

Hence, $l_1+t_1 \ge p_1+m_1$. Similarly, $l_2+t_2 \ge p_2+m_2$
\end{proof}
\end{itemize}

\subsection*{Proof of Claim \ref{claim_LDC4}}
\begin{itemize}
\item $g_1+m_2 = g_2+m_1 = \max\lp n_{11}+n_{22}, n_{12}+n_{21}\rp$.
\begin{proof}
Note that 
\begin{align}
\max\lp n_{22}, n_{21}\rp - \lp n_{21}-n_{22}\rp^+ &= n_{22}\\
\max\lp n_{22}, n_{21}\rp - \lp n_{22}-n_{21}\rp^+ &= n_{21}.
\end{align}

Hence,
\begin{align}
&g_1 + m_2\\
&= \max\lbp n_{11} - \lp n_{21}-n_{22}\rp^+, n_{12} - \lp n_{22}-n_{21}\rp^+\rbp + \max\lp n_{22}, n_{21}\rp\\
&= \max\lbp n_{11}+n_{22}, n_{12}+n_{21} \rbp.
\end{align}

By symmetry, $g_2+m_1 = \max\lp n_{11}+n_{22}, n_{12}+n_{21}\rp$.
\end{proof}

\item $s_2 + m_1 =  l_1 + m_2$; $s_1 + m_2 =  l_2 + m_1$
\begin{proof}
\begin{align}
&s_2 + m_1\\
&= \max\lbp n_{21}, n_{22} - \lp n_{12} - n_{11}\rp^+\rbp + \max\lp n_{11}, n_{12}\rp\\
&= \max\lbp n_{21}+\max\lp n_{11}, n_{12}\rp, n_{22}+n_{11}\rbp\\
&= \max\lbp n_{21}+ n_{11}, n_{21}+n_{12}, n_{22}+n_{11}\rbp;\\
&l_1 + m_2\\
&= \max\lbp n_{11}, n_{12} - \lp n_{22}-n_{21}\rp^+\rbp + \max\lp n_{22}, n_{21}\rp\\
&= \max\lbp n_{11}+\max\lp n_{22}, n_{21}\rp, n_{12}+n_{21}\rbp\\
&= \max\lbp n_{11}+n_{22}, n_{11}+n_{21}, n_{12}+n_{21}\rbp.
\end{align}
Hence $s_2 + m_1 =  l_1 + m_2$. By symmetry $s_1 + m_2 =  l_2 + m_1$.
\end{proof}
\end{itemize}

\subsection*{Proof of Claim \ref{claim_LDC5}}
\begin{itemize}
\item $p_1+m_2 = p_2+m_1 = \max\lp n_{11}, n_{22}, n_{12}, n_{21}\rp$
\begin{proof}
If $n_{11} \ge n_{21} \ge n_{22}$, then $n_{11} \ge n_{12}$ (otherwise contradicts the assumption $n_{11}+n_{22} = n_{12}+n_{21}$) and
\begin{align}
p_1+m_2 = n_{11} - n_{21} + n_{21} = n_{11} = \max_{i,j\in\{1,2\}}\lbp n_{ij}\rbp.
\end{align}

If $n_{21} \ge n_{22}$ and $n_{21} \ge n_{11}$, then $n_{21} \ge n_{12}$ (contradiction otherwise) and 
\begin{align}
p_1+m_2 = 0 + n_{21} = n_{21} = \max_{i,j\in\{1,2\}}\lbp n_{ij}\rbp.
\end{align}

If $n_{21} \le n_{22}$ and $n_{21} \le n_{11}$, then $n_{12} \ge n_{11}$ and $n_{12} \ge n_{22}$ (contradiction otherwise) and 
\begin{align}
p_1+m_2 = n_{11} - n_{21} + n_{22} = n_{12} = \max_{i,j\in\{1,2\}}\lbp n_{ij}\rbp.
\end{align}

If $n_{11} \le n_{21} \le n_{22}$, then $n_{22} \ge n_{12}$ (contradiction otherwise) and 
\begin{align}
p_1+m_2 = 0 + n_{22} = n_{22} = \max_{i,j\in\{1,2\}}\lbp n_{ij}\rbp.
\end{align}

Hence, $p_1+m_2 = \max_{i,j\in\{1,2\}}\lbp n_{ij}\rbp$. Similarly,  $p_2+m_1 = \max_{i,j\in\{1,2\}}\lbp n_{ij}\rbp$.
\end{proof}

\item $p_1+t_2 = p_2+n_{11}$; $p_2+t_1 = p_1+n_{22}$
\begin{proof}
Note that $t_2 = \max\lbp n_{21}, \lp n_{22} - n_{12} \rp^+\rbp = n_{21}$, since $n_{21} \ge 0$ and 
\begin{align}
n_{21} = n_{11}+n_{22} - n_{12} \ge n_{22} - n_{12}.
\end{align}

Hence,
\begin{align}
p_1 + t_2 = \lp n_{11} - n_{21}\rp^+ + n_{21} = \max\lp n_{11}, n_{21}\rp
\end{align}

On the other hand, 
\begin{align}
p_2 + n_{11} = \lp n_{22}-n_{12}\rp^+ + n_{11} =  \lp n_{21}-n_{11}\rp^+ + n_{11} = \max\lp n_{11}, n_{21}\rp
\end{align}

Hence, $p_1+t_2 = p_2+n_{11}$. Similarly, $p_2+t_1 = p_1+n_{22}$.
\end{proof}
\end{itemize}

\section{Proof of Theorem \ref{thm_Gap}: Achievable Rate Region}\label{app_GICAchieve}

Plug in Theorem \ref{thm_GeneralAcieve} and evaluate, we obtain the following achievable rates:

{\flushleft \ul{Constraints at Receiver 1}}:
\begin{align}
R_{1p} &\le \log\lp 1+\frac{\SNR_{1p}}{1+\sigma_1^2+\INR_{1p}}\rp \\	
\wtild{R}_{1h} &\le \log\lp 1+ \frac{K_{u_1|\ul{x}_o}}{1+\sigma_1^2+\INR_{1p}}\rp\\						
\wtild{R}_{1h} + R_{1p} &\le \log\lp 1+ \frac{K_{u_1|\ul{x}_o}+\SNR_{1p}}{1+\sigma_1^2+\INR_{1p}}\rp\\
R_{2c} + R_{1p} &\le \log\lp 1+\frac{\INR_{1c}+\SNR_{1p}}{1+\sigma_1^2+\INR_{1p}}\rp\\						
R_{1c} + R_{1p} &\le \log\lp 1+\frac{\SNR_{1c}+\SNR_{1p}}{1+\sigma_1^2+\INR_{1p}}\rp\\					
R_{2c} + \wtild{R}_{1h} &\le \log\lp 1+ \frac{K_{u_1|\ul{x}_o}+\INR_{1c}}{1+\sigma_1^2+\INR_{1p}}\rp\\			
R_{2c} + \wtild{R}_{1h} + R_{1p} &\le \log\lp 1+ \frac{K_{u_1|\ul{x}_o}+\INR_{1c}+\SNR_{1p}}{1+\sigma_1^2+\INR_{1p}}\rp\\		
R_{1c} + \wtild{R}_{1h} + R_{1p} &\le \log\lp 1+ \frac{K_{u_1|\ul{x}_o}+\SNR_{1c}+\SNR_{1p}}{1+\sigma_1^2+\INR_{1p}}\rp\\		
R_{1c} + R_{2c} + R_{1p} &\le \log\lp 1+ \frac{\SNR_{1c}+\INR_{1c}+\SNR_{1p}}{1+\sigma_1^2+\INR_{1p}}\rp\\		
R_{1c} + R_{2c} + \wtild{R}_{1h} + R_{1p} &\le \log\lp 1+ \frac{K_{u_1|\ul{x}_o}+\SNR_{1c}+\INR_{1c}+\SNR_{1p}}{1+\sigma_1^2+\INR_{1p}}\rp\\	
R_o + \wtild{R}_{1h} &\le \log\lp 1+ \frac{K_{u_1}}{1+\sigma_1^2+\INR_{1p}}\rp\\
R_o + \wtild{R}_{1h} + R_{1p} &\le \log\lp 1+ \frac{K_{u_1}+\SNR_{1p}}{1+\sigma_1^2+\INR_{1p}}\rp\\
R_o + R_{2c} + \wtild{R}_{1h} &\le \log\lp 1+ \frac{K_{u_1}+\INR_{1c}+\SNR_{1p}}{1+\sigma_1^2+\INR_{1p}}\rp\\
R_o + R_{2c} + \wtild{R}_{1h} + R_{1p} &\le \log\lp 1+ \frac{K_{u_1}+\SNR_{1c}+\INR_{1c}+\SNR_{1p}}{1+\sigma_1^2+\INR_{1p}}\rp\\
R_o + R_{1c} + \wtild{R}_{1h} + R_{1p} &\le \log\lp 1+ \frac{K_{u_1}+\SNR_{1c}+\SNR_{1p}}{1+\sigma_1^2+\INR_{1p}}\rp\\
R_o + R_{1c} + R_{2c} + \wtild{R}_{1h} + R_{1p} &\le \log\lp 1+ \frac{K_{u_1}+\SNR_{1c}+\INR_{1c}+\SNR_{1p}}{1+\sigma_1^2+\INR_{1p}}\rp
\end{align}

{\flushleft \ul{Constraints at Receiver 2}}: Above with index 1 and 2 exchanged.

{\flushleft \ul{Constraints at Transmitters}}:
\begin{align}
R_{1h} &\le \wtild{R}_{1h}\\
R_{2h} &\le \wtild{R}_{2h}\\
R_{1o}+ R_{2o} &= R_o\\
R_{1o} + R_{1h} &\le \C_{12}\\
R_{2o} + R_{2h} &\le \C_{21}\\
\wtild{R}_{1h} + \wtild{R}_{2h} - R_{1h} -R_{2h} &\ge 0,
\end{align}
for some nonnegative $(\wtild{R}_{1h}, \wtild{R}_{2h})$.

Notice that $\SNR_{ic}+\SNR_{ip} = \SNR_i/4$, $\INR_{ic}+\INR_{ip} = \INR_i/4$, and $K_{u_i}\ge \SNR_i/4+\INR_i/4$ for $i=1,2$. For simplicity, we consider the subset of the above region:
{\flushleft \ul{Constraints at Transmitters}}: The same as above.

{\flushleft \ul{Constraints at Receiver 1}}:
\begin{align}
R_{1p} &\le p_1\\ 
\wtild{R}_{1h} &\le g_1 & R_o + \wtild{R}_{1h} &\le m_1\\
\wtild{R}_{1h} + R_{1p} &\le g_1 & R_o + \wtild{R}_{1h} + R_{1p} &\le m_1\\
R_{2c} + R_{1p} &\le t_1\\ 
R_{1c} + R_{1p} &\le n_{11}\\ 
R_{2c} + \wtild{R}_{1h} &\le s_1 & R_o + R_{2c} + \wtild{R}_{1h} &\le m_1\\
R_{2c} + \wtild{R}_{1h} + R_{1p} &\le s_1 & R_o + R_{2c} + \wtild{R}_{1h} + R_{1p} &\le m_1\\
R_{1c} + \wtild{R}_{1h} + R_{1p} &\le l_1 & R_o + R_{1c} + \wtild{R}_{1h} + R_{1p} &\le m_1\\
R_{1c} + R_{2c} + R_{1p} &\le m_1\\ 
R_{1c} + R_{2c} + \wtild{R}_{1h} + R_{1p} &\le m_1 & R_o + R_{1c} + R_{2c} + \wtild{R}_{1h} + R_{1p} &\le m_1,
\end{align}
where
\begin{align}
p_1 &:= \log\lp 1+\frac{\SNR_{1p}}{1+\sigma_1^2+\INR_{1p}}\rp, & g_1 &:= \log\lp 1+ \frac{K_{u_1|\ul{x}_o}}{1+\sigma_1^2+\INR_{1p}}\rp\\
t_1 &:= \log\lp 1+\frac{\INR_{1c}+\SNR_{1p}}{1+\sigma_1^2+\INR_{1p}}\rp, & n_{11} &:=  \log\lp 1+\frac{\SNR_1/4}{1+\sigma_1^2+\INR_{1p}}\rp\\
s_1 &:= \log\lp 1+ \frac{K_{u_1|\ul{x}_o}+\INR_{1c}}{1+\sigma_1^2+\INR_{1p}}\rp, & l_1 &:= \log\lp 1+ \frac{K_{u_1|\ul{x}_o}+\SNR_{1}/4}{1+\sigma_1^2+\INR_{1p}}\rp\\
m_1 &:= \log\lp 1+ \frac{\SNR_{1}/4+\INR_{1c}}{1+\sigma_1^2+\INR_{1p}}\rp
\end{align}

{\flushleft \ul{Constraints at Receiver 2}}: Above with index 1 and 2 exchanged.

Notice now the rate region is symbolically identical to that in LDC when the system matrix is full rank. Hence, after the Fourier-Motzkin procedure, we have the following achievable rates, which are also symbolically identical to those in LDC. The only difference is that, ``redundancy" is replaced by ``approximate redundancy". Proof of the claims to show the approximate redundancy will be given later in this section.

{\flushleft (1) $R_1$ and $R_2$}:
{\flushleft \underline{$R_1$ constraints}}:
\begin{align}
R_1 &\le n_{11} + \C_{12}\\
R_1 &\le m_1\\
R_1 &\le p_1+t_2 + \C_{12} & & \textrm{approx. redundant}\\
R_1 &\le p_1+s_2 + \C_{12} & & \textrm{approx. redundant}\\
R_1 &\le g_1+t_2 + \C_{12} & & \textrm{approx. redundant}\\
R_1 &\le g_1+s_2 + \C_{12} & & \textrm{approx. redundant}
\end{align}
\underline{$R_2$ constraints}: Above with index 1 and 2 exchanged.


To show the approximate redundancy, we need to prove the following claim:
\begin{claim}\label{claim_1}
$ $
\begin{itemize}
\item $p_1+t_2 \ge n_{11} - \log\lp 9/4\rp$, $p_2+t_1 \ge n_{22} - \log\lp 9/4\rp$
\item $p_1+s_2 \ge n_{11} - \log\lp 9/4\rp$, $p_2+s_1 \ge n_{22} - \log\lp 9/4\rp$
\item $g_1+t_2 \ge n_{11} - \log9$, $g_2+t_1 \ge n_{22} - \log9$
\item $g_1+s_2 \ge n_{11} - \log9$, $g_2+s_1 \ge n_{22} - \log9$
\end{itemize}
\end{claim}


{\flushleft (2) $R_1+R_2$}:
\begin{align}
R_1+R_2 &\le t_1+t_2 + \C_{12}+\C_{21}\\
R_1+R_2 &\le p_1+m_2 + \C_{12}\\
R_1+R_2 &\le p_2+m_1 + \C_{21}\\
R_1+R_2 &\le g_1+m_2\\
R_1+R_2 &\le g_2+m_1\\
R_1+R_2 &\le s_1+t_2 + \C_{12}+\C_{21} & & \textrm{approx. redundant}\\
R_1+R_2 &\le s_2+t_1 + \C_{12}+\C_{21} & & \textrm{approx. redundant}\\
R_1+R_2 &\le s_1+s_2 + \C_{12}+\C_{21} & & \textrm{approx. redundant}
\end{align}
To show the approximate redundancy, we need to prove the following claim:

\begin{claim}\label{claim_2}
$s_1 \ge t_1 - \log5$, $s_2 \ge t_2 - \log5$
\end{claim}

{\flushleft (3) $2R_1+R_2$ and $R_1+2R_2$}:
\begin{align}
2R_1+R_2 &\le p_1+m_1+t_2 + \C_{12}+\C_{21}\\
2R_1+R_2 &\le g_1+m_1+t_2 + \C_{12}+\C_{21}& & \textrm{approx. redundant}\\
2R_1+R_2 &\le p_1+m_1+s_2 + \C_{12}\\
2R_1+R_2 &\le g_1+m_1+s_2 + \C_{12}& & \textrm{approx. redundant}\\
2R_1+R_2 &\le p_1+l_1+m_2 + \C_{12}\\
R_1+2R_2 &\le p_2+m_2+t_1 + \C_{12}+\C_{21}\\
R_1+2R_2 &\le g_2+m_2+t_1 + \C_{12}+\C_{21}& & \textrm{approx. redundant}\\
R_1+2R_2 &\le p_2+m_2+s_1 + \C_{21}\\
R_1+2R_2 &\le g_2+m_2+s_1 + \C_{21}& & \textrm{approx. redundant}\\
R_1+2R_2 &\le p_2+l_2+m_1 + \C_{21}.
\end{align}

To prove the approximate redundancy, we need to show the following claim:
\begin{claim}\label{claim_3}
$g_1 \ge p_1 - \log 5$, $g_2 \ge p_2 - \log 5$
\end{claim}


{\flushleft (4) $2R_1+2R_2$}:
\begin{align}
2R_1+2R_2 &\le p_1+s_1+t_2+m_2 + \C_{12}+\C_{21}& & \textrm{approx. redundant}\\
2R_1+2R_2 &\le l_1+t_1+p_2+m_2 + \C_{12}+\C_{21}& & \textrm{approx. redundant}\\
2R_1+2R_2 &\le p_1+s_1+s_2+m_2 + \C_{12}+\C_{21}& & \textrm{approx. redundant}\\
2R_1+2R_2 &\le l_1+t_1+g_2+m_2 + \C_{12}+\C_{21}& & \textrm{approx. redundant}\\
2R_1+2R_2 &\le t_1+m_1+p_2+s_2 + \C_{12}+\C_{21}& & \textrm{approx. redundant}\\
2R_1+2R_2 &\le s_1+m_1+p_2+s_2 + \C_{12}+\C_{21}& & \textrm{approx. redundant}\\
2R_1+2R_2 &\le p_1+m_1+l_2+t_2 + \C_{12}+\C_{21}& & \textrm{approx. redundant}\\
2R_1+2R_2 &\le g_1+m_1+l_2+t_2 + \C_{12}+\C_{21}& & \textrm{approx. redundant}
\end{align}

To show the approximate redundancy, we need to prove the following claim:
\begin{claim}\label{claim_4}
$ $
\begin{itemize}
\item $s_1+t_2 \ge p_2+m_1-\log18$, $s_2+t_1 \ge p_1+m_2-\log18$
\item $l_1+t_1 \ge p_1+m_1-\log12$, $l_2+t_2 \ge p_2+m_2-\log12$
\end{itemize}
\end{claim}

We summarize in the lemma below an achievable rate region:
\begin{lemma}\label{lem_GICAchieve}
If $\lp R_1, R_2\rp$ satisfies the following, it is achievable.
\begin{align}
R_1 &\le \min\lbp m_1, n_{11} + \C_{12} - 2\log3\rbp\\
R_2 &\le \min\lbp m_2, n_{22} + \C_{21} - 2\log3\rbp\\
R_1+R_2 &\le t_1+t_2 + \C_{12}+\C_{21}- 2\log5\\
R_1+R_2 &\le \min\lbp p_1+m_2 + \C_{12},  p_2+m_1 + \C_{21}\rbp - \lp\log90\rp/2\\
R_1+R_2 &\le \min\lbp g_1+m_2, g_2+m_1\rbp\\
2R_1+R_2 &\le p_1+m_1+t_2 + \C_{12}+\C_{21} - \log5\\
2R_1+R_2 &\le \min\lbp p_1+m_1+s_2 + \C_{12} - \log5, p_1+l_1+m_2 + \C_{12}\rbp\\
R_1+2R_2 &\le p_2+m_2+t_1 + \C_{12}+\C_{21} - \log5\\
R_1+2R_2 &\le \min\lbp p_2+m_2+s_1 + \C_{21} - \log5, p_2+l_2+m_1 + \C_{21}\rbp.
\end{align}
\end{lemma}

\subsection{Proof of the Claims}
Prior to the proof of the above claims, we give a bunch of useful lemmas.

\begin{lemma}\label{lem_Ineq1}
\begin{align}
&\log\lp 1+\SNR_1+\INR_1+\SNR_2+\INR_2+|h_{11}h_{22} - h_{12}h_{21}|^2\rp\\
&\ge \log\lp 1+\frac{\SNR_1}{1+\INR_2}\rp + \log\lp 1+\SNR_2+\INR_2\rp.
\end{align}
\end{lemma}
\begin{proof}
Consider the Gaussian interference channel without cooperation. We take independent Gaussian input signals. Note that 
\begin{align}
&\log\lp 1+\frac{\SNR_1}{1+\INR_2}\rp + \log\lp 1+\SNR_2+\INR_2\rp\\
&= I\lp x_1;y_1,y_2| x_2\rp + I\lp x_2; y_2\rp\\
&\le I\lp x_1;y_1,y_2| x_2\rp + I\lp x_2; y_2, y_1\rp\\
&= I\lp x_1,x_2; y_1, y_2\rp\\
&= \log\lp 1+\SNR_1+\INR_1+\SNR_2+\INR_2+|h_{11}h_{22} - h_{12}h_{21}|^2\rp.
\end{align}
\end{proof}

\begin{corollary}\label{cor_Ineq1}
\begin{align}
K_{u_1|\ul{x}_o} \ge \frac{\SNR_1}{4\lp1+\INR_2\rp},\ K_{u_2|\ul{x}_o} \ge \frac{\SNR_2}{4\lp1+\INR_1\rp}
\end{align}
\end{corollary}
\begin{proof}
\begin{align}
1+K_{u_1|\ul{x}_o} &= \frac{3}{4} + \frac{1+\SNR_2+\INR_2+|h_{11}h_{22} - h_{12}h_{21}|^2+\SNR_1+\INR_1}{4\lp 1+\SNR_2+\INR_2\rp}\\
&\overset{\aaaa}{\ge} \frac{3}{4} + \frac{1+\frac{\SNR_1}{1+\INR_2}}{4} = 1+ \frac{\SNR_1}{4\lp1+\INR_2\rp},
\end{align}
where (a) is due to Lemma \ref{lem_Ineq1}. Hence $K_{u_1|\ul{x}_o} \ge \frac{\SNR_1}{4\lp1+\INR_2\rp}$. Similarly $K_{u_2|\ul{x}_o} \ge \frac{\SNR_2}{4\lp1+\INR_1\rp}$.
\end{proof}

\begin{lemma}\label{lem_Ineq2}
\begin{align}
&2|h_{11}h_{22}-h_{12}h_{21}|^2 + 4\SNR_1\SNR_2 \ge \SNR_1\SNR_2+\INR_1\INR_2\\
&2|h_{11}h_{22}-h_{12}h_{21}|^2 + 4\INR_1\INR_2 \ge \SNR_1\SNR_2+\INR_1\INR_2.
\end{align}
\end{lemma}
\begin{proof}
\begin{align}
|h_{11}h_{22}-h_{12}h_{21}|^2 &\ge \SNR_1\SNR_2 + \INR_1\INR_2 - 2\sqrt{\SNR_1\SNR_2 \INR_1\INR_2}\\
&:= x+y-2\sqrt{xy},
\end{align}
where $x = \SNR_1\SNR_2$ and $y = \INR_1\INR_2$. Hence,
\begin{align}
&2|h_{11}h_{22}-h_{12}h_{21}|^2 + 4\SNR_1\SNR_2 - \lp\SNR_1\SNR_2+\INR_1\INR_2\rp\\
&\ge 2x+2y-4\sqrt{xy}+3x-y = y\lp 5u^2 - 4u +1\rp \ge 0,
\end{align}
where $u:=\sqrt{x/y}$. 

Similarly, $2|h_{11}h_{22}-h_{12}h_{21}|^2 + 4\INR_1\INR_2 \ge \SNR_1\SNR_2+\INR_1\INR_2$.
\end{proof}

\subsection*{Proof of Claim \ref{claim_1}}
\begin{itemize}
\item $p_1+t_2 \ge n_{11} - \log\lp 9/4\rp$, $p_2+t_1 \ge n_{22} - \log\lp 9/4\rp$
\begin{proof}
\begin{align}
p_1+t_2 &= \log\lp 1+\frac{\SNR_{1p}}{1+\sigma_1^2+\INR_{1p}}\rp + \log\lp 1+\frac{\INR_{2c}+\SNR_{2p}}{1+\sigma_2^2+\INR_{2p}}\rp\\
&= \log\lp \frac{\lp1+\sigma_1^2+\SNR_{1p}+\INR_{1p}\rp\lp 1+\sigma_2^2+\INR_2/4+\SNR_{2p}\rp}{\lp1+\sigma_1^2+\INR_{1p}\rp\lp 1+\sigma_2^2+\INR_{2p}\rp}\rp\\
&\ge \log\lp \frac{1+\sigma_1^2 + \SNR_1/4+\INR_{1p}}{1+\sigma_1^2+\INR_{1p}}\rp - \log\lp 1+\sigma_2^2+\INR_{2p}\rp\\
&\ge n_{11} - \log\lp 9/4\rp.
\end{align}
Similarly, $p_2+t_1 \ge n_{22} - \log\lp 9/4\rp$.
\end{proof}
\item $p_1+s_2 \ge n_{11} - \log\lp 9/4\rp$, $p_2+s_1 \ge n_{22} - \log\lp 9/4\rp$
\begin{proof}
\begin{align}
p_1+s_2 &= \log\lp 1+\frac{\SNR_{1p}}{1+\sigma_1^2+\INR_{1p}}\rp +  \log\lp 1+ \frac{K_{u_2|\ul{x}_o}+\INR_{2c}}{1+\sigma_2^2+\INR_{2p}}\rp\\
&= \log\lp \frac{\lp1+\sigma_1^2+\SNR_{1p}+\INR_{1p}\rp\lp 1+\sigma_2^2+\INR_2/4+K_{u_2|\ul{x}_o}\rp}{\lp1+\sigma_1^2+\INR_{1p}\rp\lp 1+\sigma_2^2+\INR_{2p}\rp}\rp\\
&\ge n_{11} - \log\lp 9/4\rp.
\end{align}
Similarly, $p_2+s_1 \ge n_{22} - \log\lp 9/4\rp$.
\end{proof}


\item $g_1+t_2 \ge n_{11} - \log9$, $g_2+t_1 \ge n_{22} - \log9$
\begin{proof}
\begin{align}
g_1+t_2 &= \log\lp \frac{\lp1+\sigma_1^2+\INR_{1p}+K_{u_1|\ul{x}_o}\rp\lp 1+\sigma_2^2+\INR_2/4+\SNR_{2p}\rp}{\lp1+\sigma_1^2+\INR_{1p}\rp\lp 1+\sigma_2^2+\INR_{2p}\rp}\rp\\
&\ge \log\lp \frac{\sigma_1^2 + \INR_{1p} + \lp 1+K_{u_1|\ul{x}_o}\rp\lp 1+\INR_2/4\rp}{1+\sigma_1^2+\INR_{1p}}\rp - \log\lp 1+\sigma_2^2+\INR_{2p}\rp\\
&\overset{\aaaa}{\ge} \log\lp \frac{1+\sigma_1^2 + \INR_{1p}+\SNR_1/16}{1+\sigma_1^2+\INR_{1p}}\rp - \log\lp 1+\sigma_2^2+\INR_{2p}\rp\\
&\ge n_{11} - \log 9,
\end{align}
where (a) is due to Corollary \ref{cor_Ineq1}. Similarly, $g_2+t_1 \ge n_{22} - \log9$.
\end{proof}

\item $g_1+s_2 \ge n_{11} - \log9$
\begin{proof}
\begin{align}
g_1+s_2
&= \log\lp \frac{\lp1+\sigma_1^2+\INR_{1p}+K_{u_1|\ul{x}_o}\rp\lp 1+\sigma_2^2+\INR_2/4+K_{u_2|\ul{x}_o}\rp}{\lp1+\sigma_1^2+\INR_{1p}\rp\lp 1+\sigma_2^2+\INR_{2p}\rp}\rp\\
&\ge \log\lp \frac{\sigma_1^2 + \INR_{1p} + \lp 1+K_{u_1|\ul{x}_o}\rp\lp 1+\INR_2/4\rp}{1+\sigma_1^2+\INR_{1p}}\rp - \log\lp 1+\sigma_2^2+\INR_{2p}\rp\\
&\ge \log\lp \frac{1+\sigma_1^2 + \INR_{1p}+\SNR_1/16}{1+\sigma_1^2+\INR_{1p}}\rp - \log\lp 1+\sigma_2^2+\INR_{2p}\rp\\
&\ge n_{11} - \log 9.
\end{align}
Similarly, $g_2+s_1 \ge n_{22} - \log9$.
\end{proof}
\end{itemize}

\begin{remark}
If we want to follow the proofs in LDC closely, we can also prove the approximate redundancy by making use of the fact (to be proved later)
\begin{align}
s_2 \ge t_2 - \log5,\ s_1 \ge t_1 - \log5,\ g_1 \ge p_1 - \log 5,\ g_2 \ge p_2 - \log 5,
\end{align}
which results in a looser upper bound on the gap to outer bounds. 
\end{remark}

\subsection*{Proof of Claim \ref{claim_2}}
\begin{proof}
\begin{align}
s_1 &= \log\lp 1+ \frac{K_{u_1|\ul{x}_o}+\INR_{1c}}{1+\sigma_1^2+\INR_{1p}}\rp =  \log\lp\frac{1+\sigma_1^2+\INR_{1}/4 + K_{u_1|\ul{x}_o}}{1+\sigma_1^2+\INR_{1p}}\rp\\
&\overset{\aaaa}{\ge} \log\lp\frac{1+\sigma_1^2+\INR_{1}/4 + \frac{\SNR_1}{4\lp1+\INR_2\rp}}{1+\sigma_1^2+\INR_{1p}}\rp
\overset{\bbbb}{\ge} \log\lp\frac{1+\sigma_1^2+\INR_{1}/4 + \frac{\SNR_{1p}}{5}}{1+\sigma_1^2+\INR_{1p}}\rp\\
&\ge t_1 - \log5,
\end{align}
where (a) is due to Corollary \ref{cor_Ineq1}, and (b) is due to the fact that 
\begin{align}\frac{\SNR_1}{1+\INR_2} \ge \frac{4\SNR_{1p}}{5}.\end{align}

Similarly $s_2 \ge t_2 - \log 5$.
\end{proof}

\subsection*{Proof of Claim \ref{claim_3}}
\begin{proof}
\begin{align}
g_1 &= \log\lp\frac{1+\sigma_1^2+\INR_{1p} +K_{u_1|\ul{x}_o}}{1+\sigma_1^2+\INR_{1p}}\rp\\
&\overset{\aaaa}{\ge} \log\lp\frac{1+\sigma_1^2+\INR_{1p} + \frac{\SNR_1}{4\lp1+\INR_2\rp}}{1+\sigma_1^2+\INR_{1p}}\rp
\overset{\bbbb}{\ge} \log\lp\frac{1+\sigma_1^2+\INR_{1p} + \frac{\SNR_{1p}}{5}}{1+\sigma_1^2+\INR_{1p}}\rp\\
&\ge p_1 - \log5,
\end{align}
where (a) is due to Corollary \ref{cor_Ineq1}, and (b) is due to the fact that 
\begin{align}\frac{\SNR_1}{1+\INR_2} \ge \frac{4\SNR_{1p}}{5}.\end{align}

Similarly $g_2 \ge p_2 - \log 5$.
\end{proof}

\subsection*{Proof of Claim \ref{claim_4}}
\begin{itemize}
\item $s_1+t_2 \ge p_2+m_1-\log18$, $s_2+t_1 \ge p_1+m_2-\log18$
\begin{proof}
\begin{align}
s_1 &= \log\lp\frac{1+\sigma_1^2+\INR_{1}/4 + K_{u_1|\ul{x}_o}}{1+\sigma_1^2+\INR_{1p}}\rp\\
t_2 &= \log\lp\frac{1+\sigma_2^2+\INR_2/4+\SNR_{2p}}{1+\sigma_2^2+\INR_{2p}}\rp\\
p_2 &= \log\lp \frac{1+\sigma_2^2+\INR_{2p}+\SNR_{2p}}{1+\sigma_2^2+\INR_{2p}}\rp\\
m_1 &= \log\lp \frac{1+\sigma_1^2+\SNR_{1}/4+\INR_{1}/4}{1+\sigma_1^2+\INR_{1p}}\rp
\end{align}

Hence, it suffice to compare
\begin{align}
L &= \lp1+\sigma_1^2+\INR_{1}/4 + K_{u_1|\ul{x}_o}\rp \lp1+\sigma_2^2+\INR_2/4+\SNR_{2p}\rp
\end{align}
and
\begin{align}
R &= \lp1+\sigma_1^2+\SNR_{1}/4+\INR_{1}/4\rp \lp1+\sigma_2^2+\INR_{2p}+\SNR_{2p}\rp
\end{align}

Note that from Lemma \ref{lem_Ineq2}, if $\SNR_2\ge \INR_2$,
\begin{align}
&\frac{\INR_{1}}{4} + K_{u_1|\ul{x}_o}\\ &= \frac{|h_{11}h_{22}-h_{12}h_{21}|^2+\INR_1\INR_2+\SNR_2\INR_1 + \SNR_1 + 2\INR_1}{4\lp1+\SNR_2+\INR_2\rp}\\
&\ge \frac{\SNR_1\SNR_2+\INR_1\INR_2+4\SNR_2\INR_1 + 4\SNR_1 + 8\INR_1}{16\lp1+\SNR_2+\INR_2\rp}\\
&\ge \frac{\SNR_1\max\lp\SNR_2,1\rp + 4\INR_1\max\lp\SNR_2,1\rp}{48\max\lp\SNR_2,1\rp} \ge \frac{\SNR_1+\INR_1}{48}
\end{align}
Also, $\INR_2 / 4 \ge \INR_{2p}$. Hence, $s_1+t_2 \ge p_2+m_1 - \log12$.

If $\SNR_2 < \INR_2$,
\begin{align}
R &= \lp1+\sigma_1^2+\INR_{1}/4 \rp \lp1+\sigma_2^2+\SNR_{2p}\rp + (\SNR_1/4)\lp1+\sigma_2^2+\SNR_{2p}\rp\\
&\quad + \INR_{2p}\lp1+\sigma_1^2+\INR_{1}/4 \rp + (\SNR_1/4)\INR_{2p}\\
&\le 2\lp1+\sigma_1^2+\INR_{1}/4 \rp \lp1+\sigma_2^2+\SNR_{2p}\rp + (\SNR_1/4)\lp5/4+\SNR_{2p}\rp\\&\quad + \SNR_1/4\\
&= 2\lp1+\sigma_1^2+\INR_{1}/4 \rp \lp1+\sigma_2^2+\SNR_{2p}\rp + \frac{9}{16}\SNR_1+\frac{\SNR_1\SNR_{2p}}{4},
\end{align}
and
\begin{align}
L &= \lp1+\sigma_1^2+\INR_{1}/4 \rp \lp1+\sigma_2^2+\SNR_{2p}\rp + K_{u_1|\ul{x}_o}\lp1+\sigma_2^2+\SNR_{2p}\rp\\
&\quad + \lp1+\sigma_1^2+\INR_{1}/4 \rp(\INR_2/4) + K_{u_1|\ul{x}_o}(\INR_2/4)\\
&\ge \frac{\lp1+\sigma_1^2+\INR_{1}/4 \rp \lp1+\sigma_2^2+\SNR_{2p}\rp}{2} + \frac{K_{u_1|\ul{x}_o}\max\lp \INR_2, 1\rp}{4}\\
&\quad + \frac{\INR_1\INR_2}{16} + \frac{K_{u_1|\ul{x}_o}}{2} + \frac{\INR_1}{8}\\
&\ge \frac{\lp1+\sigma_1^2+\INR_{1}/4 \rp \lp1+\sigma_2^2+\SNR_{2p}\rp}{2} + \frac{\SNR_{1}}{96}\\
&\quad + \frac{|h_{11}h_{22}-h_{12}h_{21}|^2+2\INR_1\INR_2+\SNR_1+\INR_1}{48}\\
&\ge \frac{\lp1+\sigma_1^2+\INR_{1}/4 \rp \lp1+\sigma_2^2+\SNR_{2p}\rp}{2} + \frac{\SNR_1\SNR_2}{96} +\frac{\SNR_1}{32}.
\end{align}
Hence, $s_1+t_2 \ge p_2+m_1-\log18$.

In summary, $s_1+t_2 \ge p_2+m_1-\log18$, and similarly, $s_2+t_1 \ge p_1+m_2-\log18$.
\end{proof}

\item $l_1+t_1 \ge p_1+m_1-\log12$, $l_2+t_2 \ge p_2+m_2-\log12$
\begin{proof}
\begin{align}
l_1 &= \log\lp\frac{1+\sigma_1^2+\INR_{1p}+\SNR_{1}/4 + K_{u_1|\ul{x}_o}}{1+\sigma_1^2+\INR_{1p}}\rp\\
t_1 &= \log\lp\frac{1+\sigma_1^2+\INR_1/4+\SNR_{1p}}{1+\sigma_1^2+\INR_{1p}}\rp\\
p_1 &= \log\lp \frac{1+\sigma_1^2+\INR_{1p}+\SNR_{1p}}{1+\sigma_1^2+\INR_{1p}}\rp\\
m_1 &= \log\lp \frac{1+\sigma_1^2+\SNR_{1}/4+\INR_{1}/4}{1+\sigma_1^2+\INR_{1p}}\rp
\end{align}

Hence, it suffice to compare
\begin{align}
L &= \lp1+\sigma_1^2+\INR_{1p}+\SNR_{1}/4 + K_{u_1|\ul{x}_o}\rp \lp1+\sigma_1^2+\INR_1/4+\SNR_{1p}\rp
\end{align}
and
\begin{align}
R &= \lp1+\sigma_1^2+\SNR_{1}/4+\INR_{1}/4\rp \lp1+\sigma_1^2+\INR_{1p}+\SNR_{1p}\rp
\end{align}

Note that from Lemma \ref{lem_Ineq2}, if $\SNR_2\le \INR_2$,
\begin{align}
&\frac{\SNR_{1}}{4} + K_{u_1|\ul{x}_o}\\ &= \frac{|h_{11}h_{22}-h_{12}h_{21}|^2+\SNR_1\SNR_2+\SNR_1\INR_2 + 2\SNR_1 + 1\INR_1}{4\lp1+\SNR_2+\INR_2\rp}\\
&\ge \frac{\SNR_1\SNR_2+\INR_1\INR_2+4\SNR_1\INR_2 + 8\SNR_1 + 4\INR_1}{16\lp1+\SNR_2+\INR_2\rp}\\
&\ge \frac{\INR_1\max\lp\INR_2,1\rp + 4\SNR_1\max\lp\INR_2,1\rp}{48\max\lp\INR_2,1\rp} \ge \frac{\INR_1+\SNR_1}{48}
\end{align}
Also, $\INR_1 / 4 \ge \INR_{1p}$. Hence, $l_1+t_1 \ge p_1+m_1 - \log12$.

If $\SNR_2 > \INR_2$,
\begin{align}
R &= \lp1+\sigma_1^2+\SNR_{1}/4 \rp \lp1+\sigma_1^2+\SNR_{1p}\rp + (\INR_1/4)\lp1+\sigma_1^2+\SNR_{1p}\rp\\
&\quad + \INR_{1p}\lp1+\sigma_1^2+\SNR_{1}/4 \rp + (\INR_1/4)\INR_{1p}\\
&\le 2\lp1+\sigma_1^2+\SNR_{1}/4 \rp \lp1+\sigma_1^2+\SNR_{1p}\rp + (\INR_1/4)\lp5/4+\SNR_{1p}\rp\\&\quad + \INR_1/4\\
&= 2\lp1+\sigma_1^2+\SNR_{1}/4 \rp \lp1+\sigma_1^2+\SNR_{1p}\rp + \frac{9}{16}\INR_1+\frac{\INR_1\SNR_{1p}}{4},
\end{align}
and
\begin{align}
L &= \lp1+\sigma_1^2+\SNR_{1}/4 \rp \lp1+\sigma_1^2+\SNR_{1p}\rp + \lp\INR_{1p}+K_{u_1|\ul{x}_o}\rp(\INR_1/4)\\
&\quad + \lp1+\sigma_1^2+\SNR_{1}/4 \rp(\INR_1/4) + \lp \INR_{1p}+K_{u_1|\ul{x}_o}\rp\lp1+\sigma_1^2+\SNR_{1p}\rp\\
&\ge \lp1+\sigma_1^2+\SNR_{1}/4 \rp \lp1+\sigma_1^2+\SNR_{1p}\rp + \frac{\INR_1}{4}\lp1+ K_{u_1|\ul{x}_o}+\frac{\SNR_1}{4}\rp\\
&\ge \lp1+\sigma_1^2+\SNR_{1}/4 \rp \lp1+\sigma_1^2+\SNR_{1p}\rp + \frac{\INR_1}{4}\lp 1+\frac{\SNR_1}{48}\rp\\
&\ge \lp1+\sigma_1^2+\SNR_{1}/4 \rp \lp1+\sigma_1^2+\SNR_{1p}\rp + \frac{\INR_1}{4}\lp 1+\frac{\SNR_{1p}}{12}\rp
\end{align}
Hence, $l_1+t_1 \ge p_1+m_1-\log12$.

In summary, $l_1+t_1 \ge p_1+m_1-\log12$, and similarly, $l_2+t_2 \ge p_2+m_2-\log12$.
\end{proof}
\end{itemize}

\section{Proof of Theorem \ref{thm_Gap}: Constant Gap to Outer Bounds}\label{app_PfThmGap}
{\flushleft (1) Bounds on $R_1$:}\par
\begin{itemize}
\item
Consider the outer bound 
\begin{align}
R_1 &\le \log\lp 1+\SNR_1+\INR_1+2\sqrt{\SNR_1\INR_1}\rp
\end{align}
and the inner bound
\begin{align}
R_1 &\le m_1 = \log\lp 1+ \frac{\SNR_{1}/4+\INR_{1c}}{1+\sigma_1^2+\INR_{1p}}\rp = \log\lp \frac{1+\sigma_1^2+\frac{\SNR_1+\INR_1}{4}}{1+\sigma_1^2+\INR_{1p}}\rp
\end{align}

Note that
\begin{align}
&\log\lp 1+\SNR_1+\INR_1+2\sqrt{\SNR_1\INR_1}\rp \le \log\lp1+\SNR_1+\INR_1\rp+1,\\
&\log\lp \frac{1+\sigma_1^2+\frac{\SNR_1+\INR_1}{4}}{1+\sigma_1^2+\INR_{1p}}\rp \ge \log\lp1+\SNR_1+\INR_1\rp - \log9.
\end{align}
Hence the gap is at most $\log 9+1= 2\log3+1$.

\item
Consider the outer bound
\begin{align}
R_1 &\le \log\lp 1+\SNR_1\rp +\C_{12}
\end{align}
and the inner bound
\begin{align}
R_1 &\le n_{11}+\C_{12} - 2\log3= \log\lp 1+\frac{\SNR_1/4}{1+\sigma_1^2+\INR_{1p}}\rp + \C_{12} - 2\log3
\end{align}

Note that
\begin{align}
\log\lp 1+\frac{\SNR_1/4}{1+\sigma_1^2+\INR_{1p}}\rp \ge \log\lp1+\SNR_1\rp - \log9.
\end{align}
Hence, the gap is at most $2\log3 + \log9 = 4\log3 \approx 6.34$
\end{itemize}

In summary, the gap is at most $4\log3 \approx 6.34$.

{\flushleft (2) $R_2$:} similar to $R_1$, the gap is at most $4\log3 \approx 6.34$.

{\flushleft (3) $R_1+R_2$:}
\begin{itemize}
\item
Consider the outer bound
\begin{align}
&R_1+R_2\\
&\le \log\lp 1+\frac{\SNR_1}{1+\INR_2}\rp + \log\lp 1+\SNR_2+\INR_2+2\sqrt{\SNR_2\INR_2}\rp + \C_{12}
\end{align}
and the inner bound
\begin{align}
R_1+R_2 &\le p_1+m_2+\C_{12} - \lp \log90\rp/2.
\end{align}

Note that 
\begin{align}
&\log\lp 1+\SNR_2+\INR_2+2\sqrt{\SNR_2\INR_2}\rp \le \log\lp1+\SNR_2+\INR_2\rp +1,\\
&m_2 \ge \log\lp1+\SNR_2+\INR_2\rp - \log9,\\
&p_1 = \log\lp1+\frac{\SNR_{1p}}{1+\sigma_1^2+\INR_{1p}}\rp \ge \log\lp1+\frac{\SNR_1}{1+\INR_2}\rp - \log(9/4).
\end{align}
Hence the gap is at most $5\log3-1+(\log10)/2 \approx 8.586$.

\item
Consider the outer bound
\begin{align}
&R_1+R_2\\ &\le
\lbp\begin{array}{l}
\log\lp 1+\frac{\SNR_1+2\sqrt{\SNR_1\INR_1}}{1+\INR_2} + \INR_1\rp\\
+\log\lp 1+\frac{\SNR_2+2\sqrt{\SNR_2\INR_2}}{1+\INR_1} + \INR_2\rp
\end{array}\rbp +\C_{12}+\C_{21}
\end{align}
and the inner bound
\begin{align}
R_1+R_2 &\le t_1+t_2+\C_{12}+\C_{21} -\lp 2\log5\rp.
\end{align}

Note that
\begin{align}
&\log\lp 1+\frac{\SNR_1+2\sqrt{\SNR_1\INR_1}}{1+\INR_2} + \INR_1\rp\\
&\le \log\lp 1+\frac{2\SNR_1+\INR_1}{1+\INR_2} + \INR_1\rp\\
&\le \log\lp 1+\frac{\SNR_1}{1+\INR_2} + \INR_1\rp +1,
\end{align}
and 
\begin{align}
t_1 &= \log\lp\frac{1+\sigma_1^2+\INR_1/4+\SNR_{1p}}{1+\sigma_1^2+\INR_{1p}}\rp\\
&\ge \log\lp 1+\frac{\SNR_1}{1+\INR_2} + \INR_1\rp - \log9.
\end{align}
Hence, the gap is at most $4\log3+2+2\log5 \approx 12.984$.

\item
Consider the outer bound
\begin{align}
&R_1+R_2\\
&\le \log\lp \begin{array}{l}1+\SNR_1+\INR_1+\SNR_2+\INR_2+2\sqrt{\SNR_1\INR_1}\\+2\sqrt{\SNR_2\INR_2}+|h_{11}h_{22}-h_{12}h_{21}|^2\end{array}\rp
\end{align}
and the inner bounds
\begin{align}
R_1+R_2 &\le \min\lbp g_1+m_2, g_2+m_1\rbp.
\end{align}

Note that
\begin{align}
&\log\lp \begin{array}{l}1+\SNR_1+\INR_1+\SNR_2+\INR_2+2\sqrt{\SNR_1\INR_1}\\+2\sqrt{\SNR_2\INR_2}+|h_{11}h_{22}-h_{12}h_{21}|^2\end{array}\rp\\
&\le \log\lp 1+2\SNR_1+2\INR_1+2\SNR_2+2\INR_2+|h_{11}h_{22}-h_{12}h_{21}|^2\rp\\
&\le \log\lp 1+\SNR_1+\INR_1+\SNR_2+\INR_2+|h_{11}h_{22}-h_{12}h_{21}|^2\rp + 1,
\end{align}
and 
\begin{align}
&g_1+m_2\\ 
&= \log\lp\frac{\lp1+\sigma_1^2+\INR_{1p}+K_{u_1|\ul{x}_o}\rp\lp 1+\sigma_2^2+\frac{\SNR_2+\INR_2}{4}\rp}{\lp1+\sigma_1^2+\INR_{1p}\rp\lp1+\sigma_2^2+\INR_{2p}\rp}\rp\\
&\ge \log\lp \lp1+K_{u_1|\ul{x}_o}\rp \lp1+\SNR_2+\INR_2\rp\rp - \log(81/4)\\
&= \log\lp 1+\SNR_1+\INR_1+\SNR_2+\INR_2+|h_{11}h_{22}-h_{12}h_{21}|^2\rp -\log81.
\end{align}
Hence, the gap is at most $4\log3+1\approx 7.34$.
\end{itemize}

In summary, the gap is at most $4\log3+2+2\log5 \approx 12.984$.

{\flushleft (4) $2R_1+R_2$:}
\begin{itemize}
\item
Consider the outer bound
\begin{align}
2R_1+R_2 &\le \lbp\begin{array}{l}\log\lp 1+\SNR_1+\INR_1+2\sqrt{\SNR_1\INR_1}\rp + \log\lp1+\frac{\SNR_1}{1+\INR_2}\rp\\ + \log\lp 1+\frac{\SNR_2+2\sqrt{\SNR_2\INR_2}}{1+\INR_1}+\INR_2\rp + \C_{12}+\C_{21}\end{array}\rbp\end{align}
and the inner bound
\begin{align}
2R_1+R_2 &\le p_1+m_1+t_2+\C_{12}+\C_{21} - \lp\log5\rp.
\end{align}

From previous arguments, one can directly see that the gap is at most
\begin{align}
\log(9/4) + (2\log3+1) + (2\log3+1) + \log5= 6\log3 + \log5 \approx 11.832.
\end{align}

\item
Consider the outer bound
\begin{align}
2R_1+R_2 &\le \lbp\begin{array}{l} \log\lp \begin{array}{l}1+\SNR_1+\INR_1+\SNR_2+\INR_2+\SNR_1\SNR_2\\+\INR_1\INR_2+\SNR_1\INR_2 + 2\lp1+\INR_2\rp\sqrt{\SNR_1\INR_1}\end{array}\rp\\
+\log\lp 1+\frac{\SNR_1}{1+\INR_2}\rp + 1+ \C_{12}\end{array}\rbp
\end{align}
and the inner bounds
\begin{align}
2R_1+R_2 &\le p_1+m_1+s_2+\C_{12}-\lp \log5\rp, & 2R_1+R_2 &\le p_1+l_1+m_2+\C_{12}.
\end{align}

Note that
\begin{align}
&\log\lp \begin{array}{l}1+\SNR_1+\INR_1+\SNR_2+\INR_2+\SNR_1\SNR_2\\+\INR_1\INR_2+\SNR_1\INR_2 + 2\lp1+\INR_2\rp\sqrt{\SNR_1\INR_1}\end{array}\rp\\
&\le \log\lp \begin{array}{l}1+2\SNR_1+2\INR_1+\SNR_2+\INR_2+\SNR_1\SNR_2\\+2\INR_1\INR_2+2\SNR_1\INR_2\end{array}\rp.
\end{align}

For the inner bounds,
\begin{align}
&m_1+s_2\\
&\ge \log\lp \lp1+\SNR_1+\INR_1\rp \lp1+\INR_2/4+K_{u_2|\ul{x}_o}\rp \rp - (4\log3-2)\\
&= \log\lp\begin{array}{l} 1+\SNR_1+\INR_1+\frac{\SNR_2}{4} + \frac{\INR_2}{2}\\ + \frac{\SNR_1\INR_2}{4} + \frac{|h_{11}h_{22}-h_{12}h_{21}|^2+\INR_1\INR_2}{4}\end{array}\rp - (4\log3-2)\\
&\ge \log\lp\begin{array}{l} 1+\SNR_1+\INR_1+\frac{\SNR_2}{4} + \frac{\INR_2}{2}\\ + \frac{\SNR_1\INR_2}{4} + \frac{\SNR_1\SNR_2+\INR_1\INR_2}{16}\end{array}\rp - (4\log3-2),
\end{align}
and 
\begin{align}
&l_1+m_2\\
&\ge \log\lp \lp1+\SNR_2+\INR_2\rp \lp1+\SNR_1/4+K_{u_1|\ul{x}_o}\rp \rp - (4\log3-2)\\
&= \log\lp\begin{array}{l} 1+\SNR_2+\INR_2+\frac{\SNR_1}{2} + \frac{\INR_1}{4}\\ + \frac{\SNR_1\INR_2}{4} + \frac{|h_{11}h_{22}-h_{12}h_{21}|^2+\SNR_1\SNR_2}{4}\end{array}\rp - (4\log3-2)\\
&\ge \log\lp\begin{array}{l} 1+\SNR_2+\INR_2+\frac{\SNR_1}{2} + \frac{\INR_1}{4}\\ + \frac{\SNR_1\INR_2}{4} + \frac{\SNR_1\SNR_2+\INR_1\INR_2}{16}\end{array}\rp - (4\log3-2).
\end{align}

Hence, the gap is at most $(4\log3+3)+\log(9/4)+\log5 = 6\log3 + \log5 +1 \approx 12.832$.
\end{itemize}

In summary, the gap is at most $6\log3 + \log5 + 1 \approx 12.832$.

{\flushleft (5) $R_1+2R_2$:} similar to $2R_1+R_2$, the gap is at most $6\log3 + \log5 +1 \approx 12.832$.

Combining the results, we characterize the capacity region to within a constant gap, which is 
\begin{align}
\max\lbp 4\log3, \frac{4\log3+2+2\log5}{2}, \frac{6\log3 + \log5 + 1}{3}\rbp = 2\log3+1+\log5 = \log90 \approx 6.5.
\end{align}

\section{Proof of Lemma \ref{OutBd_GIC}}\label{app_PfOutBd}
We first state a useful fact \cite{Willems_83}:
\begin{fact}[Conditional Independence among Messages]\label{fact_Markov}
The following Markov relations hold:
\begin{align}
&m_1 - \lp v_{12}^N,v_{21}^N\rp - m_2;\ m_1 - \lp v_{12}^N,v_{21}^N\rp - x_2^N;\ m_2 - \lp v_{12}^N,v_{21}^N\rp - x_1^N.
\end{align}
\end{fact}
The proof can be found in \cite{Willems_83}.

Below we start the proof of the outer bounds stated in Lemma \ref{OutBd_GIC}.
\begin{proof}
{\flushleft (1) $R_1$ bound \eqref{bd_R1}:}\par
If $R_1$ is achievable, by Fano's inequality,
\begin{align}
&N\lp R_1-\epsilon_N\rp\\ 
&\le I\lp m_1; y_1^N\rp \le I\lp m_1; y_1^N| m_2\rp\\
&\le I\lp m_1; y_1^N| m_2,v_{12}^N\rp + I\lp m_1;v_{12}^N|m_2\rp\\
&= h\lp y_1^N| m_2,v_{12}^N\rp - h\lp y_1^N| m_2,v_{12}^N,m_1\rp + H\lp v_{12}^N|m_2\rp - H\lp v_{12}^N|m_2,m_1\rp\\
&\overset{\aaaa}{=} h\lp h_{11}x_1^N+z_1^N| m_2,v_{12}^N\rp - h\lp z_1^N| m_2,v_{12}^N,m_1 \rp + H\lp v_{12}^N|m_2\rp\\
&\overset{\bbbb}{\le} h\lp h_{11}x_1^N+z_1^N\rp - h\lp z_1^N\rp + H\lp v_{12}^N\rp\\
&\le N\log\lp 1+|h_{11}|^2\rp + N\C_{12} = N\lbp \log\lp1+\SNR_1\rp+\C_{12}\rbp,
\end{align}
where $\epsilon_N\rightarrow 0$ as $N\rightarrow \infty$. (a) is due to the fact that $x_2^N$ is a function of $\lp m_2,v_{12}^N\rp$, and $\lp x_1^N,x_2^N,v_{12}^N\rp$ are all functions of $\lp m_1,m_2\rp$. (b) is due to conditioning reduces entropy and the fact that $z_1^N$ is independent of everything else.

On the other hand, if $R_1$ is achievable
\begin{align}
&N\lp R_1-\epsilon_N\rp\\ 
&\le I\lp m_1; y_1^N| m_2\rp = h\lp y_1^N| m_2\rp - h\lp y_1^N| m_2,m_1\rp\\
&\le h\lp y_1^N\rp - h\lp z_1^N|m_2,m_1\rp = h\lp y_1^N\rp - h\lp z_1^N\rp\\
&\le \max_{\|\rho|\le 1}\lbp N\log\lp 1+|h_{11}|^2+|h_{12}|^2+2\Re\lbp h_{11}h_{12}^*\rho\rbp\rp \rbp\\
&= N\log\lp 1+|h_{11}|^2+|h_{12}|^2+2 |h_{11}||h_{12}|\rp\\
&= N\log\lp 1+\SNR_1+\INR_1+2\sqrt{\SNR_1\INR_1}\rp,
\end{align}
where $\epsilon_N\rightarrow 0$ as $N\rightarrow \infty$.

{\flushleft (2) $R_2$ bound \eqref{bd_R2}:} They follow the same line as the $R_1$ bounds.

{\flushleft (3) $R_1+R_2$ bound \eqref{bd_Sum1} and \eqref{bd_Sum2}:}\par
Let $s_1:= h_{21}x_1+z_2$, and $s_2 := h_{12}x_2+z_1$. If $\lp R_1,R_2\rp$ is achievable, by Fano's inequality,
\begin{align}
&N\lp R_1+R_2-\epsilon_N\rp\\ 
&\le I\lp m_1; y_1^N\rp + I\lp m_2; y_2^N\rp\\
&\le I\lp m_1; y_1^N,s_1^N,v_{12}^N|m_2\rp + I\lp m_2,v_{12}^N;y_2^N\rp\\
&= I\lp m_1; y_1^N,s_1^N|v_{12}^N,m_2\rp + I\lp m_1; v_{12}^N|m_2\rp + h\lp y_2^N\rp - h\lp y_2^N| m_2,v_{12}^N\rp\\
&= h\lp y_1^N,s_1^N|v_{12}^N,m_2\rp - h\lp z_1^N,z_2^N\rp +H\lp v_{12}^N|m_2\rp + h\lp y_2^N\rp - h\lp s_1^N| m_2,v_{12}^N\rp\\
&= h\lp y_1^N|s_1^N,v_{12}^N,m_2\rp + h\lp y_2^N\rp - h\lp z_1^N,z_2^N\rp +H\lp v_{12}^N|m_2\rp\\
&\le h\lp h_{11}x_1^N+z_1^N|h_{21}x_1^N+z_2^N\rp + h\lp y_2^N\rp - h\lp z_1^N,z_2^N\rp +H\lp v_{12}^N\rp\\
&\le N\log\lp 1+\frac{|h_{11}|^2}{1+|h_{21}|^2}\rp + N\log\lp 1+|h_{21}|^2+|h_{22}|^2+2|h_{21}||h_{22}|\rp + N\C_{12}\\
&= N\lbp \log\lp 1+\frac{\SNR_1}{1+\INR_2}\rp + \log\lp 1+\SNR_2+\INR_2+2\sqrt{\SNR_2\INR_2}\rp + \C_{12}\rbp,
\end{align}
where $\epsilon_N\rightarrow 0$ as $N\rightarrow \infty$.

Similarly, 
\begin{align}
&R_1+R_2\\ 
&\le \log\lp 1+\frac{\SNR_2}{1+\INR_1}\rp + \log\lp 1+\SNR_1+\INR_1+2\sqrt{\SNR_1\INR_1}\rp + \C_{21}.
\end{align}

{\flushleft (4) $R_1+R_2$ bound \eqref{bd_Sum3}:}\par
If $\lp R_1,R_2\rp$ is achievable, by Fano's inequality,
\begin{align}
&N\lp R_1+R_2-\epsilon_N\rp\\ 
&\le I\lp m_1; y_1^N\rp + I\lp m_2; y_2^N\rp\\
&\le I\lp m_1; y_1^N| v_{12}^N,v_{21}^N\rp + I\lp m_2; y_2^N| v_{12}^N,v_{21}^N\rp + I\lp m_1; v_{12}^N,v_{21}^N\rp + I\lp m_2; v_{12}^N,v_{21}^N\rp\\
&\overset{\aaaa}{\le} I\lp m_1; y_1^N| v_{12}^N,v_{21}^N\rp + I\lp m_2; y_2^N| v_{12}^N,v_{21}^N\rp + I\lp m_1,m_2; v_{12}^N,v_{21}^N\rp\\
&\le I\lp m_1; y_1^N,s_1^N| v_{12}^N,v_{21}^N\rp + I\lp m_2; y_2^N,s_2^N| v_{12}^N,v_{21}^N\rp + I\lp m_1,m_2; v_{12}^N,v_{21}^N\rp\\
&= h\lp y_1^N,s_1^N| v_{12}^N,v_{21}^N\rp - h\lp y_1^N,s_1^N| v_{12}^N,v_{21}^N,m_1\rp\\&\quad + h\lp y_2^N,s_2^N| v_{12}^N,v_{21}^N\rp - h\lp y_2^N,s_2^N| v_{12}^N,v_{21}^N,m_2\rp + H\lp v_{12}^N,v_{21}^N\rp\\
&= h\lp y_1^N,s_1^N| v_{12}^N,v_{21}^N\rp - h\lp s_2^N,z_2^N| v_{12}^N,v_{21}^N,m_1\rp\\&\quad + h\lp y_2^N,s_2^N| v_{12}^N,v_{21}^N\rp - h\lp s_1^N,z_1^N| v_{12}^N,v_{21}^N,m_2\rp + H\lp v_{12}^N,v_{21}^N\rp\\
&\overset{\bbbb}{=} h\lp y_1^N,s_1^N| v_{12}^N,v_{21}^N\rp - h\lp s_2^N| v_{12}^N,v_{21}^N,m_1\rp - h\lp z_2^N\rp\\&\quad + h\lp y_2^N,s_2^N| v_{12}^N,v_{21}^N\rp - h\lp s_1^N| v_{12}^N,v_{21}^N,m_2\rp - h\lp z_1^N\rp + H\lp v_{12}^N,v_{21}^N\rp\\
&\overset{\cccc}{=} h\lp y_1^N,s_1^N| v_{12}^N,v_{21}^N\rp - h\lp s_2^N| v_{12}^N,v_{21}^N\rp - h\lp z_2^N\rp\\&\quad + h\lp y_2^N,s_2^N| v_{12}^N,v_{21}^N\rp - h\lp s_1^N| v_{12}^N,v_{21}^N\rp - h\lp z_1^N\rp + H\lp v_{12}^N,v_{21}^N\rp\\
&= h\lp y_1^N| s_1^N,v_{12}^N,v_{21}^N\rp + h\lp y_2^N| s_2^N,v_{12}^N,v_{21}^N\rp - h\lp z_1^N\rp - h\lp z_2^N\rp + H\lp v_{12}^N,v_{21}^N\rp\\
&\le h\lp y_1^N| s_1^N\rp + h\lp y_2^N| s_2^N\rp - h\lp z_1^N\rp - h\lp z_2^N\rp + H\lp v_{12}^N\rp + H\lp v_{21}^N\rp\\
&\le N \max_{|\rho| \le 1} \lbp\begin{array}{l}
\log\lp 1+\frac{\SNR_1+2\Re\lbp h_{11}h_{12}^*\rho\rbp}{1+\INR_2} + \frac{1+\lp1-|\rho|^2\rp\INR_2}{1+\INR_2}\INR_1\rp\\
+\log\lp 1+\frac{\SNR_2+2\Re\lbp h_{21}h_{22}^*\rho\rbp}{1+\INR_1} + \frac{1+\lp1-|\rho|^2\rp\INR_1}{1+\INR_1}\INR_2\rp
\end{array}\rbp  + N\C_{12} + N\C_{21},\notag{}
\end{align}
where $\epsilon_N\rightarrow 0$ as $N\rightarrow \infty$. (a) is due to the fact that $m_1$ and $m_2$ are independent. (b) is due to the fact that $z_1^N$ and $z_2^N$ are independent to everything else, respectively. (c) is due to the following fact regarding the Markov relations:

{\flushleft (5) $R_1+R_2$ bound \eqref{bd_Sum4}:}\par
If $\lp R_1,R_2\rp$ is achievable, by Fano's inequality,
\begin{align}
&N\lp R_1+R_2-\epsilon_N\rp\\ 
&\le I\lp m_1; y_1^N\rp + I\lp m_2; y_2^N\rp \le I\lp m_1; y_1^N,y_2^N\rp + I\lp m_2; y_1^N,y_2^N\rp\\
&\le I\lp m_1; y_1^N,y_2^N| m_2\rp + I\lp m_2; y_1^N,y_2^N\rp = I\lp m_1,m_2; y_1^N,y_2^N\rp\\
&= h\lp y_1^N,y_2^N\rp - h\lp z_1^N,z_2^N\rp\\
&\le  N \max_{|\rho|\le1}\lbp\log\lp \begin{array}{l}1+\SNR_1+\INR_1+\SNR_2+\INR_2+2\Re\lbp h_{11}h_{12}^*\rho\rbp\\+2\Re\lbp h_{21}h_{22}^*\rho\rbp+\lp1-|\rho|^2\rp|h_{11}h_{22}-h_{12}h_{21}|^2\end{array}\rp\rbp,
\end{align}
where $\epsilon_N\rightarrow 0$ as $N\rightarrow \infty$.

{\flushleft (6) $2R_1+R_2$ bound \eqref{bd_SlopeTwo1} and $R_1+2R_2$ bound \eqref{bd_SlopeHalf1}:}\par
If $(R_1,R_2)$ is achievable, by Fano's inequality,
\begin{align}
&N\lp 2R_1+R_2 - \epsilon_N\rp\\
&\le 2I\lp m_1; y_1^N\rp +  I\lp m_2; y_2^N\rp\\
&\le  I\lp m_1; y_1^N\rp +  I\lp m_1; y_1^N, s_1^N, v_{12}^N, v_{21}^N|m_2\rp +  I\lp m_2; y_2^N,s_2^N, v_{12}^N, v_{21}^N\rp\\
&\le I\lp m_1, v_{12}^N, v_{21}^N;  y_1^N\rp +  I\lp m_1; y_1^N, s_1^N|v_{12}^N, v_{21}^N,m_2\rp +  I\lp m_2; y_2^N,s_2^N|v_{12}^N, v_{21}^N\rp\\&\quad + I\lp m_1; v_{12}^N,v_{21}^N|m_2\rp + I\lp m_2; v_{12}^N,v_{21}^N\rp\\
&= h\lp y_1^N\rp - h\lp s_2^N| m_1,v_{12}^N, v_{21}^N\rp + h\lp h_{11}x_1^N+z_1^N,s_1^N| m_2, v_{12}^N, v_{21}^N\rp\\&\quad - h\lp z_1^N,z_2^N\rp+ h\lp y_2^N,s_2^N | v_{12}^N,v_{21}^N\rp -h\lp s_1^N,z_1^N| m_2, v_{12}^N,v_{21}^N\rp\\&\quad + I\lp m_1,m_2; v_{12}^N,v_{21}^N\rp\\
&\overset{\aaaa}{=} h\lp y_1^N\rp - h\lp s_2^N| v_{12}^N, v_{21}^N\rp + h\lp h_{11}x_1^N+z_1^N| s_1^N, m_2, v_{12}^N, v_{21}^N\rp\\&\quad + h\lp y_2^N,s_2^N | v_{12}^N,v_{21}^N\rp + H\lp v_{12}^N,v_{21}^N\rp - 2h\lp z_1^N\rp - h\lp z_2^N\rp\\
&= h\lp y_1^N\rp + h\lp h_{11}x_1^N+z_1^N| s_1^N, m_2, v_{12}^N, v_{21}^N\rp + h\lp y_2^N| s_2^N, v_{12}^N,v_{21}^N\rp\\&\quad + H\lp v_{12}^N,v_{21}^N\rp - 2h\lp z_1^N\rp - h\lp z_2^N\rp\\
&\le N\lbp\begin{array}{l} \log\lp 1+\SNR_1+\INR_1+2\sqrt{\SNR_1\INR_1}\rp + \log\lp1+\frac{\SNR_1}{1+\INR_2}\rp\\ + \log\lp 1+\frac{\SNR_2+2\sqrt{\SNR_2\INR_2}}{1+\INR_1}+\INR_2\rp + \C_{12} + \C_{21}\end{array}\rbp,
\end{align}
where $\epsilon_N\rightarrow 0$ as $N\rightarrow \infty$. (a) is due to the Markovity in Fact \ref{fact_Markov}.

Similar arguments work for $R_1+2R_2$ bound \eqref{bd_SlopeHalf1}.

{\flushleft (7) $2R_1+R_2$ bound \eqref{bd_SlopeTwo2} and $R_1+2R_2$ bound \eqref{bd_SlopeHalf2}:}\par
Using the intuition from the study of linear deterministic channel, we give the following side information to receiver 2:
\begin{align}
\wtild{y}_2^N := h_{22}x_2^N + \wtild{z}_2^N, 
\end{align}
where $\wtild{z}_2 \sim \mcal{CN}\lp 0, 1+\INR_2\rp$, i.i.d. over time and is independent of everything else.

Now, if $(R_1,R_2)$ is achievable, by Fano's inequality,
\begin{align}
&N\lp 2R_1+R_2 - \epsilon_N\rp\\
&\le 2I\lp m_1; y_1^N\rp +  I\lp m_2; y_2^N\rp\\
&\le I\lp m_1; y_1^N, s_1^N | m_2, v_{12}^N\rp + I\lp m_1; v_{12}^N|m_2\rp + I\lp m_1; y_1^N\rp + I\lp m_2; \wtild{y}_{2}^N, y_{2}^N\rp\\
&= h\lp h_{11}x_1^N + z_1^N, s_1^N| m_2, v_{12}^N\rp - h\lp z_1^N,z_2^N\rp+ I\lp m_1; y_1^N\rp + I\lp m_2; \wtild{y}_{2}^N\rp\\&\quad + I\lp m_2; y_{2}^N| \wtild{y}_{2}^N\rp + I\lp m_1; v_{12}^N|m_2\rp\\
&\overset{\aaaa}{\le} h\lp h_{11}x_1^N + z_1^N, s_1^N| m_2, v_{12}^N\rp - h\lp z_1^N,z_2^N\rp+ I\lp m_1,m_2; y_1^N, \wtild{y}_{2}^N\rp\\&\quad + I\lp m_2, v_{12}^N; y_{2}^N| \wtild{y}_{2}^N\rp + I\lp m_1; v_{12}^N|m_2\rp\\
&= h\lp h_{11}x_1^N + z_1^N, s_1^N| m_2, v_{12}^N\rp - h\lp z_1^N,z_2^N\rp+ h\lp y_1^N, \wtild{y}_{2}^N\rp - h\lp z_1^N, \wtild{z}_2^N\rp\\&\quad + h\lp y_{2}^N| \wtild{y}_{2}^N \rp - h\lp y_{2}^N| \wtild{y}_{2}^N, m_2, v_{12}^N\rp + H\lp v_{12}^N|m_2\rp\\
&= h\lp h_{11}x_1^N + z_1^N, s_1^N| m_2, v_{12}^N\rp+ h\lp y_1^N, \wtild{y}_{2}^N\rp + h\lp y_{2}^N| \wtild{y}_{2}^N \rp - h\lp s_{1}^N| \wtild{z}_{2}^N, m_2, v_{12}^N\rp\\&\quad + H\lp v_{12}^N|m_2\rp - h\lp z_1^N,z_2^N\rp - h\lp z_1^N, \wtild{z}_2^N\rp\\
&\overset{\bbbb}{=} h\lp h_{11}x_1^N + z_1^N, s_1^N| m_2, v_{12}^N\rp+ h\lp y_1^N, \wtild{y}_{2}^N\rp + h\lp y_{2}^N| \wtild{y}_{2}^N \rp - h\lp s_{1}^N| m_2, v_{12}^N\rp\\&\quad + H\lp v_{12}^N|m_2\rp - h\lp z_1^N,z_2^N\rp - h\lp z_1^N, \wtild{z}_2^N\rp\\
&= h\lp h_{11}x_1^N + z_1^N| s_1^N, m_2, v_{12}^N\rp+ h\lp y_1^N, \wtild{y}_{2}^N\rp + h\lp y_{2}^N| \wtild{y}_{2}^N \rp + H\lp v_{12}^N|m_2\rp\\&\quad - h\lp z_1^N,z_2^N\rp - h\lp z_1^N, \wtild{z}_2^N\rp\\
&\le N\lbp\begin{array}{l} \log\lp 1+\frac{\SNR_1}{1+\INR_2}\rp + \log\lp 2+\INR_2\rp -\log\lp 1+\INR_2\rp\\
+ \log\lp \begin{array}{l}1+\SNR_1+\INR_1+\SNR_2+\INR_2+\SNR_1\SNR_2\\+\INR_1\INR_2+\SNR_1\INR_2 + 2\lp1+\INR_2\rp\sqrt{\SNR_1\INR_1}\end{array}\rp + \C_{12}\end{array}\rbp,
\end{align}
where $\epsilon_N\rightarrow 0$ as $N\rightarrow \infty$. (a) is due to a simple fact that $I\lp m_1; y_1^N\rp + I\lp m_2; \wtild{y}_{2}^N\rp \le I\lp m_1,m_2; y_1^N, \wtild{y}_{2}^N\rp$ and that conditioning reduces entropy. (b) holds since $\wtild{z}_{2}^N$ is independent of $\lp m_2,v_{12}^N\rp$ and $s_1^N$.

Similar arguments work for $R_1+2R_2$ bound \eqref{bd_SlopeHalf2}.
\end{proof}

\section{Proof of Theorem \ref{thm_Reciprocity}}\label{app_PfReciprocity}

In \cite{WangTse_09}, we characterize the capacity region of Gaussian interference channel with conferencing receivers to within 2 bits per user. Hence by Theorem \ref{thm_Gap}, we only need to compare the outer bounds. Note that for the reciprocal channel, its channel parameters are
\begin{align}
&\SNR_1' = \SNR_1,\ \SNR_2' = \SNR_2;\
\INR_1' = \INR_2,\ \INR_2' = \INR_1;\
{\C_{12}}' = \C_{21},\ {\C_{21}}' = \C_{12}.
\end{align}


{\flushleft \underline{Outer bounds for the reciprocal channel} \cite{WangTse_09}}
\begin{align}
R_1 &\le \min\lbp \log\lp1+\SNR_1\rp+\C_{12},\log\lp1+\SNR_1+\INR_1\rp \rbp\\
R_2 &\le \min\lbp \log\lp1+\SNR_2\rp+\C_{21},\log\lp1+\SNR_2+\INR_2\rp \rbp\\
R_1+R_2 &\le \log\lp1+\INR_2+\frac{\SNR_1}{1+\INR_1}\right) + \log\left(1+\INR_1+\frac{\SNR_2}{1+\INR_2}\rp + \C_{21}+\C_{12} \\
R_1+R_2 &\le \log\left(1+\SNR_2+\INR_1\right) + \log\left(1+\frac{\SNR_1}{1+\INR_1}\right) + \C_{21}\\
R_1+R_2 &\le \log\left(1+\SNR_1+\INR_2\right) + \log\left(1+\frac{\SNR_2}{1+\INR_2}\right) + \C_{12}\\
R_1+R_2 &\le \log\lp 1+\SNR_1+\SNR_2+\INR_1+\INR_2 + |h_{11}h_{22} - h_{12}h_{21}|^2 \rp \\
2R_1 + R_2 &\le \lbp \begin{array}{l} \log\left(1+\INR_1+\frac{\SNR_2}{1+\INR_2}\right) + \log\left(1+\frac{\SNR_1}{1+\INR_1}\right)\\ + \log\lp 1+ \SNR_1+\INR_2\rp + \C_{21}+\C_{12}\end{array}\rbp\\
R_1 + 2R_2 & \le \lbp \begin{array}{l} \log\left(1+\INR_2+\frac{\SNR_1}{1+\INR_1}\right) + \log\left(1+\frac{\SNR_2}{1+\INR_2}\right)\\ + \log\lp 1+ \SNR_2+\INR_2\rp + \C_{12}+\C_{21}\end{array}\rbp\\
2R_1+R_2 &\le \lbp \begin{array}{l} \log\lp 1+ \frac{\SNR_2}{1+\INR_2} + \INR_1 + \SNR_1 + \frac{\INR_2}{1+\INR_2} + \frac{|h_{11}h_{22} - h_{12}h_{21}|^2}{1+\INR_2}\rp\\
+ \log\lp1+\SNR_1+\INR_2\rp + \C_{12}\end{array}\rbp\\
R_1+2R_2 &\le \lbp \begin{array}{l} \log\lp 1+ \frac{\SNR_1}{1+\INR_1} + \INR_2 + \SNR_2 + \frac{\INR_1}{1+\INR_1} + \frac{|h_{11}h_{22} - h_{12}h_{21}|^2}{1+\INR_1}\rp\\
+ \log\lp1+\SNR_2+\INR_1\rp + \C_{21}\end{array}\rbp
\end{align}


{\flushleft (1) Bounds on $R_1$ and $R_2$}:\par
Note that 
\begin{align}
&\log\lp1+\SNR_1+\INR_1\rp\\
&\le \log\lp 1+\SNR_1+\INR_1+2\sqrt{\SNR_1\INR_1}\rp\\
&\le \log\lp1+\SNR_1+\INR_1\rp+1.
\end{align}
Hence the gap is at most 1 bit.

{\flushleft (2) Bounds on $R_1+R_2$}:\par
Note that:
\begin{itemize}
\item[(a)]
\begin{align}
&\log\lp1+\INR_2+\frac{\SNR_1}{1+\INR_1}\rp +  \log\lp1+\INR_1+\frac{\SNR_2}{1+\INR_2}\rp\\
&= \log\lp \frac{\lp1+\INR_1\rp\lp1+\INR_2\rp+\SNR_1}{1+\INR_1}\rp + \log\lp \frac{\lp1+\INR_2\rp\lp1+\INR_1\rp+\SNR_2}{1+\INR_2}\rp\\
&= \log\lp \frac{\lp1+\INR_1\rp\lp1+\INR_2\rp+\SNR_1}{1+\INR_2}\rp + \log\lp \frac{\lp1+\INR_2\rp\lp1+\INR_1\rp+\SNR_2}{1+\INR_1}\rp\\
&= \log\lp1+\INR_1+\frac{\SNR_1}{1+\INR_2}\rp +  \log\lp1+\INR_2+\frac{\SNR_2}{1+\INR_1}\rp\\
&\le\begin{array}{l} 
\log\lp 1+\frac{\SNR_1+2\sqrt{\SNR_1\INR_1}}{1+\INR_2} + \INR_1\rp
+\log\lp 1+\frac{\SNR_2+2\sqrt{\SNR_2\INR_2}}{1+\INR_1} + \INR_2\rp\end{array}\\
&\le \log\lp1+\INR_1+\frac{2\SNR_1+\INR_1}{1+\INR_2}\rp +  \log\lp1+\INR_2+\frac{2\SNR_2+\INR_2}{1+\INR_1}\rp\\
&\le \log\lp1+\INR_1+\frac{\SNR_1}{1+\INR_2}\rp +  \log\lp1+\INR_2+\frac{\SNR_2}{1+\INR_1}\rp + 2
\end{align}

\item[(b)]
\begin{align}
&\log\left(1+\SNR_2+\INR_1\right) + \log\left(1+\frac{\SNR_1}{1+\INR_1}\right)\\
&= \log\left(1+\frac{\SNR_2}{1+\INR_1}\right) + \log\left(1+\SNR_1+\INR_1\right)\\
&\le \log\lp 1+\frac{\SNR_2}{1+\INR_1}\rp + \log\lp 1+\SNR_1+\INR_1+2\sqrt{\SNR_1\INR_1}\rp\\
&\le \log\lp 1+\frac{\SNR_2}{1+\INR_1}\rp + \log\lp 1+\SNR_1+\INR_1\rp+1
\end{align}

\item[(c)]
\begin{align}
&\log\lp 1+\SNR_1+\SNR_2+\INR_1+\INR_2 + |h_{11}h_{22} - h_{12}h_{21}|^2 \rp\\
&\le \log\lp \begin{array}{l}1+\SNR_1+\INR_1+\SNR_2+\INR_2+2\sqrt{\SNR_1\INR_1}\\+2\sqrt{\SNR_2\INR_2}+|h_{11}h_{22}-h_{12}h_{21}|^2\end{array}\rp\\
&\le \log\lp 1+2\SNR_1+2\INR_1+2\SNR_2+2\INR_2+|h_{11}h_{22}-h_{12}h_{21}|^2\rp\\
&\le \log\lp 1+\SNR_1+\INR_1+\SNR_2+\INR_2+|h_{11}h_{22}-h_{12}h_{21}|^2\rp + 1
\end{align}
\end{itemize}
Hence the gap is at most 2 bits.

{\flushleft (3) Bounds on $2R_1+R_2$ and $R_1+2R_2$}:\par
Note that:
\begin{itemize}
\item[(a)]
\begin{align}
&\log\left(1+\INR_1+\frac{\SNR_2}{1+\INR_2}\right) + \log\left(1+\frac{\SNR_1}{1+\INR_1}\right) + \log\lp 1+ \SNR_1+\INR_2\rp\\
&= \log\left(1+\INR_2+\frac{\SNR_2}{1+\INR_1}\right) + \log\left(1+\SNR_1+\INR_1\right) + \log\lp 1+ \frac{\SNR_1}{1+\INR_2}\rp\\
&\le \log\lp 1+\frac{\SNR_2+2\sqrt{\SNR_2\INR_2}}{1+\INR_1}+\INR_2\rp + \log\lp1+\frac{\SNR_1}{1+\INR_2}\rp\\ &\quad + \log\lp 1+\SNR_1+\INR_1+2\sqrt{\SNR_1\INR_1}\rp\\
&\le \log\left(1+\INR_2+\frac{\SNR_2}{1+\INR_1}\right) + 1 + \log\lp 1+ \frac{\SNR_1}{1+\INR_2}\rp\\
&\quad + \log\left(1+\SNR_1+\INR_1\right) +1
\end{align}

\item[(b)]
\begin{align}
&\log\lp 1+ \frac{\SNR_2}{1+\INR_2} + \INR_1 + \SNR_1 + \frac{\INR_2}{1+\INR_2} + \frac{|h_{11}h_{22} - h_{12}h_{21}|^2}{1+\INR_2}\rp\\
&\quad + \log\lp1+\SNR_1+\INR_2\rp\\
&= \log\lp \begin{array}{l}1+\SNR_1+\INR_1+\SNR_2+2\INR_2\\+\SNR_1\INR_2+\INR_1\INR_2+|h_{11}h_{22}-h_{12}h_{21}|^2\end{array}\rp + \log\lp1+\frac{\SNR_1}{1+\INR_2}\rp\\
&\le \log\lp \begin{array}{l}1+\SNR_1+\INR_1+\SNR_2+2\INR_2\\+\SNR_1\INR_2+\INR_1\INR_2+2\lp\SNR_1\SNR_2+\INR_1\INR_2\rp\end{array}\rp\\&\quad + \log\lp1+\frac{\SNR_1}{1+\INR_2}\rp\\
&\le \log\lp \begin{array}{l}1+\SNR_1+\INR_1+\SNR_2+\INR_2+\SNR_1\SNR_2\\+\INR_1\INR_2+\SNR_1\INR_2 + 2\lp1+\INR_2\rp\sqrt{\SNR_1\INR_1}\end{array}\rp\\
&\quad +\log\lp 1+\frac{\SNR_1}{1+\INR_2}\rp + 1\\
&\le \log\lp \begin{array}{l}1+2\SNR_1+2\INR_1+\SNR_2+\INR_2+\SNR_1\SNR_2\\+2\INR_1\INR_2+2\SNR_1\INR_2\end{array}\rp\\
&\quad +\log\lp 1+\frac{\SNR_1}{1+\INR_2}\rp + 1\\
&\le \log\lp \begin{array}{l}1+2\SNR_1+2\INR_1+\SNR_2+\INR_2+2\SNR_1\INR_2\\+4|h_{11}h_{22}-h_{12}h_{21}|^2+8\INR_1\INR_2\end{array}\rp\\
&\quad +\log\lp 1+\frac{\SNR_1}{1+\INR_2}\rp + 1\\
&\le \log\lp \begin{array}{l}1+\SNR_1+\INR_1+\SNR_2+2\INR_2\\+\SNR_1\INR_2+\INR_1\INR_2+|h_{11}h_{22}-h_{12}h_{21}|^2\end{array}\rp\\
&\quad +\log\lp 1+\frac{\SNR_1}{1+\INR_2}\rp + 4
\end{align}
\end{itemize}

Hence the gap is at most 4 bits.

In summary, we have
\begin{align}
\ol{\mcal{C}}_{\rm{Rx}} \subset \ol{\mcal{C}}_{\rm{Tx}} \subset \ol{\mcal{C}}_{\rm{Rx}}\oplus [0,\tau]\times[0,\tau],
\end{align}
where $\tau = 4/3$ bits.


\end{document}